\documentclass[twocolumn,numberedappendix,twocolappendix,appendixfloats]{openjournal_dhayaa}

\usepackage{amsmath}
\usepackage{fontawesome5}
\usepackage{color}
\usepackage{xcolor, multirow}

\usepackage{natbib}
\setcitestyle{aysep={}}

\usepackage[colorlinks=true
  ,urlcolor=blue
  ,anchorcolor=blue
  ,citecolor=blue
  ,filecolor=blue
  ,linkcolor=blue
  ,menucolor=blue
  ,linktocpage=true
  ,pdfproducer=medialab
  ,pdfa=true
]{hyperref}

\usepackage{etoolbox}
\makeatletter

\usepackage{graphicx}	
\usepackage{amsmath}	

\usepackage{fontawesome5}
\usepackage{color}
\usepackage{xcolor}

\newcommand{\eg}{{\sl e.g.}, }

\newcommand{\Rtwohc}{R_{\rm 200c}}
\newcommand{\Mtwohc}{M_{\rm 200c}}

\newcommand{\Mstar}{M_{\rm star}}
\newcommand{\Mgas}{M_{\rm gas}}

\newcommand{\Lproj}{L_{\rm p}}
\newcommand{\rp}{r_{\rm p}}

\newcommand{\msol}{\ensuremath{\, {\rm M}_\odot}}    
\newcommand{\msun}{\ensuremath{\, {\rm M}_\odot}} 
         
\newcommand{\mpc}{\ensuremath{\, {\rm Mpc}}}         
\newcommand{\gpc}{\ensuremath{\, {\rm Gpc}}}
     
\newcommand{\dln}{\ensuremath{{\rm d \ln}}}
\newcommand{\GHz}{\,\, {\rm GHz}}
\newcommand{\nhat}{\mathbf{\hat{n}}}

\newcommand{\ctwohc}{\ensuremath{c_{\rm 200c}}}

\definecolor{orcidlogocol}{HTML}{A6CE39}
\definecolor{purple}{RGB}{128, 0, 128}

\newcommand{\OrcidID}[1]{ \href[urlcolor = red]{https://orcid.org/#1}{\textcolor{lightgray}{\faOrcid}}}
\newcommand{\OrcidIDName}[2]{\href{https://orcid.org/#1}{#2}}

\defcitealias{Pandey2024godmax}{P24}
\defcitealias{Schneider2019Baryonification}{S19}

\newcommand*{\vcenteredhbox}[1]{\begingroup
\setbox0=\hbox{#1}\parbox{\wd0}{\box0}\endgroup}

\begin{document}

\title[Map-level lensing and tSZ modelling]{Map-level baryonification: Efficient modelling of higher-order correlations in the weak lensing and thermal Sunyaev-Zeldovich fields}

\author{\OrcidIDName{0000-0003-3312-909X}{Dhayaa Anbajagane}
(\vcenteredhbox{\includegraphics[height=1.2\fontcharht\font`\B]{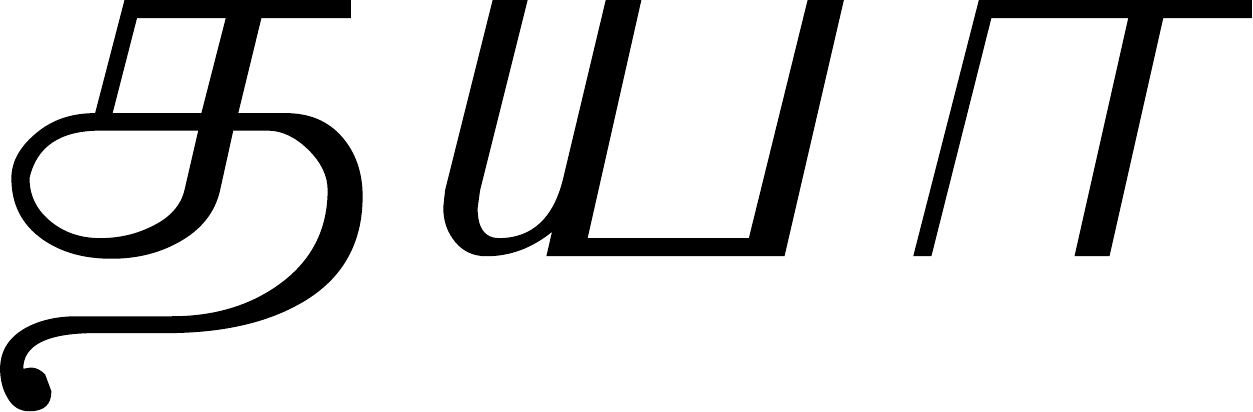}})$^{1,\,2}$}
\author{\OrcidIDName{000-0001-5780-637X}{Shivam Pandey}$^{3,\,4}$}
\author{\OrcidIDName{0000-0002-7887-0896}{Chihway Chang}$^{1,\,2}$}

\email{$^{\star}$dhayaa@uchicago.edu}

\affiliation{$^{1}$ Department of Astronomy and Astrophysics, University of Chicago, Chicago, IL 60637, USA}
\affiliation{$^{2}$ Kavli Institute for Cosmological Physics, University of Chicago, Chicago, IL 60637, USA}
\affiliation{$^{3}$ Department of Physics, Columbia University, 538 West 120th Street, New York, NY, USA 10027, USA}
\affiliation{$^{4}$ Columbia Astrophysics Laboratory, Columbia University, 550 West 120th Street, New York, NY 10027, USA\\}

\begin{abstract}
Semi-analytic methods can generate baryon-corrected fields from N-body simulations (``baryonification'') and are rapidly becoming a ubiquitous tool in modeling structure formation on non-linear scales. We extend this formalism to consistently model the weak lensing and thermal Sunyaev-Zeldovich (tSZ) fields directly on the full-sky, with an emphasis on higher-order correlations. We use the auto- and cross- $N$th-order moments, with $N \in \{2, 3, 4\}$, as a summary statistic of the lensing and tSZ fields, and show that our model can jointly fit these statistics measured in \textsc{IllustrisTNG} to within measurement uncertainties, for scales above $\gtrsim 1 \mpc$ and across multiple redshifts. The model predictions change only minimally when including additional information from secondary halo properties, such as halo concentration and ellipticity. Each individual moment is dependent on halos of different mass ranges and has different sensitivities to the model parameters. A simulation-based forecast on the \textsc{Ulagam} suite shows that the combination of all moments, measured from current and upcoming lensing and tSZ surveys, can jointly constrain cosmology and baryons to high precision. The lensing and tSZ field are sensitive to different combinations of the baryonification parameters, with degeneracy directions that are often orthogonal, and the combination of the two fields leads to significantly better constraints on both cosmology and astrophysics. Our pipeline for map-level baryonification is publicly available at \url{https://github.com/DhayaaAnbajagane/BaryonForge}.
\end{abstract}


\section{Introduction}

Measurements of weak lensing, which is a direct probe of the cosmic density field, have provided some of the best constraints on the properties of the Universe, both on the initial conditions and on their subsequent evolution \citep{DES2022Y3, Asgari2021KidsWL, Secco2022Shear, Amon2022Y3shear, More2023HSCY3, Omori:2023:CMBLensing, Chang:2023:CMBLensing, Madhavacheril:2024:DR6_lensing}. These constraints are expected to significantly improve with the advent of surveys like the Dark Energy Spectroscopic Instrument \cite{DESI2016Science}, Vera C. Rubin Observatory \citep{LSST2018SRD}, the Euclid mission \citep{Euclid}, the Roman space telescope \citep{Spergel:2015:Roman} and more, all of which will provide vastly richer datasets and therefore enable measurements of the small-scale density field with improved statistical precision. 

However, an ongoing limitation in using such small-scale measurements is modeling the impact of baryons (``baryonic imprints'') on the density field. Baryons, which constitute $\approx 15\%$ of the total matter in our Universe, with the rest being accounted for by dark matter (DM), experience a vast set of astrophysical process across a range of scales. Such processes alter the distribution and thermodynamics of these baryons, and also the DM phase space via the gravitational coupling between the two components \citep[\eg][]{Gnedin2004AdiabaticContraction, Abadi2010ShapesBaryons, Duffy2010BaryonDmProfileDensity, Anbajagane2022Baryons, Shao2023Baryons}. The signatures of these processes can be prevalent even on quasi-linear scales (roughly a few times a halo radius) and will be a significant component of any small-scale signal (i.e. the density near/within a halo). Their presence can bias any cosmological analyses that do not account for this effect \citep[\eg][]{Secco2022Shear, Amon2022S8Baryons, Gatti2022MomentsDESY3, Anbajagane2023CDFs}. This limitation in scales is further accentuated by the fact that many novel cosmological signatures are imprinted onto these small scales. Thus, efficient models of these baryonic imprints have the potential to not only improve the constraining power for current models of interest, such as $\Lambda$CDM and $w$CDM, but also open the opportunity to constrain extended models like those of primordial non-gaussianities \citep{Coulton2022QuijotePNG, Anbajagane2023Inflation, Jung2023fNLHMFQuijote, Sam:2024:LensingFNL}.

The most precise model of baryonic imprints comes from studying the formation of structure in various hydrodynamical simulations which model the evolution of gas, stars, black holes etc. in addition to the DM \citep[\eg][]{McCarthy2017BAHAMAS, Springel2018FirstClustering}. The simulations have a finite resolution scale that is larger than many relevant astrophysical processes (such as supernovae explosions, or supermassive black hole accretion, to name a few) and must therefore implement these processes as ``sub-grid'' physics that approximate their effects \citep[see][for a review]{Vogelsberger2020Hydro}. The subgrid models of each simulation vary in the equations that are solved, and in how the equations are parameterized. Studies of these simulations, however, have showcased the variety of predictions manifesting from these different, but often equally realistic, choices in the models of galaxy formation \citep[\eg][]{Anbajagane2020StellarProp, Lim2021GasProp, Lee2022rSZ, Cui2022GIZMO, Stiskalek2022TNGHorizon, Anbajagane2022Baryons, Anbajagane2022GalaxyVelBias, Shao2022Baryons, Gebhardt2023CamelsAGNSN}. The type and amplitude of these differences vary significantly depending on the exact properties --- such as the gas mass/temperature/pressure, the stellar mass/age/metallicity etc. --- being studied and the regime in mass, redshift, environment etc. being explored. Thus, while the simulations are valuable in capturing the range of possible baryonic imprints, one cannot use them to robustly model these imprints in the density field. Furthermore, studies on the thermodynamic properties of gas also find differences between the measurements from data and the predictions from these hydrodynamic simulations \citep[\eg][]{Hill2018tSZxGroups, Amodeo2021ACTxBOSS, Pandey2021DESxACT, Anbajagane2022Shocks}.

Given the current challenges in modelling baryons in simulations, it is advantageous to instead use a phenomenological model. Such a model does not require specifying any sub-grid physics and instead requires inputs that are observables, or are quantities directly obtained from observables, of the different baryonic matter components. \textit{Baryonification} is a method through which N-body simulations --- which evolve collisionless matter (dark matter) under gravity and accurately capture all non-linear evolution --- are modified through phenomonological models such that the new density field mimics one that has baryonic imprints \citep{Schneider:2015:Baryons, Schneider2019Baryonification}. This method has been used extensively to model the matter power spectrum with baryonic imprints, and has been shown to have the flexibility needed to capture the behaviors ranging across many different simulations \citep{Schneider2019Baryonification, Giri2021Baryon, Arico:2021:Bacco}. It has already been used to analyse 2-point correlation functions measured in widefield surveys \citep{Chen:2023:BaryonsWL, Arico2023BaryonY3, Bigwood:2024:BaryonsWLkSZ}.

The fiducial baryonification method is performed by displacing the positions of particles in the N-body simulations to alter the overall density distribution in the volume. This procedure has been computationally tractable for building emulators that translate the model parameters to prediction a given statistic in real space, \eg the 3D power spectrum such as done in \citet{Giri2021Baryon, Arico:2021:Bacco}. These emulators can then be used in standard analysis pipelines to predict, for example, the weak lensing correlation functions \citep[\eg][]{Arico2023BaryonY3}. However, surveys are rapidly adopting simulation-based analyses as a way to accurately forward model systematic effects, and also to incorporate more information from the field \citep[\eg][]{Gatti2023SC, Gatti:2024:LFIValidation, Jeffrey:2024:LFIResult, Joachim:2024:Kids, Cheng:2024:HSC_WST}. Some work has been done to show the baryonification model generates fields that accurately model baryonic imprints on higher-order statistics, such as the density field bispectrum \citep{Arico:2021:BkBaryons} and the peaks of the weak lensing field \citep{Lee2023Peaks}. These higher-order statistics extend upon the simpler, two-point statistics --- which describe the correlation between two points in space, \eg the power spectrum --- by also extracting correlations between three or more points \citep[\eg][]{Fluri2019DeepLearningKIDS, Gatti2020Moments, Zurcher2021WLForecast, Fluri2022wCDMKIDS, Euclid2023NGCov, Anbajagane2023CDFs, Gatti:2024:LFIResults, Jeffrey:2024:LFIResult}.

The use of baryonification in simulation-based analyses of surveys is limited by the computational expense of the method. Survey datasets frequently require lightcone maps covering a large sky fraction for their predictions (upcoming surveys will need $\approx 15,000 \deg^2$ coverage) and these maps are constructed from $\mathcal{O}(10^2)$ 3D snapshots/boxes in a simulation \citep[\eg][]{Kacprzak2023Cosmogrid, Anbajagane2023Inflation, Jeffrey:2024:LFIResult}. Applying baryonification on each box is an computationally expensive procedure --- and also a memory intensive one, since it requires storing the particles of each snapshot --- given analyses often require $\mathcal{O}(10^3 - 10^4)$ such simulations. An attractive alternative then is to perform baryonification directly on pixels of 2D maps. This has been explored by \citet{Fluri2019DeepLearningKIDS} for performing simulation-based inference of weak lensing data from the Kilo-Degree Survey \citep{Kuijken:2015:KiDs}. An aspect of our work focuses on updating this method, by robustly accounting for the pixel window functions and line-of-sight projections and then validating its performance on simulations.

We have so far only discussed the total density field, whereas other probes --- often more closely linked to the baryonic matter components --- are actively used to constrain astrophysics and cosmology. Of these, one of the most commonly used and actively studied ones is the thermal Sunyaev-Zeldovich (tSZ) field \citep{Sunyaev1972SZEffect}, observed using millimeter surveys such as the South Pole Telescope \citep{Carlstrom2011, Benson2014SPT3G}, the Atacama Cosmology Telescope \citep{ACT:2007, ACT:2016}, and the \textit{Planck} satellite \citep{Planck:2020:LegacyOverview}. The tSZ effect is the inverse Compton scattering of CMB photons with energetic electrons along the line of sight. It probes the integrated gas pressure and therefore the thermodynamics of baryons in and around the halo \citep[see][for reviews]{Carlstrom2002SZReview, Mroczkowski2019SZreview}. As a result, it is advantageous to include the tSZ probe in analyses, both for its cosmological information --- as it, like the total matter field, is sensitive to the initial conditions and their subsequent evolution --- and even more so for its information on the gas thermodynamics. However, pursuing this direction requires a consistent, common model that predicts the impact of astrophysical processes on both the distribution of matter and the thermodynamics of baryons. Studies of baryonification have thus far been exclusively on the density field \citep{Schneider2019Baryonification, Giri2021Baryon, Arico:2021:Bacco, Lee2023Peaks, Arico2023BaryonY3}, with the inclusion of tSZ only recently being explored \citep{Arico:2024:tSZ}.\footnote{See Section \ref{sec:baryonify:tSZ} for differences between the tSZ model of this work and that of \citet{Arico:2024:tSZ}, and the specific advantages presented in this work.}

\citet[][henceforth, \citetalias{Pandey2024godmax}]{Pandey2024godmax} show that the halo profiles used in the existing baryonification model can be extended, with simple modifications, to also predict the gas pressure profiles (and gas number density profile) in addition to the total matter density profiles. \citet[][see their Figure 1]{To2024DMBCluster} have further demonstrated that this model can consistently predict the tSZ--halo mass scaling relation alongside the density two-point correlations. Thus, the original baryonification model of \citet[][henceforth, \citetalias{Schneider2019Baryonification}]{Schneider2019Baryonification} can be built on to now also predict the tSZ observables (as well as other observables that trace the gas number density, such as the kinematic SZ effect or the X-ray flux). While \citetalias{Pandey2024godmax} focus on using the halo model approach to predict two-point measurements of weak lensing and the tSZ, the model remains to be explored and validated at the map-level for different summary statistics that combine information from the lensing and tSZ fields.

This work explores exactly this aspect, by both building and validating a map-level baryonification model that can be used to predict higher-order correlations. We do this in three ways: (i) we first validate the map-level baryonification method in modelling the density and tSZ fields, using the moments of the fields up to fourth order as our summary statistic, (ii) then we test the sensitivity of our predictions to the inclusion of secondary halo properties such as concentration and ellipticity, as well as additional features in the input halo profiles such as cosmological shocks and non-thermal pressure, and (iii) finally, we showcase the ability of existing and upcoming surveys to constrain these baryonic effects, either separately or in a joint analysis with cosmological parameters varied as well.

Of particular note is that our modelling pipeline is built atop the open-source \textsc{Core Cosmology Library} suite \citep[CCL, ][]{Chisari2019CCL} developed for the Rubin LSST Dark Energy Science Collaboration (DESC). This choice both allows our methods to be easily accessible/usable by the community, by borrowing the many user accesibility features built for the suite, and more importantly allows the halo profiles used in this method to also be propagated through other halo model tools already built with \textsc{CCL}.

We organize this paper as follows: Section \ref{sec:baryonify} describes the baryonification model used in this work. Section \ref{sec:sims&stats} describes the simulations, the summary statistics used to model and validate our method, and the quantification of measurement uncertainty and model best-fit. Section \ref{sec:ModelValidate} describes a subset of the tests we perform to validate our modelling pipeline and characterize its behavior. Finally, section \ref{sec:Forecast} details the power of current and upcoming surveys in constraining the parameters associated with these baryonic effects. We conclude in Section \ref{sec:Conclusions}. 

A number of appendices provide further details on the methods we used: the lensing and tSZ forward modelling approach in Appendix \ref{sec:Forecast:ForwardModel}, the sensitivity of the predictions to additional model and methodology choices in Appendix \ref{sec:ModelValidate:Method:projscale} and \ref{sec:ModelValidate:Profile}, the redshift evolution of the baryonification predictions in Appendix \ref{sec:higher_z}, the dependence between Fisher information and baryonification model-complexity in Appendix \ref{appx:fisher:params}, and finally, some salient computation details on the method, include the runtimes, in Appendix \ref{sec:baryonify:Compute}.

\section{Baryonification}\label{sec:baryonify}

We first describe the halo model used in this work in Section \ref{sec:baryonify:dmb}. Then, we detail the procedures used to include baryon signatures into the different large-scale structure fields we consider: the matter density field (Section \ref{sec:baryonify:density}) and the tSZ field (Section \ref{sec:baryonify:tSZ}). Our model follows from that presented in \citetalias{Pandey2024godmax}, which builds on the density field-only baryonification model of \citetalias{Schneider2019Baryonification} and \citet{Giri2021Baryon}. Following the CCL convention, all distance scales used in defining the profiles are written in \textit{comoving} $\mpc$ units, and all masses as in $\msol$. We will frequently use $\Mtwohc$, which is a spherical overdensity mass defined as the mass contained within a halo-centric sphere of average density $\langle \rho \rangle = 200\rho_c(z)$, with $\rho_c(z)$ being the critical density of the universe at redshift $z$. The radius of the sphere is $\Rtwohc$.

A number of salient computational details about the method are described in Appendix \ref{sec:baryonify:Compute}, including the characteristic runtimes of the pipeline. The baryonified lensing and tSZ maps for a simulation\footnote{Our fiducial simulation contains around 2 million halos, and we baryonify around 90 density shells from $0 < z \lesssim 3.5$, each shell a \textsc{HealPix} map of \texttt{NSIDE} = 1024. See Section \ref{sec:sims:Ulagam} for more details.} can be generated on the order of minutes on an Intel broadwell chip with 40 cores; see Appendix \ref{sec:baryonify:Compute} for more details.

\subsection{Dark matter baryon (DMB) halo model}\label{sec:baryonify:dmb}

The DMB halo model is a sum over multiple different components. Each component is parameterized in a different way, with the parameterization informed by prior work in simulations and/or observations. We now describe each component below:

\textbf{The dark matter profile} is modelled by a simple NFW form \citep{Navarro1997NFWProfile} with an additional truncation term that scales as $r^{-4}$,
\begin{equation} \label{eqn:NFW}
    \rho_{\rm NFW}(r) = \rho_0\bigg(\frac{r}{r_s}\bigg)^{-1}\bigg(1 +\frac{r}{r_s}\bigg)^{-2} \bigg(1 + \frac{r^2}{r_t^2}\bigg)^{-2},
\end{equation}
where $\rho_0$ is a normalization coefficient set by requiring that the mass enclosed within $\Rtwohc$ is $\Mtwohc$,
\begin{equation} \label{eqn:NFW_norm}
    \rho_0 = M_{\rm 200c} \bigg[\int_0^{\Rtwohc} 4\pi r^2 \rho_{\rm NFW}(r) dr\bigg]^{-1}.
\end{equation}
Note that this profile is implicitly a function of $r_s$, or alternatively the halo concentration $\ctwohc = \Rtwohc/r_s$. The truncation term is necessary as the baryonification model requires as estimate of the total enclosed mass at $r \rightarrow \infty$ and this is a divergent quantity for the standard NFW profile as the profile scales as $r^{-3}$ at $r \gg r_s$. We follow \citetalias{Schneider2019Baryonification} and set $r_t = 4 \Rtwohc$ (see Table \ref{tab:params}).

\textbf{The two halo profile}, which describes the extended matter distribution around a halo contributed to by its neighboring structures, is defined similar to \citetalias{Schneider2019Baryonification}, as
\begin{equation} \label{eqn:Twohalo}
    \rho_{\rm 2h} = \rho_m(z)\bigg[1 + \xi_{mm}b(\nu_{\rm 200c})\bigg],
\end{equation}
with $\rho_m(z)$ being the comoving matter density at redshift $z$, $\xi_{mm}$ the \textit{linear} matter correlation function, and $b(\nu_{\rm 200c})$ is the halo bias given by
\begin{equation} \label{eqn:halobias}
    b(\nu_{\rm 200c}) = 1 + \frac{q\nu_{\rm 200c}^2 - 1}{\delta_c} + \frac{2p}{\delta_c(1 + q\nu_{\rm 200c}^2)^p} .
\end{equation}

Here, $\delta_c = 1.686 / D(z)$, with $D$ being the growth factor normalized to $D(z = 0) = 1$, and $\nu_{\rm 200c}$ is the peak height of the halo, and is computed as,
\begin{equation}
    \nu_{\rm 200c} = \delta_c / \sigma(R_{\rm 200c}),
\end{equation}
with $\sigma(R_{\rm 200c})$ being the root-mean square of the density field smoothed with a tophat of radius $\Rtwohc$. We fix $q = 0.707$ and $p = 0.3$, following \citetalias{Schneider2019Baryonification} which in turn is based on the findings of \citet{Sheth:1999:bias}.

The NFW profile and the two halo profile can be summed to predict the total matter distribution in the dark matter-only (DMO) case. Now, we describe the profiles of the baryonic components.

\textbf{The stellar profile} model follows \citet{Mohammed:2014:Baryonification, Schneider:2015:Baryons} and is a simple powerlaw with an exponential cutoff,
\begin{equation} \label{eqn:star_norm}
    \rho_{\rm star} = \frac{M_{\rm star}}{4\pi^{3/2}R_h}\frac{1}{r^2} \exp\bigg[-\bigg(\frac{r}{2R_h}\bigg)^2\bigg],
\end{equation}
where $R_h = \epsilon_h \Rtwohc$ is the half-light radius of the stellar halo, and $\Mstar$ is the normalization of the profile obtained as,
\begin{equation} \label{eqn:Mstar}
    \Mstar = f_{\rm cga} M_{\rm tot},
\end{equation}
where we have defined a theoretical total halo mass, $M_{\rm tot}(r \rightarrow \infty)$, through the expression
\begin{equation}\label{eqn:Menc}
    M_X(r) = \int_0^{r} {\rm d}r 4\pi r^2 \rho_X(r).
\end{equation}
The central galaxy fraction, $f_{\rm cga}$, is then given by
\begin{equation}\label{eqn:cga_fraction}
    f_{\rm cga} = 2A\bigg[\bigg(\frac{\Mtwohc}{M_1}\bigg)^{\tau_{\rm cga}} + \bigg(\frac{\Mtwohc}{M_1}\bigg)^{\eta_{\rm cga}}\bigg]^{-1}.
\end{equation}
Here, the two power-law exponents are given by $\tau_{\rm cga} = \tau + \tau_{\rm \delta}$ and $\eta_{\rm cga} = \eta + \eta_{\rm \delta}$, where $\tau$ and $\eta$ correspond to the scaling of the total stellar fraction, and $\tau_\delta$ and $\eta_\delta$ are a correction to convert the total stellar fraction scaling to the central galaxy stellar fraction scaling. The factor of 2 ensures that for $\Mtwohc = M_1$, we have $f_{\rm cga} = A$. The original baryonification model of \citetalias{Schneider2019Baryonification} used a different version of Equation \eqref{eqn:cga_fraction}, which included $\eta$ and $\eta_\delta$ but did not have $\tau$ and $\tau_\delta$. These latter parameters control the slope of the stellar-mass halo-mass relation for low-mass halos and is included in this work.\footnote{\citetalias{Schneider2019Baryonification} note that analyses of weak lensing only use halos above $\Mtwohc > 10^{12} \msun$, for which their single power-law model is accurate and adequate. We extend this to the double power-law model traditionally used in observational and theoretical analyses of galaxies \citep[\eg][see their Figure 9]{Behroozi2019UniverseMachine}. This extension allows the baryonification model to be used consistently with halos of lower masses.} In practice, the $\tau$ and $\tau_\delta$ parameters have no impact in our work as the signal for both weak lensing and tSZ is generated by halos that are above the $M_1 = 3\times 10^{11} \msol$ mass scale (see Table \ref{tab:params} for the fiducial values of different parameters), and such halos are sensitive only to $\eta, \eta_\delta$ and not $\tau, \tau_\delta$. However, we include this parameter here as it is a simple extension that generalizes the model for smaller mass halos.

As discussed before, the quantity $M_{\rm tot}$ does not diverge now due to our addition of a truncation radius, $r_t$, in Equation \eqref{eqn:NFW}. Without this factor, $M_{\rm tot} = M_{\rm NFW}(r \rightarrow \infty) \rightarrow \infty$. This is relevant as we use this total mass, $M_{\rm tot}$, alongside the relevant mass fractions discussed above, to define the total mass of stars/gas mass in a halo. The baryonification model assumes that $M_{\rm tot}$ for a given halo is the same in Universes with and without the presence of baryons. This assumption is valid as the redistribution of matter is mostly localized around the halo, and for sufficiently large radii (we use $r \rightarrow \infty$ in definition $M_{\rm tot}$) the enclosed mass is constant even after this redistribution process \citep[\eg][]{Ayromlou2022ClosureRadius, Gebhardt2023CamelsAGNSN}.

\textbf{The gas profile} is a modified version of the generalized NFW profile \cite[][see their Equation A1]{Nagai:2007:GNFW}, with two distinct length scales,\footnote{The modification is in using $(1 + r/R_{\rm co})$ rather than $(r/R_{\rm co})$. The former is used in \citetalias{Schneider2019Baryonification} as it defines a cored gas component (i.e. the density does not fall with radius) below $R_{\rm co}$, whereas the latter does not.}
\begin{equation} \label{eqn:gas}
    \rho_{\rm gas} = \rho_{\rm gas, 0}\bigg(1 + \frac{r}{R_{\rm co}}\bigg)^{-\beta} \bigg(1 + \bigg(\frac{r}{R_{\rm ej}}\bigg)^{\gamma}\bigg)^{-\frac{\delta - \beta}{\gamma}}.
\end{equation}
Here, $R_{\rm co} = \theta_{\rm co}\Rtwohc$ is the length scale of the gas core, and $R_{\rm ej} = \theta_{\rm ej}\Rtwohc$ is that of the ejected gas. The scaling $\beta$, which controls the slope between $R_{\rm co} < r < R_{\rm ej}$, has an additional mass dependence as,
\begin{equation}\label{eqn:Beta}
    \beta = \frac{3 (\Mtwohc/M_c)^{\mu_{\rm \beta}}}{1 + (\Mtwohc/M_c)^{\mu_{\rm \beta}}},
\end{equation}
which asymptotes to $3$ for $M \gg M_c$, and $0$ for $M \ll M_c$. Equation \eqref{eqn:Beta} is the updated parameterization from \citet{Giri2021Baryon} and guarantees positivitity of $\beta$ for all masses. The other power-law indices, $\gamma$ and $\delta$, control the slopes at $r \sim R_{\rm ej}$ and $r \gg R_{\rm ej}$, respectively.

The normalization, $\rho_{\rm gas, 0}$ is set in an analagous manner to the stellar profile normalization of Equation \eqref{eqn:Mstar}, and is given as
\begin{equation} \label{eqn:gas_norm}
    \rho_{\rm gas, 0} = f_{\rm gas} M_{\rm tot} \bigg[\int_0^\infty 4\pi r^2 \rho_{\rm gas}(r)dr\bigg]^{-1},
\end{equation}
with $f_{\rm gas}$ derived from the baryon and stellar mass fraction,
\begin{align}
    f_{\rm gas} & = f_{\rm b} - f_{\rm star}, \label{eqn:fgas} \\
    f_{\rm b} & = \Omega_{\rm b}/\Omega_{\rm m}, \label{eqn:fb} \\
    f_{\rm star} & = 2A\bigg[\bigg(\frac{\Mtwohc}{M_1}\bigg)^{\tau} + \bigg(\frac{\Mtwohc}{M_1}\bigg)^{\eta}\bigg]^{-1},\label{eqn:star_scaling}
\end{align}
The baryon fraction within an isolated, toy-model halo is simply the cosmological baryon fraction, and once we specify the stellar fraction --- which contains in it the rich physics of star formation --- we can obtain the gas fraction of the halo. Note that $f_{\rm star}$ denotes the fraction of \textit{all} stellar matter, including those in satellite galaxies, and we will use this fact in Equation \eqref{eqn:rho_clm}. Next, following \citetalias{Pandey2024godmax}, we add additional mass and redshift dependence to multiple gas profile parameters through the following promotion for each parameter,
\begin{equation}\label{eqn:gasMzcdep}
    X \rightarrow X \bigg(\frac{\Mtwohc}{M_X}\bigg)^{\mu_X}(1 + z)^{\nu_X}(\ctwohc)^{\alpha_X}.
\end{equation}
where $M_X$ is a pivot mass, $\mu_X, \nu_X, \alpha_X$ control the scaling with mass, redshift, and halo concentration. Note that in practice, most of these additional scaling parameters are not included in the final model; see Section \ref{sec:baryonify:params} and Table \ref{tab:params} for more details.

\textbf{The collisionless matter profile} is the final component of the baryonification density profile model. This profile constitutes both the DM as well as the subhalos/galaxies in the halo, as both components are collisionless and interact only gravitationally (and not hydrodynamically) with the gas distribution. We follow previous baryonification models in ignoring any explicit modelling of gas within individual subhalos/galaxies. For the observables we wish to model --- the tSZ field, and the baryonic imprints in the density field --- the gas distribution in substructure is irrelevant compared to the gas of the host halo. The collisionless matter profile is modelled by accounting for the adiabatic, gravitational contraction/expansion of the original matter distribution \citep{Blumenthal1986AdiabaticContraction, Gnedin2004AdiabaticContraction} --- where this original distribution is taken to be an NFW profile; see Equation \eqref{eqn:NFW} --- due to the presence of baryonic components, namely gas and stars.

This adiabatic relaxation is performed via the transformation $r \rightarrow r/\zeta$, where $\zeta$ is obtained by solving
\begin{equation}\label{eqn:zeta:angmom}
    \zeta(r) - 1 = a\bigg[\bigg(\frac{M_i(r)}{M_f(r)}\bigg)^n - 1\bigg],
\end{equation}
where $a$ and $n$ are phenomenological parameters. For $a = n = 1$, the expression corresponds to exact angular momentum conservation. We instead take $a = 0.3$ and $n = 0.2$ (see Table \ref{tab:params}), which provides a better match to simulations \citep{Abadi2010ShapesBaryons}. The deviation from $a = n = 1$ arises because galaxy growth is not an instantaneous process \citep{Gnedin2004AdiabaticContraction, Abadi2010ShapesBaryons}.

The initial (final) mass distribution $M_i$ ($M_f$) is given by,
\begin{align}
    M_i & = M_{\rm nfw}(r) \\
    M_f & = f_{\rm CLM}M_{\rm nfw}(r) + \Mgas(r \zeta) + \Mstar(r\zeta).
\end{align}
We solve for $\zeta$ iteratively, and find that in practice it converges ($<1\%$ differences) in under 10 iterations. Upon obtaining a solution, we compute the final $\rho_{\rm CLM}$ profile as,

\begin{equation}\label{eqn:rho_clm}
    \rho_{\rm CLM}(r) = \frac{f_{\rm CLM}}{4\pi r^2} \frac{d}{dr} M_{\rm nfw}(r),
\end{equation}
with the definition $f_{\rm CLM} = (\Omega_{\rm m} - \Omega_{b})/\Omega_{\rm m} + (f_{\rm star} - f_{\rm cga})$. The latter two fractions are defined in Equation \eqref{eqn:cga_fraction} and \eqref{eqn:star_scaling}, and their difference $f_{\rm star} - f_{\rm cga}$ corresponds to the mass fraction of stars in satellite galaxies. This quantity, $f_{\rm CLM}$, corresponds to the fraction of collisionless matter in the total matter distribution.

We have now described all the profile components needed for modelling the baryonic imprints. We can write the total density distribution in the DMO and DMB halo models as
\begin{align}
    \rho_{\rm DMO} & = \rho_{\rm NFW} + \rho_{\rm 2h}\label{eqn:rho_DMO}\\
    \rho_{\rm DMB} & = \rho_{\rm CLM} + \rho_{\rm gas} + \rho_{\rm star} + \rho_{\rm 2h}\label{eqn:rho_DMB}
\end{align}

\begin{figure*}
    \centering
    \includegraphics[width = 2\columnwidth]{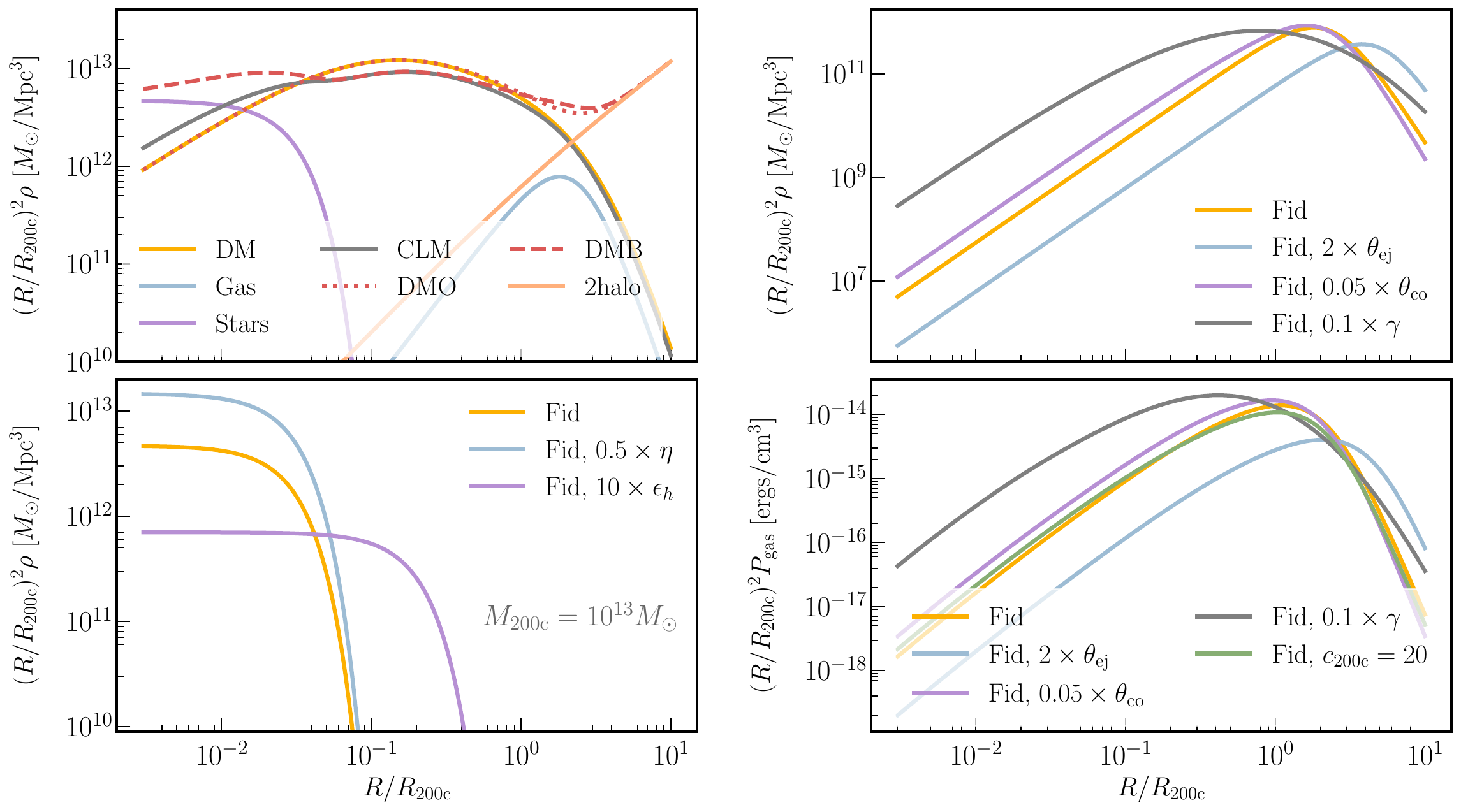}
    \caption{An illustration of the different profiles used in the baryonification pipeline. The top left shows the dark-matter only (DMO) and dark matter baryon (DMB) models, including all the components that constitute each profile. The top right, bottom left, and bottom right, show the fiducial gas, star, and pressure profile, respectively. In addition to the fiducial parameter choice, listed in Table \ref{tab:params}, we show a handful of variations (see legends) to illustrate the impact of certain parameters on the profiles.}
    \label{fig:ProfExample}
\end{figure*}

Figure \ref{fig:ProfExample} shows examples of the different profiles we use in this work. We see that the collisionless matter profile (CLM) is amplified at small scales and suppressed at larger scales, when compared to the dark matter NFW profile, as is expected from the process of adiabatic relaxation. The DMB model is further amplified at small scales due to the presence of the stellar component. We have verified that our pipeline reproduces the profiles shown in Figure 1 of \citetalias{Schneider2019Baryonification}.

\subsection{Matter density field} \label{sec:baryonify:density}

The novelty of the baryonification model is in taking results of an N-body simulation --- which includes accurate predictions for the gravitationally induced correlations of non-linear structure --- and \textit{perturbing} it to induce baryon-like features in the density field. Such an approach requires a displacement function, $\Delta d(r)$, which specifies the distance a particle must be perturbed from its current location. 

To define this function, we first compute the enclosed mass corresponding to the density distributions in Equations \eqref{eqn:rho_DMB} and \eqref{eqn:rho_DMB}. Then, the displacement function is given by
\begin{equation}\label{eqn:displ_3D}
    \Delta d (r) = M_{\rm DMB}^{-1}(M_{\rm DMO}(r)) - r,
\end{equation}
where $M_{\rm DMB}^{-1}(M)$ is the functional inversion of enclosed mass-radius function\footnote{In practice, we perform the functional inversion by generating $M_{\rm DMB}(r)$ on a grid of radius, and using an interpolator to build $r(M_{\rm DMB})$.} and outputs a radius given a mass, and $M_{\rm DMO}(r)$ is the enclosed mass in the DMO case. For a given halo, we evaluate this function at the location of every particle within some maximum halo-centric distance (nominally set to $20\Rtwohc$), and perform a radial offset to the particle position with respect to the halo. In practice, we do not immediately offset the particle locations. Instead, we loop over all halos in the volume, and accumulate the offsets per particle. Once the loop is finished, we now shift the particles by the total, accumulated offset. In this way, the algorithm and its predictions are invariant to changes in the ordering of halos in the loop. The offset positions obey periodic boundary conditions of the simulation box; for example, any pixel that is offset beyond the left edge of the simulation volume will reappear on the right edge of the volume.

Thus far, we have used the formalism of \citetalias{Schneider2019Baryonification}. However, as discussed previously, we require a faster method for applying this displacement function to large suites of full-sky simulations. \citet{Fluri2019DeepLearningKIDS} introduced the \textit{shell} baryonification method, where the displacement is performed on individual map pixels rather than on individual matter particles. In this case, the displacement is written as,

\begin{equation}\label{eqn:displ_2D}
    \Delta d (\rp) = M_{\rm DMB, p}^{-1}(M_{\rm DMO, p}(\rp)) - \rp,
\end{equation}
where we have substituted the 3D distance $r$ with the projected distance $\rp$. The enclosed masses are all now masses enclosed within the projected radius,
\begin{equation} \label{eqn:Mass_proj}
    M_{X, p}(\rp) = \int_0^{\Lproj/2} 2dl \int_0^{\rp} dr_{\rm p} 2\pi r \rho_X\bigg(\sqrt{l^2 + r_{\rm p}^2}\bigg),
\end{equation}
where we now integrate over the line-of-sight distance $l$. The factor of 2 in the integral over $dl$ arises from the symmetry in $l$ used to set integration limits to $0 < l < \Lproj/2$ instead of $-\Lproj/2 < l < \Lproj/2$. The choice of $\Lproj/2$ as the limit, where $\Lproj$ is the thickness of the shell (or simulation box, when considering a 3D snapshot), is different from \citet{Fluri2019DeepLearningKIDS}, where the integration limits were set to $50r_{\rm p}$ instead. We motivate the former choice by recognizing that the mass in a given map pixel is a line-of-sight integration overing the distance $-\Lproj/2 < l < \Lproj/2$. Thus, a similar choice must be made in the baryonification model.

This choice of projection scale can also be motivated by taking the limit $\Lproj \rightarrow \infty$, where $M_{\rm DMB, p} \approx M_{\rm DMO, p}$ since the large-scale matter distribution (which is much less affected by baryons than the small-scale distribution, and is modelled by the two-halo term in our work) dominates the mass in the pixel. In this limit, the displacement function is simply $\Delta d(r_{\rm p}) \approx 0$. By setting $\Lproj$ as the integration distance, our model predictions are formally consistent in these extreme limits. Appendix \ref{sec:ModelValidate:Method:projscale} shows the changes in the predicted displacement due to variations in the assumed projection scale. We note that the projection integral in Equation \eqref{eqn:Mass_proj} assumes the halo is spherically symmetric. We test the anisotropic baryonification corrections in Section \ref{sec:ModelValidate:Ell}, but the baryonification model in those cases will still use the integral in Equation \eqref{eqn:Mass_proj}.

Similar to the case with the particle baryonification, we loop over all halos in a given shell and accumulate offsets to the pixel locations. Once the loop is done, we shift the pixel locations by the sum of the accumulated offsets. For each offset \texttt{parent} pixel, we use the \textsc{Healpy} interpolation routine \texttt{get\_interp\_weights} to (i) determine the nearest four \textsc{HealPix} \texttt{child} pixels that the \texttt{parent} pixel contributes to, and (ii) determine the contribution the \texttt{parent} pixel makes to each \texttt{child} pixel, which is determined by the weights provided by the function. Thus, we regrid the offset pixel back onto the \textsc{HealPix} grid using the neighboring pixels and corresponding contributions as determined by the \textsc{HealPix} routines. This procedure is guaranteed to preserve the total mass/density in the shell. 

In Section \ref{sec:ModelValidate}, we perform a number of tests to evaluate the model's accuracy and these tests use 2D rectilinear simulation maps rather than curved sky maps. In this case, the baryonification-based mass reassignment is done similar to the \textsc{HealPix} case: for each offset \texttt{parent} pixel, we find its nearest four \texttt{child} pixels, calculate the area overlap with each \texttt{child} pixel, and re-assign the \texttt{parent} pixel's values to the four \texttt{child} pixels.

We have verified that our particle baryonification pipeline, operating on the \textsc{IllustrisTNG} DMO simulation, can reproduce the 3D power spectrum results from Figure 2 of \citetalias{Schneider2019Baryonification}. We have also checked --- using the shell baryonification pipeline --- that the angular power spectrum of the baryon-corrected density fields shows the same scale-dependent variation with model parameters as in that Figure; for example, using $\theta_{\rm ej} = 2$ leads to an excess of power on intermediate scales.\footnote{For brevity, we do not include plots of these validation tests, but they are found (and can be easily reproduced by any user) in Jupyter notebooks in our public repository; see the Data Availability Section below for more details on accessing the repository.}

\subsection{Thermal Sunyaev-Zeldovich (tSZ) effect} \label{sec:baryonify:tSZ}

The tSZ effect is sourced by the thermal pressure of the electrons in the cosmic distribution of gas. Given we already have a model for the total density profile, in Equation \eqref{eqn:rho_DMB}, and for the gas density profile, in Equation \eqref{eqn:gas}, we can now assume the halo is in hydrostatic equilibrium and trivially obtain an estimate of the gas pressure,
\begin{equation} \label{eqn:dPdr_HE}
    \frac{dP}{dr} = -\frac{GM^{\rm 1-halo}_{\rm DMB}(r)}{r^2}\rho_{\rm gas}(r).
\end{equation}
This expression can be integrated to obtain the pressure, $P = -\int_\infty^r \frac{dP}{dr}dr$ , using the boundary condition of $P(r\rightarrow\infty) = 0$. Note that Equation \eqref{eqn:dPdr_HE} uses the enclosed mass corresponding to only the one-halo term, denoted above as $M^{\rm 1-halo}_{\rm DMB}(r)$, and ignores the two-halo contribution. This is so our pressure profile corresponds to only the one-halo component. The tSZ field is generated by pasting tSZ profiles around all halos in the simulation. Under such a method, the two-halo term of the tSZ field is naturally modelled as the clustering of halo positions includes this two-halo component. Thus, we avoid double-counting the two-halo term by explicitly ignoring it in Equation \eqref{eqn:dPdr_HE}. 

Equation \eqref{eqn:dPdr_HE} is derived using hydrostatic equilibrium, which assumes that the gravitational pull on the gas is balanced entirely by the thermal pressure --- random motions of the ions --- of the gas. While this is generally true for the gas closer to the core of the halo, it is not accurate for the gas at the outskirts \citep[\eg][]{Nelson2014Nth}. Thus, Equation \eqref{eqn:dPdr_HE} only represents the \textit{total} pressure required to support the gas, and we must estimate the fraction contributed separately by the thermal and the ``non-thermal'' pressure, where the latter primarily constitutes turbulent motions of the gas (i.e. the gas' velocity dispersion).

We follow \citetalias{Pandey2024godmax} by including in our predictions a parameteric model, based on \citet{Shaw2010tSZ}, of the non-thermal pressure,
\begin{align}\label{eqn:f_nth}
    f_{\rm nt} & = \alpha_{\rm nt} f_z \bigg(\frac{r}{\Rtwohc}\bigg)^{\gamma_{\rm nt}},\\
    f_z & = \min[(1 + z)^{\nu_{\rm nt}}, (f_{\rm max} - 1)\tanh(\nu_{\rm nt}z) + 1],\label{eqn:f_nth_z}\\ 
    f_{\rm max} & = 6^{-\gamma_{\rm nt}}/\alpha_{\rm nt},
\end{align}
where $\alpha_{\rm nt}$ is the amplitude of non-thermal pressure, $\gamma_{\rm nt}$ its scaling with radius, and $\nu_{\rm nt}$ its scaling with redshift. We set $\gamma_{\rm nt} = 0.5$ and $\nu_{\rm nt} = 0.3$ following \citetalias{Pandey2024godmax}. The quantities $f_z$ and $f_{\rm max}$ are defined so $f_{\rm nt}$ is between $0$ and $1$ for $r < 6\Rtwohc$. We also explicitly clip its value between $0 < f_{\rm nt} < 1$. The non-thermal pressure is varied in all our predictions. We also show in Section \ref{sec:ModelValidate:Nth} the predictions of this model compared to others in the literature and show the model is adequately flexible.

\citet{Arico:2024:tSZ} also introduce an extension of their baryonification method \citep{Arico:2021:Bacco} to include a tSZ field consistent with the baryon-corrected density field. They first assign ``synthetic'' gas and star particles to their simulations using the DM particle dataset. The pressure profile is then modelled using the prescription of \citet[][]{Komatsu:2002:SZ}. That prescription connects the gas pressure and density by assuming the gas is a polytrope ($P \propto \rho^\Gamma$), which is allowed under the condition of hydrostatic equilibrium, and then by assuming the total matter distribution follows an NFW-like profile \citep[][see their Equation 7]{Komatsu:2001:tSZ}. Note that $\Gamma$ is furthermore assumed to be independent of scale, whereas it has been found to be a function of radius in both simulations \citep[\eg][see their Figure 4]{Battaglia:2012:Gamma} and observations \citep{Capelo:2012:Polytrope}; it does, however, depend explicitly on halo concentration and therefore halo mass. Finally, a particle at a given radial distance, $r$, is assigned a value given by the pressure profile evaluated at that distance.

\noindent In comparison, our work predicts the pressure field directly on the sky, avoiding the use of particles to reduce computational cost and relying instead on the halo positions. Our pressure profile is obtained using the assumptions of hydrostatic equilibrium as well, but we do not assume an NFW profile for the total matter distribution. Instead, we directly use  $M_{\rm dmb}(r)$ predicted by the baryonification model; see Equation \eqref{eqn:rho_DMB}. Our model also does not assume the gas is a polytrope or that it has a scale-independent polytropic index. This allows our predictions to be more easily generalized to both the outskirts of halos and to lower-mass halos. Our profile painting method can capture anisotropic features in the field via the use of elliptical profiles (see Section \ref{sec:ModelValidate:Ell}), but it does not capture more fine-grained morphological features, as would be possible in the particle-based method of \citet{Arico:2024:tSZ}.

\subsection{Pixelization} \label{sec:baryonify:pixel}

Our predictions are pixelized maps rather than a discrete, particle/point. There is therefore a window function needed due to this the pixelization, and must be accounted for during the theory modelling. We do so by modifying the profiles as,
\begin{equation} \label{eqn:prof_conv}
    X(r) \rightarrow \bigg[W_p \star X\bigg](r),
\end{equation}
where $[W_{\rm pix} \star X]$ represents a convolution between the pixel window function, $W_{\rm pix}$, and the radial profile, $X$. This convolution is performed entirely in Fourier space. In the case of 2D maps, which have square pixels, we assume a circular tophat with an area equivalent to that of the square pixel.\footnote{A circular window function can be applied as a simple 1D convolution, using techniques like FFTLog, whereas a square window function requires a 2D convolution that is significantly more expensive.} We have tested that this approximation makes a negligible difference in the convolved profiles: the results are identical for scales smaller/larger than the pixel-scale, and differ at the percent level around the pixel-scale. For the full-sky maps, we approximate the Healpix pixel window function by a Gaussian window function which we verifiy is accurate to better than $1\%$,
\begin{equation}\label{eqn:pix:Gauss}
    W_{\rm pix}(\ell) = \exp\bigg[-\frac{\sigma^2}{2}\ell(1 + \ell)\bigg],
\end{equation}
with $\sigma = \theta_{\rm pix} / (4 \sqrt{\ln2})$, $\theta_{\rm pix}$ the pixel resolution scale obtained from the \texttt{nside2resol} function of \textsc{HealPy}, and $\ell$ is the angular multipole. This analytic function is advantageous as it can be evaluated at any $\ell$ --- which is necessary for performing FFTlog-based Fourier transforms to high numerical precision --- whereas the default Healpix window function is only computed/provided up to an $\ell_{\rm max}$ set by the Healpix map's resolution.\footnote{Note that this latter function (which depends on only $\ell$) is already an approximation of the true pixel window function, which depends on both harmonic variables, $\ell$ and $m$.}

\begin{figure}
    \centering
    \includegraphics[width = \columnwidth]{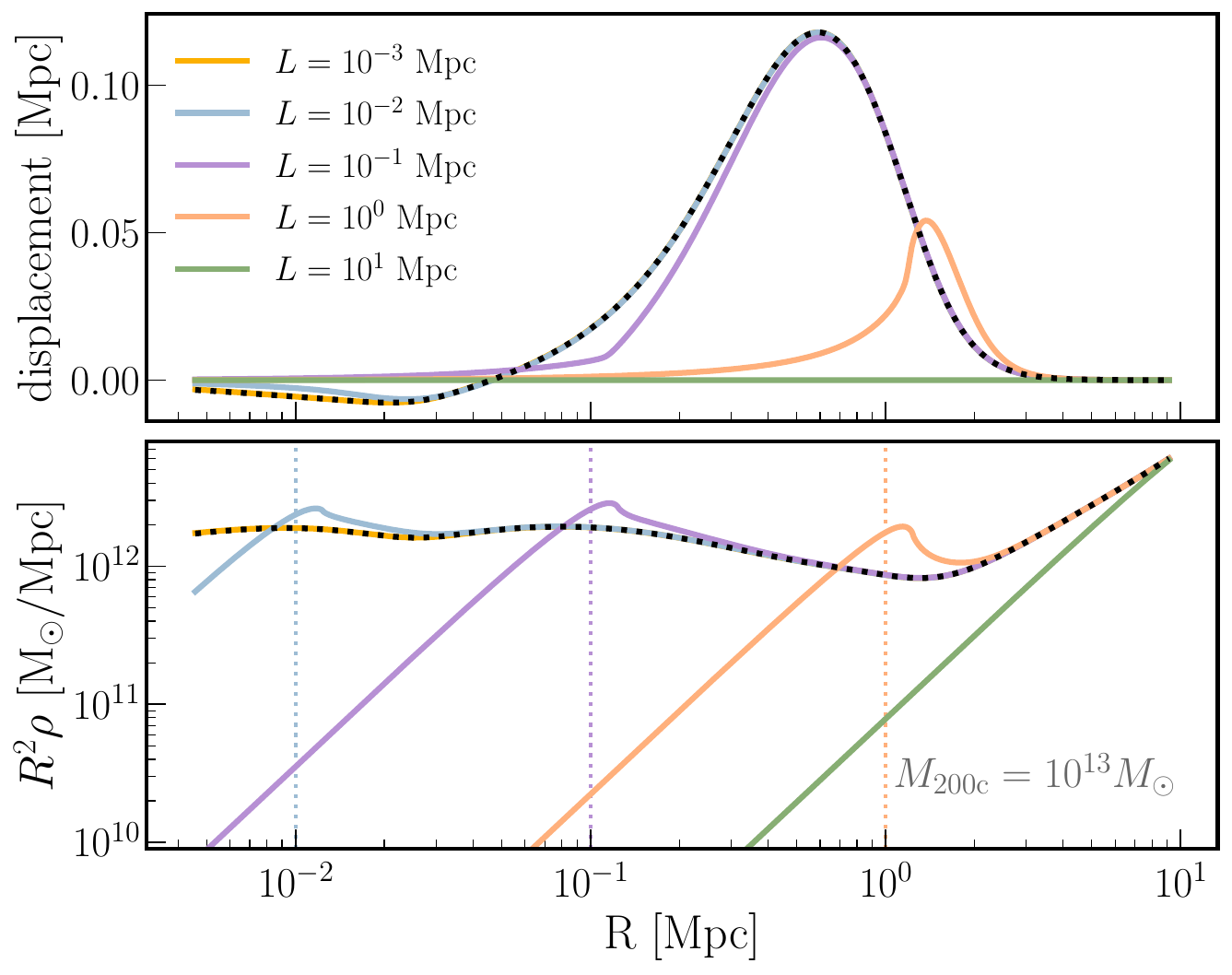}
    \caption{The displacement function (top) and the DMB profile (bottom) corresponding to different pixel/map resolutions (pixel length scale denoted in the legend). The black dotted line is the result from no convolutions, and it is consistent with the result from convolving the predictions with a vanishingly small pixel. The pixel window function smooths the mass on small scales, thereby leading to (i) nearly zero displacement at small scales, and (ii) a pileup of mass at the scale of the pixel (denoted by vertical lines).}
    \label{fig:PixelConv}
\end{figure}

Figure \ref{fig:PixelConv} shows an example of the pixelization's impact on the density profiles and therefore the displacement function. For larger pixelizations, we smooth over baryonic imprints and so the displacement function approaches zero. It is at zero for $L = 10 \mpc$, which is about 20 times larger than the radius $\Rtwohc = 0.45 \mpc$ for the halo mass we show. There is a pileup of mass (identified as the ``bump'' in the profiles) at the scale of pixel, where this scale is shown in vertical dotted lines. The black dotted line in Figure \ref{fig:PixelConv} shows the unsmoothed result, and there is numerical consistency between it and the result from using a very small pixel window function. We have verified that the FFTlog calculations,\footnote{We directly use the FFTlog routines provided within \textsc{ccl}, which are used for high precision calculations such as converting power spectra to correlation functions.} which are used to convert the profiles to Fourier space and perform the convolution, preserve the profiles down to the $10^{-5}\%$ level, which is adequate for our modelling. We use these pixel convolutions for all steps in our work. The pixel scales vary depending on the exact simulation/analysis being performed, and we will denote the exact values where relevant.

\subsection{Model parameterization}\label{sec:baryonify:params}

The profile models in Section \ref{sec:baryonify:dmb} and Section \ref{sec:baryonify:tSZ} has a number of free parameters, but not all of them need to be varied. Table \ref{tab:params} summarizes the parameters of the model, and then the priors we use when performing any model comparisons/fits in the discussions to follow. The key parameters to vary are those associated with the gas profile. Note that this includes $\eta$, given the stellar fraction directly controls the gas fraction as shown in Equation \eqref{eqn:fgas}. The rest of the parameters are fixed to values discussed in \citetalias{Schneider2019Baryonification}. The parameter $\tau$, which we introduce in this work in Equation \eqref{eqn:star_scaling}, is set to $\tau = -1.5$ which is the $1\sigma$ upper bound from the findings of \citet[][see their Table 1]{Moster:2013:SMHM}.

In Equation \eqref{eqn:gasMzcdep} we modified the gas profile parameters to include additional dependence on mass, redshift, and halo concentration. We set these additional scalings to zero for most parameters; the notable exceptions are the scale radii, $\theta_{\rm ej}$ and $M_c$, where we follow \citetalias{Pandey2024godmax} in varying the relevant mass- and redshift-dependence parameters associated these quantities. We also follow \citetalias{Pandey2024godmax} in setting the concentration-dependence parameter $\alpha_X = 0$. The concentration dependence of baryonic feedback has been recently measured in many works \citep[\eg][]{Shao2022Baryons, Shao2023Baryons, Gebhardt2023CamelsAGNSN, Pandey2024godmax} but given our model parameterization is already flexible we do not include this additional dependence. 

The halo concentration is formally a free parameter in the model. In practice, for a given halo, we use the halo concentration measured directly in the simulation when available and otherwise assume the simulation-calibrated relation of  \citet{Diemer2019concentrations} --- alongside the halo mass, redshift, and assumed cosmology of the simulation --- to assign the halo a $\ctwohc$ value.

\begin{table*}[]
    \centering
    \begin{tabular}{ccccl}
       Param  & Fid. val. & Prior &  Profile & Description \\
       \hline
       \hline\\
       $\epsilon_t$ & 4 & --- & NFW & The trunction radius of the dark matter profile, set as $r_t = \epsilon_t \Rtwohc$.\\[10pt]
       
       $p$ & 0.3 & --- & Two-Halo & Relation between peak height and halo bias\\
       $q$ & 0.707 & --- & Two-Halo & Relation between peak height and halo bias\\[10pt]
       
       $\epsilon_h$ & 0.015 & --- & Stars & The half-light radius of the halo, $R_h = \epsilon_h \Rtwohc$. \\
       $A$ & 0.055 & --- & Stars & The normalization of the stellar fraction at $M_1$ \\
       $M_1$ & $3\times10^{11} \msol$ & --- & Stars & The mass-scale used in the stellar fraction -- mass relation. \\
       $\eta$ & $0.2$ & [0.001, 0.9] & Stars & The (negative) power-law scaling of stellar fraction for $\Mtwohc \gg M_1$. \\
       $\tau$ & $-1.5$ & --- & Stars & The (negative) power-law scaling of stellar fraction for $\Mtwohc \ll M_1$. \\
       $\eta_\delta$ & $0.1$ & --- & Stars & Difference between scaling of central galaxy and total stellar fraction, $\eta_{\rm cga} = \eta + \eta_\delta$ \\
       $\tau_\delta$ & $0$ & --- & Stars & Similar to $\eta_\delta$ but for $\tau_{\rm cga}$ \\[10pt]

       $\theta_{\rm ej}$ & $4$ & [0.8, 20] & Gas & The radial scale of the ejected gas, $R_{\rm ej} = \theta_{\rm ej}\Rtwohc$. \\
       $\theta_{\rm co}$ & $0.1$ & [$10^{-4}$, 0.8] & Gas & The radial scale of the collapsed gas, $R_{\rm ej} = \theta_{\rm co}\Rtwohc$. \\
       $M_{\rm c}$ & $10^{14.5} \msun$ & [$10^{10}$, $10^{18} \msun$] & Gas & Mass-scale where the characteristic gas slope has value, $\beta = 1.5$\\
       $\gamma$ & $2.5$ & [0.2, 8] & Gas & The slope of the GNFW profile at $r \sim R_{\rm ej}$\\
       $\delta$ & $7$ & --- & Gas & The slope of the GNFW profile at $r \gg R_{\rm ej}$\\[5pt]

       $\mu_{\theta_{\rm ej}}$ & 0 & [-3, 3] & Gas & The mass scaling of the ejection radius. \\
       $\mu_{\theta_{\rm co}}$ & 0 & [-3, 3] & Gas & The mass scaling of the collapsed radius. \\
       $\mu_{\beta}$ & $0.1$ & [-3, 3] & Gas & The mass scaling of $\beta$, in Equation \eqref{eqn:Beta}\\[5pt]

       $\nu_{\theta_{\rm ej}}$ & $0$ & [-5, 5] & Gas & The redshift scaling of the ejection radius. \\
       $\nu_{M_{\rm c}}$ & $0$ & [-5, 5] & Gas & The redshift scaling of the gas mass scale. \\[5pt]

       $a$ & $0.3$ & --- & CLM & Parameter controlling adiabatic relaxation of halo; Equation \eqref{eqn:zeta:angmom} \\
       $n$ & $0.2$ & --- & CLM & Parameter controlling adiabatic relaxation of halo; Equation \eqref{eqn:zeta:angmom} \\[5pt]

       $\alpha_{\rm nt}$ & $0.2$ & [0.01, 0.5] & Pressure & The non-thermal pressure fraction at $R_{\rm 200c}$. \\
       $\gamma_{\rm nt}$ & $0.5$ & --- & Pressure & The power-law scaling of non-thermal pressure with radius, in Equation \eqref{eqn:f_nth} \\
       $\nu_{\rm nt}$ & $0.3$ & --- & Pressure & The power-law scaling of non-thermal pressure with redshift, in Equation \eqref{eqn:f_nth_z}. \\
       
    \hline
    \end{tabular}
    \caption{The input parameters to the profiles of the baryonification model, described in Section \ref{sec:baryonify:dmb} and \ref{sec:baryonify:tSZ}. For each parameter, we list its fiducial value, the prior used when fitting it (left blank if the parameter is fixed in our analysis) and a description of the parameter's physical meaning. The prior ranges are similar to, but generally broader than, the ones defined in \citetalias[][see their Table 1]{Pandey2024godmax}. Any mass, redshift, and concentration-dependent parameters --- as defined by Equation \eqref{eqn:gasMzcdep} --- are fixed to 0 if they are not shown above.}
    \label{tab:params}
\end{table*}

\section{Simulations \& Summary statistics}\label{sec:sims&stats}

We use three distinct simulation suites and one set of summary statistics to validate our model predictions against the simulation measurements. The \textsc{IllustrisTNG} suite (Section \ref{sec:sims:TNG}) and the \textsc{Quijote} suite (Section \ref{sec:sims:Quijote}) are used to validate the baryonification model. The \textsc{Ulagam} suite (Section \ref{sec:sims:Ulagam}) is used to simulate surveys as part of our Fisher forecast. All analyses of the fields are done after summarizing the fields into a data vector containing moments (from 2nd to 4th order) of the lensing and tSZ field (Section \ref{sec:stats:Moments}). All maps made from the simulations are either a 2D rectilinear grid of $512^2$ pixels or a \textsc{HealPix} curved-sky map of $\texttt{NSIDE} = 1024$.

\subsection{\textsc{IllustrisTNG} simulations}\label{sec:sims:TNG}

The \textsc{IllustrisTNG}\footnote{\url{https://www.tng-project.org}} suite \citep{Nelson2018FirstBimodality, Pillepich2018FirstGalaxies, Springel2018FirstClustering, Naiman2018FirstEuropium, Marinacci2018FirstFields} is a set of high-resolution hydrodynamical simulations that self-consistently evolve a wide set of astrophysical processes \citep{Weinberger2017Methods, Pillepich2018Methods}. The simulation volume spans a box with side $L_{\rm box} = 205 \mpc /h$. At such volumes, the particle mass resolution is high but at the cost of poor number statistics for the larger halos. The latter limits our accuracy in validating the pressure field model, as the pressure field depends more heavily on the most massive halos. The suite contains simulations of three resolution levels, and we use the third, TNG300-3, variant. This run has $625^3$ particles each of dark matter and gas, with a DM particle mass resolution of $4.5\times10^9 \msol$. While higher resolution runs are also available, we utilize only this low-resolution run as its resolution more closely mimics that of the larger simulations used for modelling full-sky lensing and tSZ fields. The pixel-scale of the \textsc{IllustrisTNG} 2D maps is $0.6 \mpc$.

Each simulation in the suite also has a DMO counterpart, that shares the exact same initial conditions. The pair of simulations can be used to make measurements whose uncertainties are cosmic variance suppressed (see Section \ref{sec:sec:uncertainty} for details) and therefore allow more precise validations than otherwise possible. We use the public data release, as introduced in \citet{Nelson2019TNGPublicData}. There also exist other, public simulation suites that span much large volumes than \textsc{IllustrisTNG} --- such as \textsc{Magneticum} \citep{Hirschmann2014MGTM} and \textsc{Bahamas} \citep{McCarthy2017BAHAMAS} --- and would therefore be better suited for our tests of the tSZ field and lensing field cross-correlations. These simulations also span a wide range of baryonic feedback prescriptions, some of which are stronger compared to the relatively weaker feedback in \textsc{IllustrisTNG}. However, unlike the case with \textsc{IllustrisTNG}, the quantities we require from these simulations (the particle snapshots) are not publicly accessible. For these reasons, we use \textsc{IllustrisTNG} for our initial validation here, but highlight the need for a more precise validation --- using a larger simulation --- before employing this model on survey data.

\subsection{\textsc{Quijote} simulations}\label{sec:sims:Quijote}

\textsc{Quijote}\footnote{\url{https://quijote-simulations.readthedocs.io}} is a large suite of DMO simulations spanning a wide range of cosmological parameters \citep{Navarro2020Quijote}. We use the high-resolution runs generated the fiducial cosmology which is based on \citet{Planck2016CosmoParams}. These run have $1024^3$ particles spanning a large volume ($L_{\rm box} = 1000 \mpc/h$), and therefore has much poorer mass/spatial resolution than \textsc{IllustrisTNG} with each DM particle of mass $\approx 10^{11} \msol$, but also have a statistical sample of halos of higher masses. In our work, we use these simulations for tests that (i) use the halo concentration, as we can utilize the $\ctwohc$ estimates from the \textsc{Rockstar} catalogs \citep{Behroozi2013Rockstar} provided with these simulations, and (ii) for checking sensitivities to higher masses that are not testable in \textsc{IllustrisTNG}. The pixel-scale of the 2D maps is $3 \mpc$.

\subsection{\textsc{Ulagam} simulations}\label{sec:sims:Ulagam}

The \textsc{Ulagam}\footnote{\url{https://ulagam-simulations.readthedocs.io}} suite is a set of full-sky N-body simulations designed for widefield survey analyses, and was introduced in \citet{Anbajagane2023Inflation}. The simulations are based on the \textsc{Quijote} and \textsc{Quijote-Png} suites \citep{Navarro2020Quijote, Coulton2022QuijotePNG} and complement them by enabling Fisher forecasts of observables from wide-field surveys. In particular, they vary four cosmology parameters, and four inflationary parameters as well. In this work, we use these simulations to perform forecasts for weak lensing and tSZ surveys (Section \ref{sec:Forecast}).

The simulations resolve $512^3$ particles in a volume with $L_{\rm box} = 1\gpc/h$, following the same configuration as the fiducial (not high-resolution) \textsc{Quijote} run. However, the halo catalogs provide friends-of-friends (FoF) mass, estimated with a linking length of $b = 0.2$ in units of the mean interparticle separation, instead of $\Mtwohc$. This catalog is used only in our forecast analysis (Section \ref{sec:Forecast}), and we make the approximation of treating $M_{\rm fof}$ masses as though they were $\Mtwohc$ ones. FoF linking lengths of $b = 0.2$ will generally correspond to halo overdensities of $200 \rho_c$, but this statement depends on the concentration of the specific halo \citep[][see their Figure 4]{More:2011:FofSO}. Even with this caveat, we can still meaningfully study the degeneracy directions of the different baryonification parameters, as well as the change in relative constraining power across different analysis configurations.

Halos of $M_{\rm fof} > 10^{14} \msun$ are adequately resolved by the simulation as these halos contain at least $100$ particles.\footnote{We only use the halo mass and position estimates from the catalog. Accurate measurements of the phase-space of matter \textit{within the halo} would require better particle resolution.} \citet[][see their Figure 12 and 13]{Anbajagane2023Inflation} show that the halo mass function and halo bias of this catalog agreed with those from higher-resolution simulations. This mass limit is also adequate for studying tSZ auto-correlations: we show in Section \ref{sec:ModelValidate:MassDep}, using the high-resolution \textsc{Quijote} suite, the sensitivity of our observable (the moments of the fields) to different mass scales. For the scales of interest to us, the tSZ-only measurements are sensitive to halos above $\Mtwohc > 10^{14.5} \msol$. For the lensing-only measurements, we have already shown in \citet[][see their Figure 10 and 11]{Anbajagane2023Inflation} that the density and lensing field are accurately predicted in these simulations.

Modelling the \textit{cross-correlation} of the lensing and tSZ field through these simulations incurs one caveat. \citet[][see their Figure 2]{Pandey2021DESxACT} compute the mass-dependence of the lensing and tSZ real-space, two-point cross-correlation function. For the smallest scales we measure ($\theta \approx 3\arcmin$) they show the peak contributing halo mass is $\Mtwohc \approx 10^{14} \msol$ but halos below that mass-scale also contribute a significant amount to the total signal. Our map will therefore underestimate this signal since we only paint pressure profiles around halos with a minimum mass of $10^{14} \msol$. Higher-order correlations push the peak contributing halo mass to higher masses, therefore making any missing contribution from lower mass halos less relevant. Regardless, we stress that the absence of lower mass halos in our analysis will \textit{reduce the impact of baryons} on our simulated signal (relative to their impact when using all halos) and therefore will \textit{weaken} our Fisher information on baryonic processes. We do not risk quoting artificially tighter bounds due to this caveat. In summary, the \textsc{Ulagam} suite still has sufficient, though not ideal, characteristics for performing our joint analysis of lensing and tSZ fields.

\subsection{Field moments statistic}\label{sec:stats:Moments}

The baryonification model has been validated for the density field on both the power spectrum \citep{Schneider2019Baryonification, Arico2023BaryonY3}, as well as the bispectrum \citep{Arico:2021:BkBaryons} and the weak lensing peaks \citep{Lee2023Peaks}. It has also been validated for the tSZ power spectrum \citep{Arico:2024:tSZ}. 
In this work, we focus on the moments of the field which have been used extensively on weak lensing data \citep[\eg][]{VanWaerbeke2013CFHTLens, Chang2018MassMap, Peel2018Moments, Petri2015MomentsMinkowski, Gatti2022MomentsDESY3, Gatti2023SC, Gatti:2024:LFIValidation, Gatti:2024:LFIResults}, and are a computationally efficient estimators of higher-order information in the fields. 

These moments are computed as,
\begin{equation}\label{eqn:Moments}
    \langle A^{(1)}A^{(2)}\ldots A^{(N)}\rangle(\theta) = \frac{1}{N_{\rm pix} - 1}\sum_{i=1}^{N_{\rm pix}}A^{(1)}_iA^{(2)}_i\ldots A^{(N)}_i\,,
\end{equation}
for some set of fields $A^{(1)}, A^{(2)}, \ldots , A^{(N)}$, where all fields are smoothed on some scale, $\theta$. We follow previous works \citep[\eg][]{Gatti2020Moments, Anbajagane2023CDFs} in performing the smoothing using a tophat-filter, which is defined in harmonic space as
\begin{equation}\label{eqn:TophatFilter}
    B(\ell) = 2 \frac{J_1(\ell\theta)}{\ell\theta}
\end{equation}
where $J_1(x)$ is a Bessel function of the first order. In this work, we will consider two fields: the density field and the tSZ field. For $N = 2$, the moments capture the same information as a power spectrum, and this has been checked extensively in many recent works \citep[\eg][]{Anbajagane2023CDFs, Gatti:2024:LFIValidation}.

An ideal validation of our model would also use other summary statistics --- such as the poly-spectra, cumulative distribution functions, the wavelet harmonics, etc. --- but we choose a single statistics, the moments, for simplicity in presenting and interpreting our results. Different statistics will probe somewhat similar types of information, even if the exact definitions vary significantly. Thus, validating our approach for one set of statistics serves as a positive sign for other statistics as well, though an explicit validation must always be performed before employing the baryonification method, with that given statistic, on data. One particular note is that the isotropic shape of the filter in Equation \eqref{eqn:Moments} means the moments are an angle-averaged quantity. Correlation functions for $N > 2$ points depend on the exact configuration (or shape) of the points being correlated; for correlations of three points, the shapes are different triangles. Statistics computed on isotropically smoothed fields, such as the moments defined here, average over all these configurations/shapes. Thus, the validation of baryonification for these moments does not guarantee the validation of shape-specific correlations.

\subsection{Estimation of measurement uncertainty and model best fit}\label{sec:sec:uncertainty}

One key validation done in this work is to compare the predictions of the baryonification model to the fields from \textsc{IllustrisTNG}. The robustness of the match is quantified in relation to the uncertainty on the simulation measurement. There are two different contributions to this uncertainty: (i) cosmic/sample variance, with quantifies the intrinsic uncertainty of our measurement due to using only a finite volume/sample of the Universe, and (ii) shot noise, which is the poission uncertainty from using a finite number of discrete objects.

In the case of \textsc{IllustrisTNG}, our key measurement is the fractional difference between the  moments measured on the true fields and those measured on the predicted ``baryonification-based'' fields. The latter fields are derived using products from the \textit{dark matter-only simulation} in \textsc{IllustrisTNG}. The DMO and hydrodynamical simulations share the same initial conditions, and therefore our predicted fields have the same initial conditions of the hydrodynamical simulations. In this case, the uncertainty on the fractional difference is cosmic variance-suppressed. Note that it is \textit{suppressed}, not \textit{cancelled}, as the late-time density distributions in the two simulations still vary slightly due to differences in the dynamical equations being solved in each (the DMO simulation does not solve any hydrodynamical equations).

The uncertainty is estimated as follows: we define the ratio $R \equiv X_{\rm dmo}/X_{\rm hydro}$, where $X$ is the $N^{\rm th}$-order moment of the density field and/or tSZ field. The uncertainties on $R$ are computed through a leave-one-out jackknife resampling of the 2D fields, with 64 patches. In each sampling, we remove a contiguous, square patch of the field and compute $R$ using the remaining area. In the case of the density field, this is straightforward as both DMO and hydrodynamical simulations produce density fields. However, this process cannot be repeated as is for the tSZ field as the DMO simulation does not have an associated tSZ field. We instead create a mock field through the following transformation,
\begin{equation}\label{eqn:tsZDMO}
    y_{\rm dmo} = {\rm iFFT} \bigg({\rm FFT}(y_{\rm hydro}) \times \frac{{\rm FFT}(\delta_{\rm dmo})}{{\rm FFT}(\delta_{\rm hydro})} \bigg).
\end{equation}
Equation \eqref{eqn:tsZDMO} uses the tSZ field of the hydrodynamical simulation and modifies it in Fourier space to now correspond to the density field of the DMO simulation. This field is then used in estimating $R$ for moments of the tSZ field. This new ``artificial'' tSZ field accounts for differences in the exact realizations of the hydrodynamical and DMO simulations, which requires information about fourier-mode phases, and also the differences in the clustering of matter on small scales, which requires information about the fourier-mode amplitudes. Note that there are many other reasonable ways to estimate a ``artificial'' tSZ field, such as the combination $y_{\rm dmo} = {\rm iFFT} ({\rm FFT}(\delta_{\rm dmo}) \times \sqrt{P_{yy, \rm hydro}/P_{\delta\delta, \rm hydro}})$. We find these alternatives result in larger covariances for the ratio $R$, and we therefore use the method described above since a tighter (or smaller) covariance estimate results in a more stringent test of our method.

Due to the small volume probed by \textsc{IllustrisTNG}, not all jackknife patches are representative volumes of the Universe (for example, only two out of 64 patches contains a halo with $\Mtwohc > 10^{15} \msun$), and therefore the jackknife provides a biased estimate of the covariance. However, this estimate is still useful in interpreting the amplitude of deviations between the baryonification model and \textsc{IllustrisTNG}.

We find the best fit baryonification model for the simulation measurement by generating predictions for a large number of points in parameter space and performing a simple minimum-$\chi^2$ search using the fractional residuals: given as $\chi^2 = (M/s - 1)^T \mathcal{C}^{-1}(M/s - 1)$, where we compare simulations, $s$, with the model, $M$. During this procedure, we set the off-diagonal terms of the covariance matrix, $\mathcal{C}$, to zero. Our numerical estimate of this matrix is noisy, and given the comparatively long datavectors ($120$ to $240$ datapoints) we can incur numerical artifacts/errors during matrix inversion. Increasing the number of jackknife resamplings will alleviate this issue at the cost of further biasing our covariance estimates.\footnote{Conservatively, we would need $\approx 500$ realizations to have a numerically stable matrix. This corresponds to a square patch size of $10 \times 10 \mpc^2$, and all scales around/larger than this scale (which is at least half the scales we use in this work) will have inaccurate covariance estimates.} For the purpose of checking that the baryonification model jointly fits the simulation measurements, this choice still enables a meaningful analysis while avoiding numerical artifacts. However, this highlights that a more precise validation on larger simulations --- for example, by comparing the baryonification predictions to full-sky simulations like \textsc{MilleniumTNG} \citep{Pakmor:2023:MTNG} or the \textsc{Flamingo} project \citep{Schaye:2023:Flamingo} --- will be needed to more robustly characterize this model.

\section{Model validation}\label{sec:ModelValidate}

A primary focus of this work is validating the baryonification model for the density and tSZ fields up to 4th order, by using auto/cross moments of the field as our summary statistic. Here, we will use the density field as an analog of the lensing field as our validation is done with the \textsc{IllustrisTNG} simulation for which we cannot create full-sky maps (such maps require simulations, like those of \textsc{Ulagam} suite, which have at least twenty to thirty times larger volumes). Validating the model for the density field will automatically validate it for the lensing field, as the latter derives directly from the former: see Equation \eqref{eqn:convergence_definition}.\footnote{The estimate of sensitivity to halo concentration (Section \ref{sec:ModelValidate:Method:Conc}) is done using the \textsc{Quijote} simulations, which \textit{do} span a large enough volume to make full-sky mocks \citep{Anbajagane2023Inflation}. However, we do not make full-sky mocks here so that our validation follows the same statistics and procedures as those done with \textsc{IllustrisTNG}.} We will use the full-sky approach for our forecasting results in Section \ref{sec:Forecast}.

We split our validation steps into two categories: (i) methodology and (ii) model. The former encapsulates all decisions related to how we move the particles or paint the tSZ field (including how we use the simulation data as inputs). The latter includes changes in the profile model that can then be propogated into the rest of the pipeline. For brevity, we discuss only the former case here. The latter analysis, which is detailed in Appendix \ref{sec:ModelValidate:Profile}, shows that our profile models are adequate in their flexibility/parameterization.

\subsection{Varying methodology choices}\label{sec:ModelValidate:Method}

\subsubsection{Comparing Particle and Shell baryonification}\label{sec:ModelValidate:Method:3Dvs2D}

An critical choice in this work is modelling the signatures of baryon astrophysics using just 2D fields and halo catalogs. As mentioned earlier, the original baryonification methods focused explicitly on 3D particle snapshots \citep[\citetalias{Schneider2019Baryonification},][]{Arico2023BaryonY3} though some work has employed a 2D approach for weak lensing analyses \citep{Fluri2019DeepLearningKIDS}. In the 3D approach, denoted as ``particle'' baryonifiction, one applies the displacement function to the particles in the 3D snapshot. This snapshot can then be compressed into a lightcone, which is the relevant dataproduct for lensing analyses. In the 2D approach, denoted ``shell'' baryonification, we apply the displacement function to the 2D pixels. The latter is computationally cheaper and the discretization step in converting particles to maps significantly reduces the memory footprint. This is an increasingly important consideration given the community is already producing simulation suites nearing $\mathcal{O}(10^4)$ individual simulations \citep[\eg][]{Navarro2020Quijote, Coulton2022QuijotePNG, Kacprzak2023Cosmogrid, Anbajagane2023Inflation}. 

For our tSZ profile painting method, we can interchangeably use the 3D and 2D approach as the difference is simply in what step the line-of-sight integration is done.\footnote{This will no longer be the case if we used the approach of \citet{Arico:2024:tSZ}, who use the 3D locations of particles in the N-body simulation to generate a gas pressure field. In this case, there will be an associated error in using 2D projections of the 3D positions.} In the first method, we paint profiles on a 3D grid (we \textit{do not} use particle information when painting profiles) and then integrate along line-of-sight by collapsing the map along one direction. In the second, the integration is done within the definition of the projected, 2D profile, before any painting happens. Thus, the tSZ field does not suffer accuracy drops between using 3D fields and 2D fields.

On the other hand, the density field predication can vary between the particle and shell baryonification methods. We test this explicitly by performing the two on the $z = 0$ snapshot from the \textsc{IllustrisTNG} simulation; in this case ``shell'' baryonification is done on the 2D rectilinear map and not the full-sky maps. We first perform both types of baryonification for 32 points in the parameter prior space (defined in Table \ref{tab:params}),  drawn using a Sobol sequence \citep{Sobol1967}. We use only 32 models given the computational expense of the particle baryonification procedure, which takes $\mathcal{O}(10^3)$ longer than the shell-based method. We will use a significantly finer sampling (1024 samples) in Section \ref{fig:TNGFit} when comparing our shell baryonification predictions to \textsc{IllustrisTNG} measurements.

\begin{figure}
    \centering
    \includegraphics[width = \columnwidth]{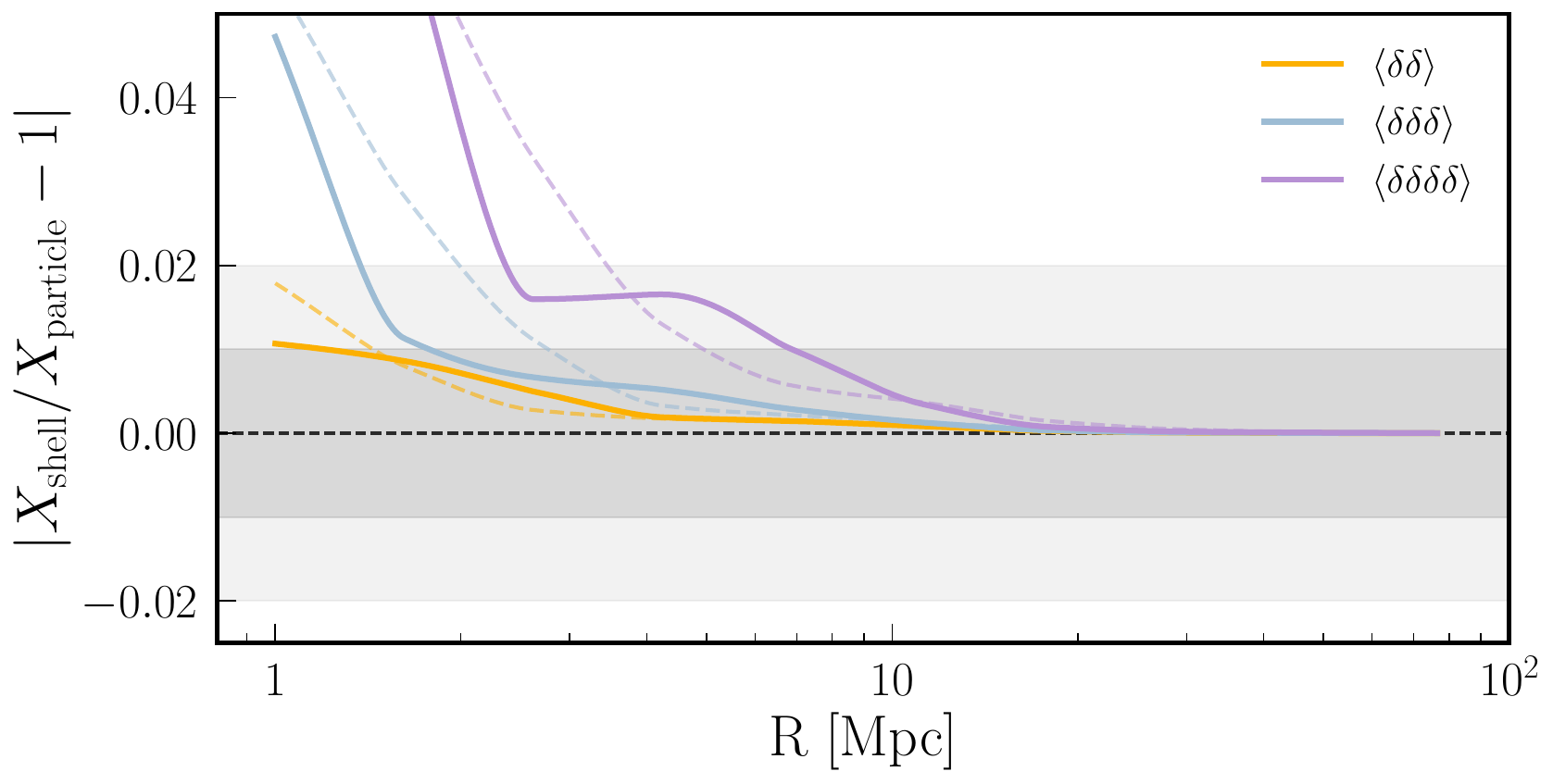}
    \caption{The absolute fractional difference in the baryonification predictions, averaged over 32 parameter choices, between the 2D shell method and the 3D particle method (see Section \ref{sec:baryonify:density} for details on both methods). The dashed lines are the standard deviation of the absolute fractional differences. The 32 models are a Sobol sequence drawn from a prior defined in Table \ref{tab:params}. The bias is within $2\%$ across all scales and increases for higher-order statistics for scales below $\lesssim 2 \mpc$. We deem this a negligible systematic; see text for more details.}
    \label{fig:PartShellBaryon}
\end{figure}

Once we have made the map for a given model, we take the fractional difference of the moments measured on the two maps from particle/shell baryonification, and then compute the average absolute ratio over all 32 models. Figure \ref{fig:PartShellBaryon} shows this average difference for the moments of the density field. In practice, the offsets are within the 1-2\% level, which is a negligible error. The error in the third moments increases at around the pixel-scale of $\approx 1 \mpc$, and the same for the fourth moments increases below $\lesssim 2\mpc$. These differences are still within the cosmic variance-suppressed uncertainties of the measurement (see gray bands in Figure \ref{fig:TNGFit}). We also show the $1\sigma$ spread of these absolute, fractional differences and find them to also be around the $1-2\%$ level for the scales mentioned above.

In general, a difference between the two methods can occur in two different ways: (i) the geometry of the offsets are fundamentally different in the particle and shell baryonification methods, as in the former some particles will be displaced just along the line-of-sight and have no impact on the final, 2D field, whereas the projected displacements are always perpendicular to the line-of-sight; and (ii) the presence of interlopers, which are halos that overlap in 2D space but are distant in 3D space, as the offsets predicted by the smaller halo will impact the matter distribution of the larger halo given both distributions will contribute to the pixel being displaced. Formally, this latter effect is indeed accounted for through the two halo profile in Equation \eqref{eqn:Twohalo}. Both effects are alleviated when the projection is done over smaller distances, i.e. if the shells are thin. The tests above are done with the projection scale set to $L = \approx 300 \mpc$ whereas in full-sky simulations --- such as the \textsc{Ulagam} we perform shell baryonification on in Section \ref{sec:Forecast} --- the shell thickness is $\approx 60-100\mpc$. Therefore, Figure \ref{fig:PartShellBaryon} is a more strict test of the method.

Following the discussions in Section \ref{sec:baryonify:density}, the model for the displacement function and projected displacement function should be equivalent as both are derived from the same input profile and are physical models of the 3D and projected density distribution, respectively. Figure \ref{fig:PartShellBaryon} shows our predictions obey this condition down to at least $\approx 1\mpc$ ($2\mpc$) for up to third-order (fourth-order) in the density field.

\subsubsection{Dependence on concentration}\label{sec:ModelValidate:Method:Conc}

\begin{figure}
    \centering
    \includegraphics[width=\columnwidth]{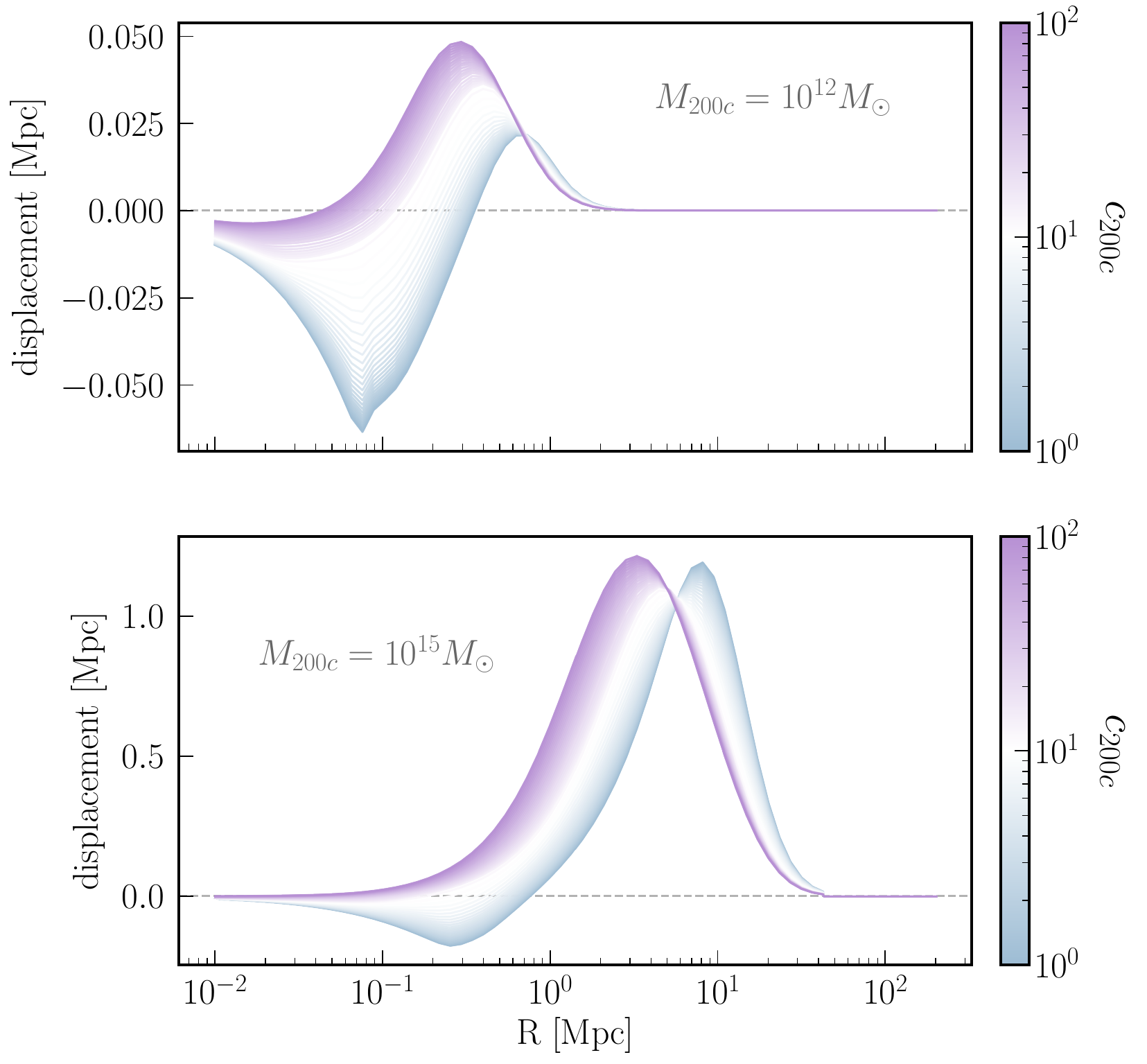}
    \caption{The impact of concentration on the displacement function for halos of two different masses. There is a strongly scale-dependent effect from varying $\ctwohc$. Note that we use a large range of $\ctwohc$ values (beyond our physical expectations for halos of these masses) for illustrative purposes. In general, lower $\ctwohc$ halos require stronger contraction (more negative displacements) on small scales, and the physical scale of maximum displacement is also shifted to slightly larger scales. This prediction is made using the fiducial values of the baryonification parameters (see Table \ref{tab:params}).}
    \label{fig:c200c:displ}
\end{figure}

\begin{figure}
    \centering
    \includegraphics[width=\columnwidth]{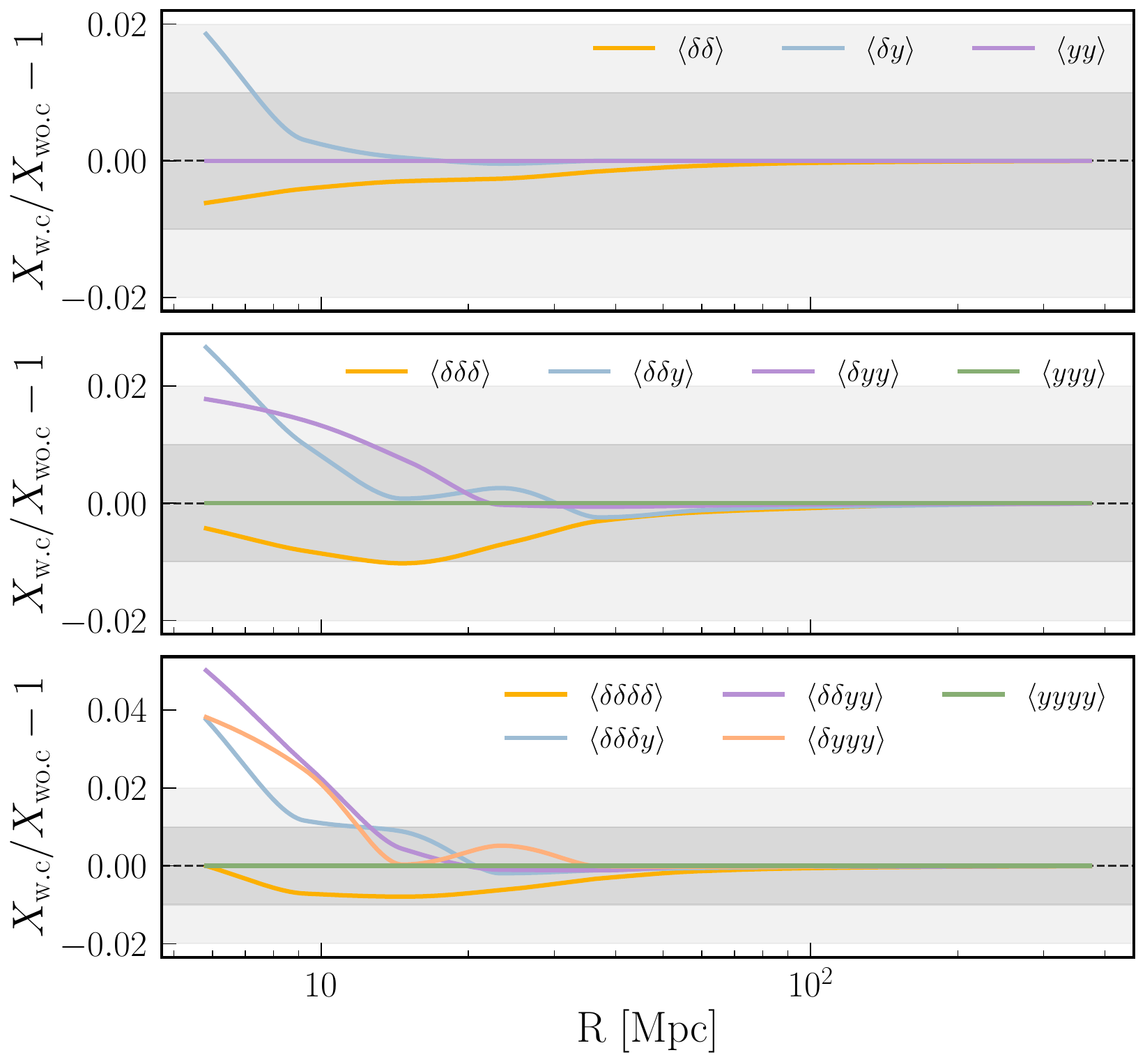}
    \caption{The fractional difference in the baryonification predictions, for the 2nd, 3rd and 4th moments of the density and tSZ field, when including/ignoring the concentration information. The difference is below 1\% for almost all correlations, and is most amplified only for the cross-correlations where varying the concentration will generate more spatially correlated features across the density and tSZ field.}
    \label{fig:c200c:stat}
\end{figure}

To first order, halos can be defined using just their mass and redshift. Once these two properties are given, many other halo properties can be approximately derived through simple scaling arguments. However, there are a number of secondary halo properties that add significant information about the halo. Of these, the halo concentration plays a key role and has been studied extensively as a result. 

In our current baryonification model, which does require an input $\ctwohc$ per halo, we simply use a precomputed concentration--mass relation --- in our case the model of \citet{Diemer2019concentrations} --- and assign each halo a $\ctwohc$. This is then used in the rest of the pipeline. This is effectively a mass-only model as the halo is still defined using just its mass; a halo of a given mass will also have a fixed concentration value. However, at fixed halo mass there is a $\sim 40\%$ scatter in the concentration values \citep[\eg][]{Wechsler2002Concentrations, Diemer2015Concentration, Diemer2019concentrations, Ragagnin2019HaloConcentration, Anbajagane2022Baryons}. Furthermore, the concentration value of a halo depends on the halo's cosmological context, i.e. its entire accretion history, its environment and more \citep[\eg][]{Wechsler2002Concentrations, Mansfield2020AssemblyBias}. Thus, the baryonification model's predictions for the cosmological correlations can be more accurate if we self-consistently include the concentration of each halo as measured in the simulation.

First, we show in Figure \ref{fig:c200c:displ} the impact of $\ctwohc$ on the displacement function for halos of the Milky Way mass-scale and the cluster mass-scale. In both cases, lower $\ctwohc$ values lead to a more negative displacement at small scales. A lower $\ctwohc$ implies a more diffuse halo and therby requires a stronger displacement value to transform the mass distribution into the more concentrated form that is generated by the presence of baryonic matter. The second effect of a lower $\ctwohc$ is increasing the radial scale corresponding to the maximum displacement. This is because for a more diffuse halo, the gas distribution starts dominating the dark matter profile at larger radii. This analysis uses the fiducial baryonification parameters as defined in Table \ref{tab:params}.

We then test the impact of this difference on the statistics of interest. Note that this test is done on \textsc{Quijote} given those simulations have a readily available concentration measurement, and therefore the range of scales we probe here is larger than that of the other tests. We generate 32 models across the parameter prior and show the median, fractional deviation in the statistics. Figure \ref{fig:c200c:stat} shows there is a $\approx 2\%$ offset in the measured moments down to $5\mpc$, which is the smallest scale we can probe in \textsc{Quijote}. In summary, while the $\ctwohc$ variation does have an impact on the statistics, it is not significant above $>5\mpc$ and is completely negligible above $> 10 \mpc$. It is likely the effect is more pronounced towards smaller scales and for higher orders of the fields. We find it is best to use the simulation-measured concentration when available.

Figure \ref{fig:c200c:displ} also shows that the difference in the displacement function is minor when varying the concentration by nearly two orders of magnitude. This implies the model is insensitive to using other halo concentration-mass relations \citep[\eg][]{Child2018ConcentratioMassRelation, Ishiyama2020UchuuConcentration, Anbajagane2022Baryons, Shao2022Baryons, Sorini:2024:Baryons} which deviate, to varying levels, from the predictions of \citet{Diemer2019concentrations}.

One can also trivially amplify the impact of concentration on the measurements by setting $\alpha_X \neq 0$ (see Equation \eqref{eqn:gasMzcdep} and Section \ref{sec:baryonify:dmb}) to enforce a relationship between $\ctwohc$ and the gas profile parameters. Thus our test above is primarily showing that the default, fiducial model is not sensitive to the choice of concentration. While \citetalias{Pandey2024godmax} find values of $\alpha_{\theta_{\rm ej}}$ that allow the model predictions to match the \textsc{IllustrisTNG} simulations, more detailed simulation studies are required to understand the correlation between the gas parameters and the halo concentration. Such a correlation could be important for the baryonification technique as the halos' concentration values are correlated with their spatial distribution, an effect commonly referred to as ``assembly bias'' \citep[\eg][]{Zentner:2014:HaloBias}. Note that some aspect of this phenomenon can already be included in the current baryonification method, through using the concentration measured for each individual halo in the simulation (as opposed to assuming a concentration--mass relation, which will not include the effects of assembly bias).

\subsubsection{Dependence on ellipticity}\label{sec:ModelValidate:Ell}

Another critical secondary property of the halo is its shape. To good approximation, halos are quasi-spherical objects. However, they are not exactly spherical, and their ellipticities/orientations are sourced by the cosmological process of structure formation; meaning the shape and orientation of halos exhibit cosmologically sourced, spatial correlations. Such correlations are closely connected to the ``intrinsic alignments'' effect studied in the weak lensing community \citep[see][for a review]{Troxel2015IAReview}.

It is therefore important to determine the impact of the ellipticity correlations in the baryonification pipeline, for modelling the statistics of interest to us. The implementation of ellipticities in our model is trivial. The halo-centric distance to a pixel is now scaled as $x^2 + y^2 \rightarrow x^2 + y^2/q^2$ where $q \leq 1$ is the ratio of the minor axis to the major axis.\footnote{We only explore the impact of ellipticity on the 2D field, where $q$ is the only relevant shape parameter.} Under this formalism, the major axis is assumed to be of length $\Rtwohc$. In practice, this scaling is done by transforming the coordinate system --- rotating the pixel grid so the major/minor axis of the halo is along the x-axis and y-axis of the grid --- and then rescaling the $y$ distance by $q$ as denoted above. The relevant quantity, a displacement or gas pressure, is computed at the location given by this \textit{rescaled} distance. We perform this test with \textsc{IllustrisTNG}, where we computed the orientations of halos using the same pipeline as \citet{Anbajagane2022Baryons}; this work computes the mass inertia tensor in 3D to compute the 3D ellipticity. In our case, we take only the 2$\times$2 subset of the tensor corresponding to the x and y directions, and compute a 2D ellipticity using the new tensor.

\begin{figure}
    \centering
    \includegraphics[width = \columnwidth]{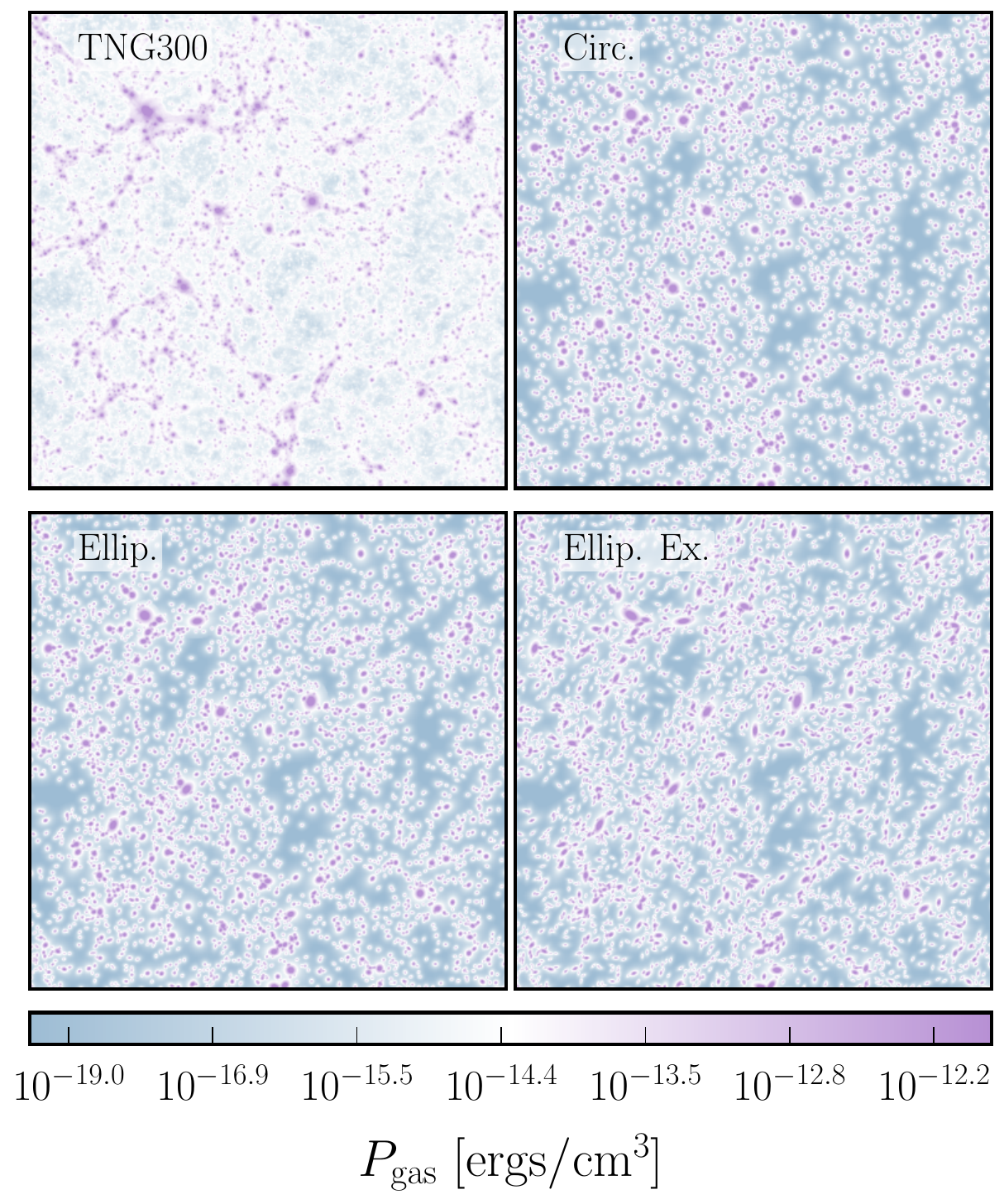}
    \caption{Different maps of the gas pressure: from the \textsc{IllustrisTNG} hydrodynamic run (top left), from the baryonification model of the DMO run assuming circular profiles (top right), elliptical profiles (bottom left), and ``extreme'' elliptical profiles (bottom right). The former elliptical model uses ellipticies computed directly in the simulation, whereas the latter is simply a scaled, more-elliptical version of the former, $q_{\rm ex} = 0.6q_{\rm TNG}$. We can clearly see the correlation between the halo orientations and the large-scale density field: the alignments are along filaments or point in the direction of the nodes.}
    \label{fig:elldep:Pgasmap}
\end{figure}

\begin{figure}
    \centering
    \includegraphics[width = \columnwidth]{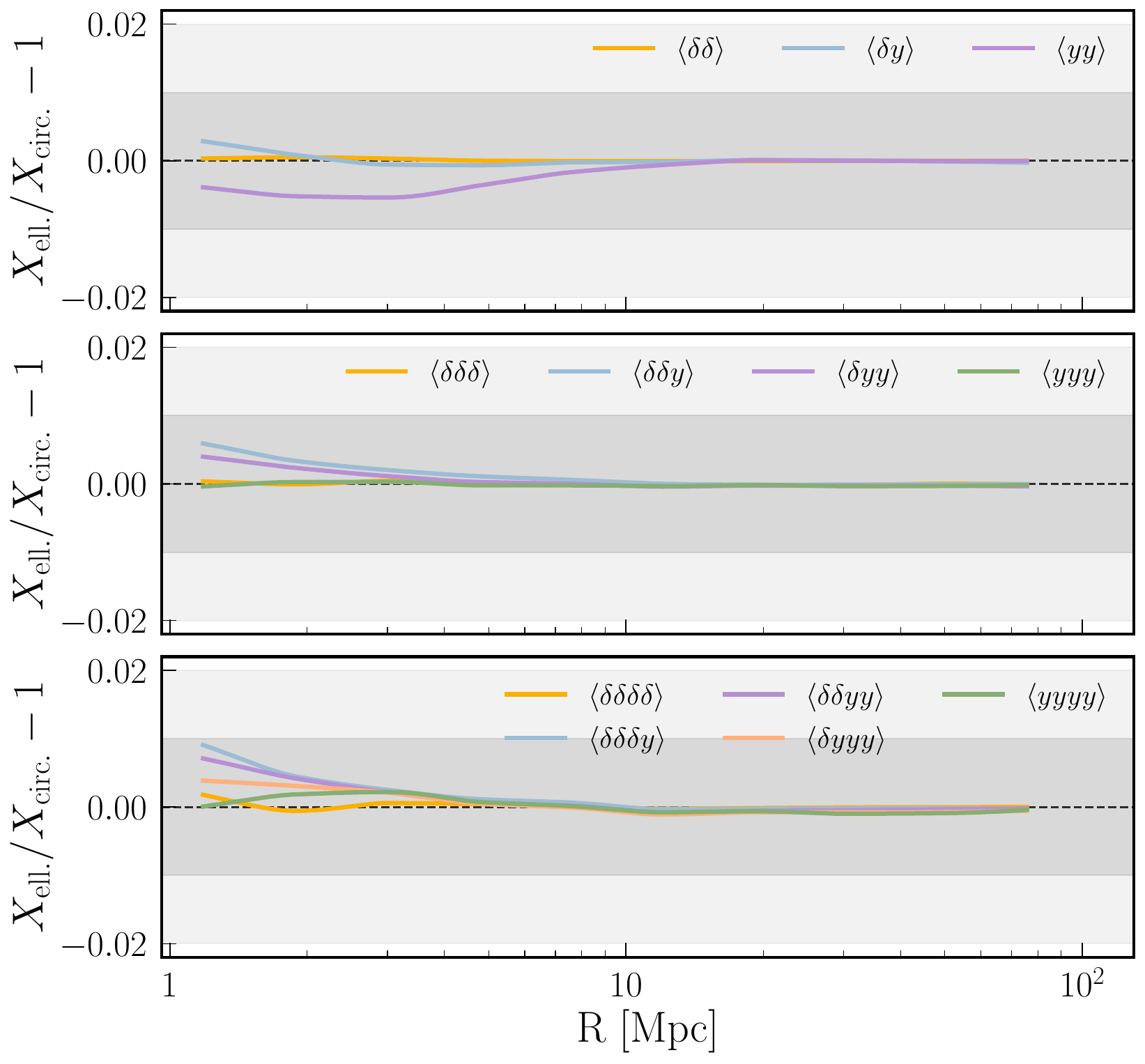}
    \caption{The fractional difference in the baryonification model predictions when excluding/including ellipticity in the model. We show all moments of the density and tSZ field, up to the 4th moment. The impact is sub-percent for all moments, and generally grows towards small-scales.}
    \label{fig:elldep:Stats}
\end{figure}

Figure \ref{fig:elldep:Pgasmap} shows a visual comparison of theoretical maps made with and without the orientation information. One can clearly see the correlation of the ellipticities with the large-scale density field. The alignments are often in the direction of filaments or are pointing towards nodes in the field. This is most prevalent in ``Ellip. Ex.'' map, which is an extreme version of the ellipticity, given as $q_{\rm ex} = 0.6q_{\rm TNG}$, and shown purely for illustrative purposes. 

Figure \ref{fig:elldep:Stats} then shows the change in the moments of the fields due to the inclusion/exclusion of the halo ellipticity in our modelling. The change in the statistics is below percent-level in almost all cases. In summary, the ellipticity of halos is not necessary for modelling the moments of the field. Other statistics --- particularly those with anisotropic information, such as the bispectrum which is a function of angle, and not just isotropic information like the moments --- may have stronger sensitivity to this choice and therefore must be separately tested. Similarly, methods that involve the explicit detection/characterization of filamentary structures, such as topological measures \citep[\eg][]{Heydenreich2021BettiNumbersWL, Heydenreich2022Y1BettiNumberWL, Euclid2023NGCov}, could have a stronger sensitivity to the choice of including/ignoring ellipticity.

\subsection{Comparison to hydro simulation}\label{sec:ModelValidate:Hydro}

\begin{figure*}
    \centering
    \includegraphics[width = 2\columnwidth]{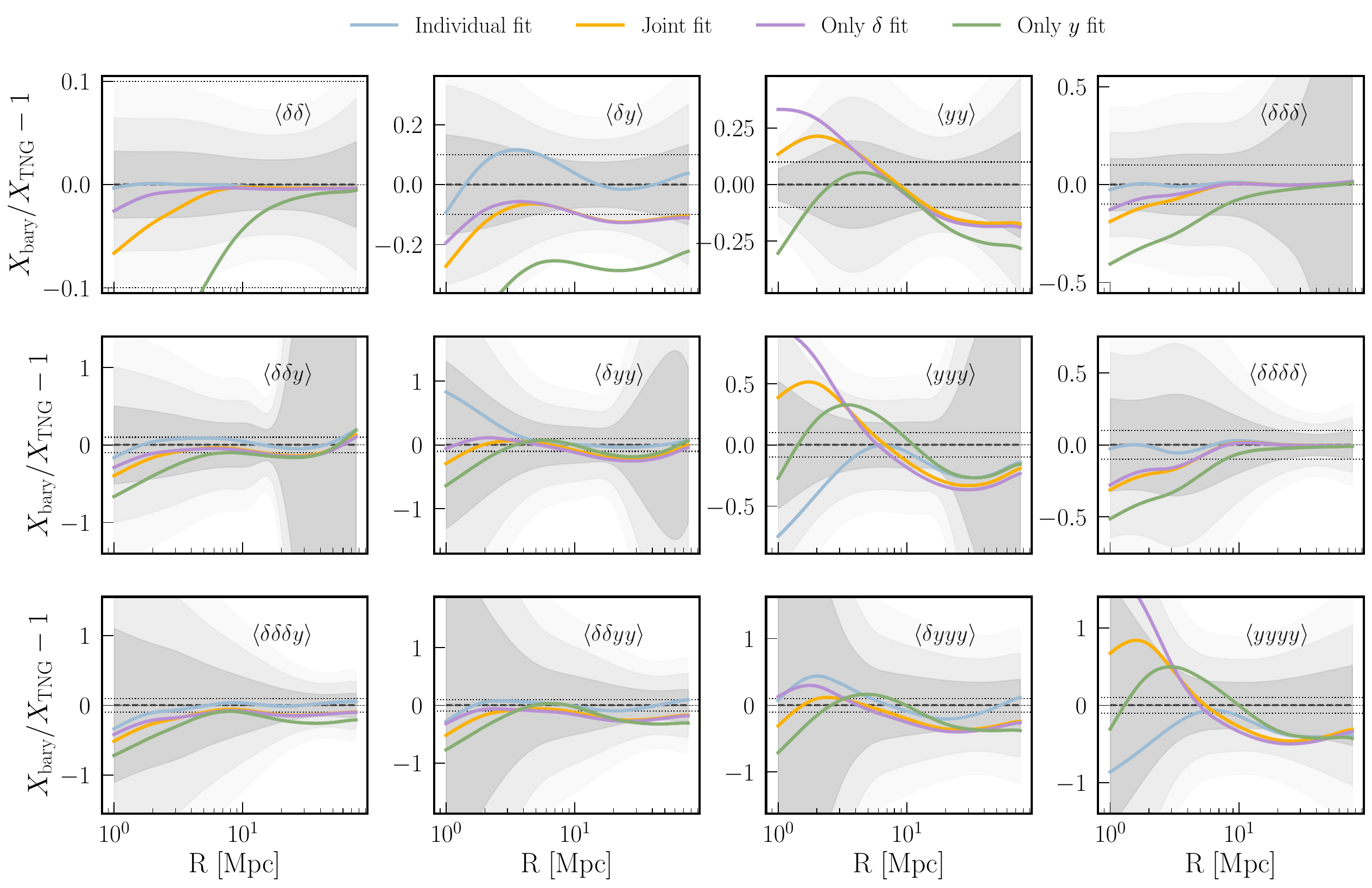}
    \caption{The fractional difference between the moments of the baryonified DMO simulations and of the TNG hydrodynamical simulation. All moments are computed on 2D maps at $z = 0$. The gray bands show the 1, 2, 3$\sigma$ uncertainties. The dotted black lines denote 10\% fractional errors. Cosmic variance is highly suppressed given both maps derive from the same initial conditions. The joint fit (yellow line) is within 1-2$\sigma$ for all different moments. We a-priori expect some irreducible deviations due to the low statistics of massive halos in \textsc{IllustrisTNG}; see Section \ref{sec:ModelValidate:Hydro} for details. For the panel with $\langle y y \rangle$, the ``Individual Fit'' line is underneath the ``Joint Fit'' line.}
    \label{fig:TNGFit}
\end{figure*}

We now check that our baryonification model can jointly fit measurements from hydrodynamical simulations. This is done by computing the fractional difference between the simulation measurements and our best-fit baryonification predictions --- for all moments of the density and tSZ field from 2nd to 4th order --- and comparing the difference against the measurement uncertainty described in Section \ref{sec:sec:uncertainty}.

Figure \ref{fig:TNGFit} compares our predictions with measurements made on \textsc{IllustrisTNG}. Our predictions use the DMO simulation whereas the measurements are made on the full hydrodynamical simulation. Each subplot shows the fractional difference for different moments. The gray-bands denote the $1,2,3\sigma$ uncertainty bounds computed using the method in Section \ref{sec:sec:uncertainty}. The baryonification model is accurate to within the cosmic variance-suppressed uncertainties of the moments; the fraction differences are always within the 1-2$\sigma$ bounds (gray bands). Note that the percent-level agreement for the large-scales of the density auto-correlations (at any order) is expected as the baryonic imprints are negligible on such scales.

We a-priori expect to poorly match some features on small-scales, and particularly for any moments that involves the tSZ field. The signal for the tSZ field comes from the most massive clusters, but the \textsc{IllustrisTNG} simulation has only a handful of such clusters (\eg only two clusters above $\Mtwohc > 10^{15} \msun$). For a larger simulation, with a larger sample of clusters contributing to the signal, any individual characteristics of the cluster would average out over the population, and the baryonification prediction --- which by construction represents only the average cluster --- would match the simulation measurement better. When the simulation contains only a few clusters, these individual characteristics can have a more dominant role in the emergent signal. For example, the most massive cluster in \textsc{IllustrisTNG} --- which will dominate the total tSZ signal, especially for the higher-order moments (see Figure \ref{fig:MassDep}) --- has undergone a recent merger. Our model will not account for this characteristic. However, a significantly larger simulation would have enough massive systems such that the individual merger history of a single object would not have a notable impact on the observed field.

Figure \ref{fig:TNGFit} also shows the results from fitting only the density field, or only the tSZ field. The density field is not a sensitive probe of all the different baryonic physics, and therefore a fit to the density alone does not correctly estimate the amplitude of the tSZ field. We also show the result from fitting a single moment at a time. Unsurprisingly, this fit is the best of all, given the large degrees of freedom of the model compared to number of datapoints used to constrain it. Note that in some moments, even the invidiual fit is in the 2$\sigma$ regime, which indicates the presence of the individualistic features in the few most massive clusters (which are not captured in the model). We have separately used the \textsc{Illustris} simulations \citep{Vogelsberger2014Illustris}, a predecessor to the \textsc{IllustrisTNG} suite we use here, and verified that our model predictions are not systematically low on small-scales in those analyses. This further indicates that the offsets seen in Figure \ref{fig:TNGFit} is a stochastic fluctuation, induced by the individual histories of the handful of the largest clusters in the simulation. 

Under the current statistical uncertainties, we find that our model is able to jointly fit the simulation data at a fractional accuracy that is at worst between $5\%$ to $\approx 50\%$, depending on the exact moment being considered, with the fractional difference increasing for moments with larger measurement uncertainties (for example, $\langle y^3 \rangle$ or $\langle y^4 \rangle$, whose signal is dominated by just a few halos in the TNG volume). Given the caveats related to TNG that we have discussed throughout the text above, a more robust analysis is needed to calibrate the baryonification model accuracy to much better precision. Such an analysis would require a much larger simulation; for example \textsc{Bahamas} \citep{McCarthy2017BAHAMAS} or \textsc{Millenium-Tng} \citep{Pakmor:2023:MTNG}. We leave these efforts to future work.

While the validation of the baryonification method for the density two-point statistics (or 2nd moments) has been done to the percent-level \citep[\eg][]{Giri2021Baryon, Arico:2021:Bacco}, the model for the other eleven measurements shown in Figure \ref{fig:TNGFit} can still be useful even if the model accuracy is larger than 1\%: the noise level in the higher-order moments increases significantly compared to the 2nd moments \citep[\eg][see their Figure 6, for the lensing-only moments]{Anbajagane2023CDFs}, while the tSZ field is generally more noise-dominated than the lensing field. These two factors add leniency to the validation requirements of these other eleven moments. The actual requirements for using baryonifcation is survey-specific and must be determined explicitly for each survey and choice of statistic.

Appendix \ref{sec:higher_z} also shows that the baryonification model can jointly fit all moments measured across multiple redshifts in the \textsc{IllustrisTNG} simulations.

\subsection{Mass dependence}\label{sec:ModelValidate:MassDep}

\begin{figure*}
    \centering
    \includegraphics[width = 2\columnwidth]{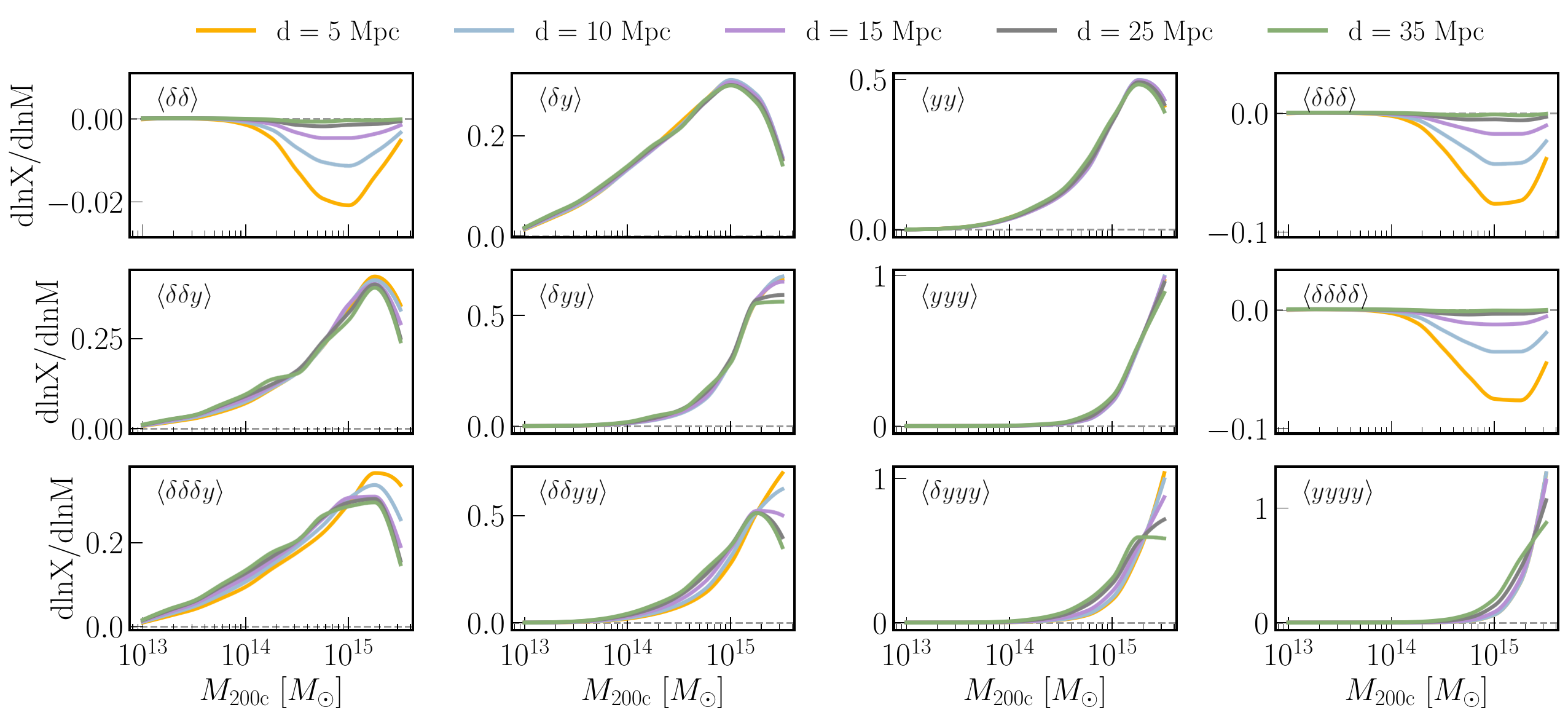}
    \caption{The derivative of the baryonification signal, with respect to the halo mass, for the moments of the density field and tSZ field (different panels, $X = \langle \delta \delta \rangle, \langle \delta y \rangle, \ldots$) measured at different smoothing scales (colors). The y-axis represents the fractional contribution, from halos of a given mass $M$, to the total signal. For the density field, the signal is the \textit{baryonic imprints} in the field rather than the density field itself. See text for details on the measurement. The mass-dependence of higher-order moments, or any moments including the tSZ field, is increased to higher masses when compared to the 2nd moments of the density field. The exact values of the derivatives depends on the baryonification parameters used, and we use the fiducial values shown in Table \ref{tab:params}.}
    \label{fig:MassDep}
\end{figure*}

In Section \ref{sec:ModelValidate:Hydro} we show that our model is flexible enough to jointly fit the simulation measurements within their statistical uncertainties. We now use this model to check the mass-dependence of the predicted signal for the different moments. This is analogous to similar plots made for the two-point function in \citet[][see their Figure 2]{Pandey2021DESxACT} or \citet[][see their Figure 3]{To2024DMBCluster}, and lend insight into the origin of the signal in different measurements. In general, we expect higher-order moments to be more sensitive to more massive halos than the 2nd moments, and similarly, for the tSZ field to be more sensitive to more massive halos than the lensing field. The former is because taking higher powers of a variable amplify the tails of its distribution, and the density field has a strongly positive-skewed tail --- occupied by massive halos --- whose presence/importance is amplified by taking higher powers of the field. The latter is because weak lensing depends linearly on the density field and therefore mass, whereas the tSZ signal follows $\propto M^{5/3}$ \citep[\eg][see their Figure 1 for a comparison across simulations]{Lee2022rSZ} and therefore is sensitive to higher masses than lensing.

We estimate the mass-dependence of the moments using the \textsc{Quijote} simulations, and in particular using their high-resolution runs, where halos of $10^{13} \msol$ are resolved by 100 particles. The contributions from halos of a given mass, $M_1$, is computed through the difference $X(> \log_{10}M_1) - X(> \log_{10}M_2 \equiv \log_{10}M_1 - \delta)$, where $\delta = 0.25$ and $X$ is a Nth-order moment measured on a density and/or tSZ field, with the field generated using only halos of mass greater than $M_1$ or  $M_2$. We use the baryonification model to make fields for different choices of $M$ and take simple numerical differences to estimate the derivative of the measurement with mass, $\dln X/\dln M$. We approximate $\dln X \approx \Delta X / X_{\rm fid}$ where $X_{\rm fid}$ is the fiducial signal from using all halos. Thus $\dln X/\dln M$ represents the fractional change --- compared to the fiducial case of using all available halos --- in the moments of the fields due to halos of a given mass. 

For the tSZ field, the change $\dln X/\dln M$ represents the contribution of a given halo mass to the total tSZ signal, but for the density field it represents the contribution to the \textit{baryonic} imprints in the field and not the density field as a whole. Note that for the density field, the $X_{\rm fid}$ used in the expression $\Delta X / X_{\rm fid}$ is the total density field signal (and not just the total baryonic imprints signal). For this reason, the fractional differences for the density field will have a lower amplitude than those for the tSZ field. For all estimates, we have used 10 high-resolution simulations and averaged the derivatives across realizations. The mass-dependence will change based on the chosen baryonification parameters; for example, changing the $\mu_X$ parameters in Table \ref{tab:params} can modify these derivatives. For our estimate of the derivative, we will use the fiducial parameter values listed in Table \ref{tab:params}.

Figure \ref{fig:MassDep} shows these derivatives across different halo masses, for five different apertures used in computing the moments. As expected, the $\langle \delta \delta\rangle$ correlation has the lowest mass-dependence, as all other correlations have derivatives that peak towards higher masses. The derivatives of the density auto-correlations are also negative as the presence of halos (and namely, the astrophysical processes within them) generally suppresses power on these scales, while for the tSZ field the derivative will be positive as the presence of these halos increases the signal. The derivatives of moments that include the tSZ field are orders-of-magnitude larger than those that include the density field alone, showcasing the dramatically higher sensitivity of the tSZ field to baryonic physics. As discussed previously, the mass-dependence of a measurement peaks at higher masses when we increase the order of the moment or if we include the tSZ field.

The mass-dependence in Figure \ref{fig:MassDep} highlights the difficulty in validating our method using simulations that span only sub-$\gpc$ volumes. If we focus on the 2nd moments of the tSZ field, the minimum contributing mass is $\Mtwohc \approx 10^{14.5}\msol$, whereas for the 4th moments this mass scale is $\Mtwohc \approx 10^{15.2}\msol$. The \textsc{IllustrisTNG} simulations have around $\mathcal{O}(100)$ halos at the former mass scale, but only a \textit{single halo} at the latter mass scale. We require hydrodynamical simulations of $\gpc$-scale volumes in order to validate these higher-order moments to good accuracy. However, we reiterate that in practical use-cases, the accuracy requirements for these higher-order moments --- especially those of the tSZ field --- will be more lenient given the larger measurement uncertainties of such moments.

\section{Forecasting}\label{sec:Forecast}

We now forecast the power of current and future surveys in constraining these baryonification parameters, and study the parameters' different degeneracy directions. This is done by forward modelling the observables from surveys; namely, the lensing convergence field and the thermal Sunyaev-Zeldovich field. For brevity, we only briefly describe the procedure below and provide a detailed discussion in Appendix \ref{sec:Forecast:ForwardModel}.

The lensing field is modelled using the same pipeline from \cite{Anbajagane2023CDFs, Anbajagane2023Inflation}. The maps are generated from the density fields of the \textsc{Ulagam} simulations (Section \ref{sec:sims:Ulagam}), and then processed to contain all the observational effects found in the data, including the relevant survey mask/area and source galaxy redshift distribution. The two lensing surveys we focus on are: (i) the Dark Energy Survey \citep[DES,][]{DES2005}, which is an optical imaging survey of 5,000 deg$^2$ of the southern sky, and is currently the largest precision photometric dataset for cosmology. The Year 3 data products and cosmology results are available \citep{Sevilla2021Y3Gold, DES2022Y3}, while the legacy Year 6 dataset is not yet available at the time of publishing this work. We focus on this Year 6 dataset. We then consider (ii) the Rubin Observatory Legacy Survey of Space and Time (LSST), which is a 14,000 deg$^2$ survey that probes higher redshifts, and is the successor to current weak lensing surveys. We forecast only for the final, year 10 dataset. The characteristics of the surveys (shape noise, and $n(z)$ distribution parameters) are given in Table \ref{tab:SurveySpecs}. More details can be found in Appendix \ref{sec:Forecast:ForwardModel:WL}

Next, the tSZ field is modelled by following the approach of \citet{Srini:2022:tSZCluster, Srini:2022:tSZNoise}. We include simple Gaussian models for a number of ``foreground'' componenets, such as the cosmic infrared background, radio background etc., as well as the thermal and atmospheric noise. We focus on two CMB surveys: (i) SPT-3G \citep{Benson2014SPT3G} is a survey of the southern sky with considerably deeper (lower noise) observations than other CMB surveys. The entire footprint covers $\mathcal{O}(10^4)\deg^2$ of the sky with different noise levels. We focus on the main $1500 \deg^2$ footprint (i.e. excluding any wide/summer fields) that also overlaps completely with DES Y6. (ii) The Simons Observatory \citep[SO, ][]{Simons:2019:Experiment} is an upcoming survey of $\approx 20,000 \deg^2$ of the sky, and has nearly complete overlap with the LSST Y10 footprint. For both surveys, we assume the analysis is supplemented by \textit{Planck} data like in existing surveys \citep{Matt:2020:tSZACTDR4, Bleem:2022:tSZ}. The characteristics of the surveys (noise properties, beam, frequency bands etc.) are given in Table \ref{tab:tSZ:surveyspec}. Further details are provided in Appendix \ref{sec:Forecast:ForwardModel:tSZ}.

We note that all our analyses use relatively ``medium''-resolution data-products: the \textsc{Ulagam} simulations are run with $512^3$ particles, and produce maps of $\texttt{NSIDE} = 1024$. These choices set a limit on the scales we can analyse. Simulations of higher resolution enable access to even smaller scales, which would place tighter constraints on the baryonic processes discussed in this work. Therefore all our constraints below could be improved further by pushing the analysis to even smaller scales. In this work, we are limited by the resolution of the simulations available for a Fisher forecast.

These results should also not be over-interpreted or generalized since our analysis explicitly limits the tSZ measurements to $\theta > 3\arcmin$, whereas the tSZ power spectrum has the highest signal-to-noise at even smaller scales, $\ell \sim 3000$ \citep[][see their Figure 4]{Srini:2022:tSZNoise}. Our tSZ maps are generated at $\texttt{NSIDE} = 1024$ --- the same choice as that of the lensing field --- whereas the maps from current SPT datasets are made at $\texttt{NSIDE} = 8192$ \citep{Bleem:2022:tSZ}. Thus, there is more usable information in the tSZ field that we have not accessed in our datavectors given we limit our maps (and therefore our analysis) to a coarser resolution. This is done primarily in the interest of minimizing the computation cost, as generating a tSZ map of $\texttt{NSIDE} = 8192$ would take around $64\times$ longer than generating one of $\texttt{NSIDE} = 1024$. Note that for the lensing analysis, the choice of $\texttt{NSIDE} = 1024$ is realistic, though it may be possible to access smaller scales ($\texttt{NSIDE} = 2048, 4096$) --- if the lensing systematics are better controlled/modelled --- which would improve our constraints. The constraints from both fields could also improve/deteriorate depending on the summary statistics being considered; for examples in the lensing field, see \citet[]{Euclid2023NGCov, Anbajagane2023CDFs, Gatti:2024:LFIResults}. We have not explored this direction, and have limited ourselves to the moments in this work, as they have already been used to constrain cosmology \citep{Gatti2022MomentsDESY3, Gatti:2024:LFIResults}.

\begin{figure*}
    \centering
    \includegraphics[width=1.85\columnwidth]{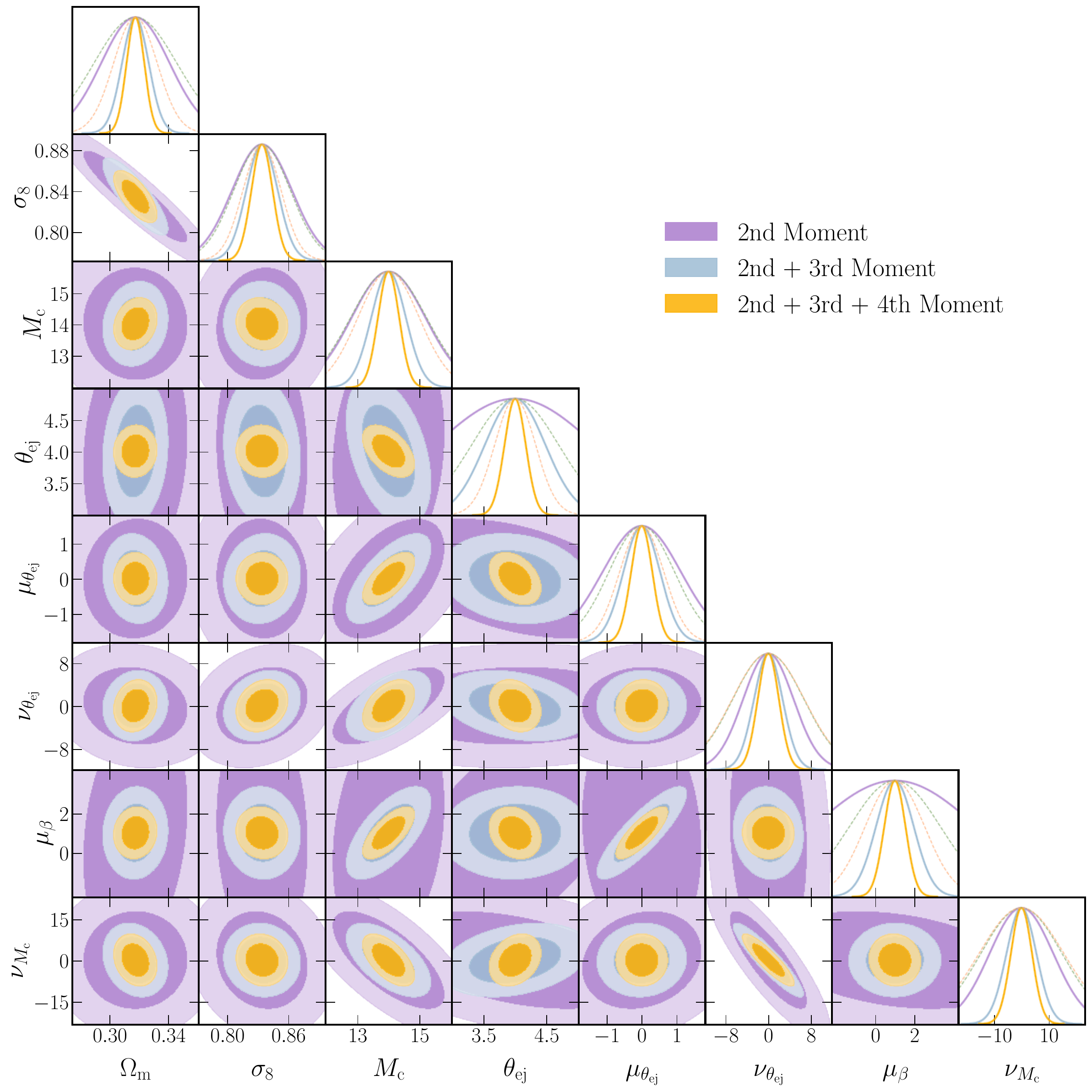}
    \caption{Fisher constraints on the baryonification/cosmology parameters (see Table \ref{tab:params} and Section \ref{sec:baryonify}) from moments of the lensing and tSZ fields constructed with DES Y6 and SPT 3G specifications, respectively. The contours show 1$\sigma$ and $2\sigma$ constraints from using the 2nd/3rd/4th moments. The dashed green/peach lines along the diagonal panels show the 1D posteriors from using only the 3rd/4th moments, respectively. We show only a subset of parameters for the sake of brevity; the full constraints are listed in Table \ref{tab:params}. The inclusion of higher-order information significantly increases constraining power.}
    \label{fig:Fisher:VaryStats}
\end{figure*}

\begin{table*}
    \centering
    \begin{tabular}{|c|ccccc|ccc|cc|}
        \hline
        \multirow{2}{*}{Param$\downarrow$ / Stat.$\rightarrow$.} & \multicolumn{5}{|c|}{$\leftarrow$  Varying statistics  $\rightarrow$} & \multicolumn{3}{c}{$\leftarrow$  Varying fields  $\rightarrow$} & \multicolumn{2}{|c|}{$\leftarrow$  Varying surveys  $\rightarrow$} \\
        & 2nd Mom. & 3rd Mom. & 4th Mom. & 2nd/3rd Mom. & 2nd/3rd/4th Mom. & $\kappa$-only & $y$-only & $\kappa \times y$ & DES$\times$SPT & LSST$\times$SO \\
        \hline
        \hline
        $\Omega_{\rm m}$ & 0.023 & 0.028 & 0.015 & 0.009 & 0.006 & 0.02 & 0.225 & 0.033 & 0.009 & 0.002 \\ 
        $\sigma_8$ & 0.029 & 0.028 & 0.019 & 0.015 & 0.01 & 0.03 & 0.297 & 0.042 & 0.015 & 0.003 \\ 
        \hline
        $M_{\rm c}$ & 1.177 & 1.213 & 0.981 & 0.554 & 0.34 & 10.302 & 9.429 & 1.101 & 0.554 & 0.209 \\ 
        $\theta_{\rm ej}$ & 1.263 & 0.686 & 0.324 & 0.478 & 0.168 & 12.478 & 4.295 & 1.416 & 0.478 & 0.178 \\ 
        $\mu_{\theta_{\rm ej}}$ & 1.114 & 0.933 & 0.628 & 0.52 & 0.302 & 4.169 & 6.769 & 0.914 & 0.52 & 0.216 \\ 
        $\nu_{\theta_{\rm ej}}$ & 4.694 & 6.634 & 6.733 & 2.663 & 1.976 & 9.887 & 34.904 & 4.657 & 2.663 & 0.654 \\ 
        $\mu_{\beta}$ & 4.163 & 2.31 & 1.509 & 0.959 & 0.54 & 20.589 & 12.388 & 2.958 & 0.959 & 0.347 \\ 
        $\nu_{M_{\rm c}}$ & 11.614 & 14.793 & 15.496 & 5.46 & 3.861 & 29.173 & 75.163 & 9.437 & 5.46 & 1.419 \\ 
        \hline
        $\theta_{\rm co}$ & 0.139 & 0.04 & 0.026 & 0.03 & 0.014 & 2.529 & 0.618 & 0.086 & 0.03 & 0.008 \\ 
        $\alpha_{\rm nth}$ & 0.389 & 0.304 & 0.517 & 0.163 & 0.127 & -- & -- & -- & 0.163 & 0.057 \\ 
        $\gamma$ & 11.175 & 6.325 & 4.61 & 2.85 & 1.842 & 38.967 & 68.416 & 6.248 & 2.85 & 1.195 \\ 
        $\delta$ & 5.348 & 5.409 & 5.786 & 2.648 & 1.985 & 15.55 & 113.361 & 7.169 & 2.648 & 0.801 \\ 
        \hline
    \end{tabular}
    \caption{The constraints presented in Figure \ref{fig:Fisher:VaryStats}, \ref{fig:Fisher:VaryFields}, and \ref{fig:Fisher:VarySurveys} for (i) different statistics measured on both the lensing and tSZ field (left columns), (ii) for different combinations of fields using the 2nd and 3rd moments (middle columns), and (iii) for different combinations of surveys using the 2nd and 3rd moments of the lensing and tSZ fields (right columns), respectively. The inclusion of higher-order statistics improves all constraints by factors of 3 to 4. Including the tSZ field in the analysis similarly improves constraints by up to factors of 10. The constraints from LSST Y10 and SO are factors of 4-5 better than that of DES Y6 and SPT-3G. The parameters are partitioned visually into cosmological parameters, baryonification parameters shown in the figures of this work, and remaining parameters that are not presented but are marginalized over. In analysis (ii), we do not use $\alpha_{\rm nth}$ due to it having no impact on the $\kappa$ field; see text in Section \ref{sec:Forecast} for details.}
    \label{tab:Fisher}
\end{table*}

We now present the Fisher information for a number of cases, each changing the statistics being measured, the fields being used, and/or the parameters being varied. The key goal of these estimates is to understand the degeneracy directions of the different parameters under different analysis choices, and also the relative constraining power of surveys. The Fisher information is estimated as,
\begin{equation}\label{eqn:Fisher}
    \boldsymbol{F}_{ij} = \sum_{m,n}\frac{d\widetilde{X}_m}{d\theta_i}\big(\mathcal{C}^{-1}\big)_{mn}\frac{d\widetilde{X}_n}{d\theta_j},
\end{equation}
where $\frac{d\widetilde{X}_m}{d\theta_i}$ is the mean derivative of point $m$ in data vector $X$ with respect to parameter $\theta_i$, where the mean is computed using 400 independent realizations of the surveys for the analysis of DES and SPT, and 300 realizations for that of LSST and SO. We compute this derivative around the fiducial parameter values, as detailed in Table \ref{tab:params}. Then, $\mathcal{C}^{-1}$ is the inverse of the numerically estimated covariance matrix and includes the Hartlap correction factor \citep{Hartlap2007},
\begin{equation}\label{eqn:invertcov}
    \mathcal{C}^{-1} \rightarrow \frac{N_{\rm sims} - N_{\rm data} - 2}{N_{\rm sims} - 1} \,\mathcal{C}^{-1}.
\end{equation}
The Hartlap factor for all analyses in this work is $\gtrsim 0.9$. The datavector, $X$, consists of the moments of the field, computed on 10 scales ranging from $3.2\arcmin < \theta < 200\arcmin$. These are the same choices as previous works on the moments of the lensing field \citep{Gatti2020Moments, Anbajagane2023CDFs, Gatti:2024:LFIResults} and we adopt it for the tSZ field as well. The covariance matrix is estimated by generating 6000 to 8000 realizations of the surveys, with the number varying depending on the survey being analysed. The $\mathcal{O}(10^4)$ realizations are enabled by the 2000 independent simulations available in the \textsc{Ulagam} simulation suite. These simulations are run at the fiducial cosmology, chosen to be the best fit values of \citet{Planck2016CosmoParams}, and the derived baryonification products use the fiducial values of the model parameters (see Table \ref{tab:params}).

The Fisher information estimated in Equation \eqref{eqn:Fisher} can be artificially increased by numerical noise as this noise can break parameter degeneracies; see \citet[][their Appendix A]{Coulton:2023:FisherBias} for a discussion of this. We quantify this amplification in our estimates by varying the number of realizations used to compute the derivatives and the covariance. The Fisher information changes by $\lesssim 2\%$ if we estimate the covariance using half the number of realizations. Doing the same exercise on the derivative estimates change the Fisher information by $\lesssim 10\%$. Our discussion below focus on constraints improving by more than factors of 2, and are therefore unaffected by any such numerical uncertainties at the $10\%$ level.

Figure \ref{fig:Fisher:VaryStats} shows the constraints for different combinations of moments, as measured on mock maps of DES Y6 and SPT 3G; the numbers are listed in Table \ref{tab:Fisher} as well. In addition to the baryonification parameters, we also include the cosmological parameters $\Omega_{\rm m}$, which is the matter energy fraction at $z = 0$, and $\sigma_8$, which is the root-mean squared fluctuations of the density field at $z = 0$, smoothed on a $8 \mpc/h$ scale. We include, and therefore marginalize over, all baryonification parameters that are listed with a prior in Table \ref{tab:params}. We however only show a subset of them in this plot for visibility.

The inclusion of higher-order moments (blue and yellow contours) greatly improves the constraints over using the 2nd moments alone (purple contours). This is expected as baryonic imprints are tied to the astrophysical processes within halos and are therefore stronger on non-linear scales. These scales are also where the field is more non-Gaussian, and parameter constraints improve from including measurements of any higher-order correlations at these scales. \citet{Arico:2021:BkBaryons} already show that a range of baryonification parameter choices can accurately fit a matter power spectrum from simulations but only a subset of those can also jointly fit the matter bispectrum, indicating that the addition of higher-order information constrains the parameter space further. We expand on this by using up to 4th order in the fields, and also by using both the lensing and tSZ fields. Figure \ref{fig:Fisher:VaryStats} also shows the marginalized constraints from using \textit{only} the 3rd moments (4th moments), presented as a green (peach) dashed line in the 1D posteriors along the diagonal panels of the triangle plot.\footnote{For the 4th moments, we have used only the \textit{connected} piece, $\langle x_1x_2x_3x_4\rangle - (\langle x_1x_2\rangle\langle x_3x_4\rangle + \text{perms.})$, so that there is a clean separation between the information in the 2nd moments and 4th moments.} In many cases, the ``only 3rd moments'' and ``only 4th moments'' constraints are more constraining than the 2nd moments alone (though this is not the case for the cosmology parameters).

For most parameters, the largest relative improvement is between using 2nd moments alone and using the combination of 2nd and 3rd moments. Including the 4th moments to this combination improves constraints as well but the relative gain is smaller. This is similar to the lensing-only analysis of \citet{Anbajagane2023CDFs}, where the 4th and 5th moments did not improve the constraining power. The one exception we find is $\theta_{\rm ej}$, which improves by nearly a factor of three when including the 4th moment.

Figure \ref{fig:Fisher:VaryStats} indicates that a number of degeneracy directions are broken by including higher-order information. It also shows that the cosmological parameters, $\Omega_{\rm m}$ and $\sigma_8$, can be constrained even when marginalizing over astrophysical systematics. Note that these analyses use both the lensing field \textit{and} the tSZ field. Limiting the analysis to the lensing field alone deteriorates the constraining power significantly (see Figure \ref{fig:Fisher:VaryFields} below). In general, this result highlights the utility in jointly constraining cosmology and astrophysics using a combination of the lensing field and tSZ field. \citetalias{Pandey2024godmax} have already shown this to be the case when using only two-point statistics in the data vector (see their Figure 8).

\begin{figure}
    \centering
    \includegraphics[width=1\columnwidth]{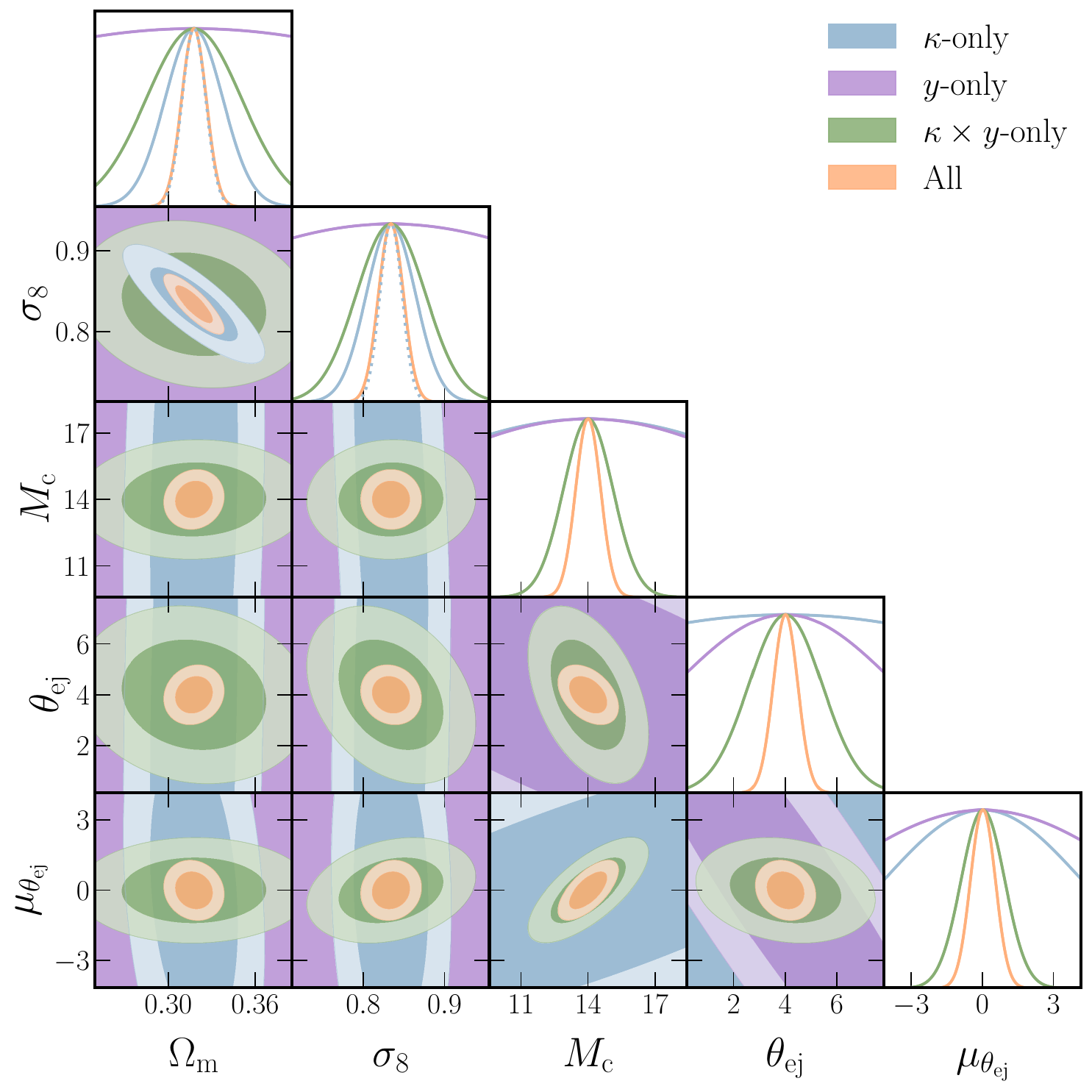}
    \caption{Fisher constraints on the baryonification and cosmological parameters from different combinations of the convergence and tSZ field as measured by DES Y6 and SPT 3G. The datavector uses the 2nd and 3rd moments. The convergence field and tSZ field have different sensitivities to the parameters, and the degeneracy directions are often orthogonal. As a result the combination of fields is significantly more constraining. The 1D marginal distributions of $\Omega_m$ and $\sigma_8$ also show a dotted blue line, which is the lensing-only constraints without marginalizing over any baryonification parameters.}
    \label{fig:Fisher:VaryFields}
\end{figure}

While the above analysis decomposes the information into different orders, we now decompose it by field. In Figure \ref{fig:Fisher:VaryFields} we split the Fisher information by convergence field and tSZ field. In the former case, we use all auto/cross-correlations between all tomographic bins of the lensing data, while in the latter we use only auto-correlations of the singular tSZ field. We then show the constraints from moments that have at least one power of the lensing field \textit{and} one of the tSZ field, and finally the constraints from all moments of the two fields. While the above analysis also used the 4th order moments, we will now use the combination of the 2nd and 3rd as our fiducial datavector, given this measurement has already been used (for the lensing field) to constrain cosmology \citep[\eg][]{Gatti2022MomentsDESY3, Gatti:2024:LFIResults}. This analysis also does not vary the non-thermal pressure parameter $\alpha_{\rm nt}$, see Equation \eqref{eqn:f_nth}, since this parameter has no impact on the $\kappa$ field, leading to zero Fisher information and therefore a singular Fisher matrix. We choose to remove this parameter for three variants (lensing-only, tSZ-only, cross correlations-only) in this analysis --- and not just for the $\kappa$-only analysis where the issue of a singular Fisher matrix arises --- to preserve consistency across analysis setups and therefore simplify the comparisons of the different constraints.

Figure \ref{fig:Fisher:VaryFields} shows that the lensing and tSZ field have different sensitivities to combinations of parameters; in many panels, the degeneracy directions of the convergence-only and tSZ-only case are nearly orthogonal. As a result, any analysis using combinations of the convergence and tSZ field is far more constraining than the individual fields. Another interesting note is that the lensing field is able to better constrain combinations of the redshift parameters, $\nu_{\rm \theta_{ej}}$ and $\nu_{M_{\rm c}}$, as shown in Table \ref{tab:params}. This is because the lensing survey has more redshift information via the availability of tomographic bins, whereas the tSZ field is a singular field integrated over all redshifts. This result highlights the highly complementary nature of combining the lensing and tSZ fields when constraining baryonification models; \citetalias{Pandey2024godmax} (see their Figure 8) have shown the same for the two-point functions. It is also interesting that the cross-correlations alone can provide a relatively good constraint on the baryonification parameters (as well as cosmology). Such cross-correlations are more robust measurements than their auto-correlation counterparts, as additive systematics in each field (that are uncorrelated across the different fields) will not be present in the measurement.

The field-by-field comparison of Figure \ref{fig:Fisher:VaryFields} also highlights the improvement in cosmology constraints ($\Omega_{\rm m}$ and $\sigma_8$) due to the addition of the tSZ field. The cross-correlation of the lensing and tSZ fields is a significant factor in this improvement. The results also indicate that the tSZ-only analysis is unable to jointly constrain baryonification models alongside cosmology. However, this result should not be over-interpreted or generalized since our analysis explicitly limits the tSZ measurements to $\theta > 3\arcmin$, whereas the tSZ power spectrum has the highest signal-to-noise at even smaller scales, $\ell \sim 3000$ \citep[][see their Figure 4]{Srini:2022:tSZNoise}. Our tSZ maps are generated at $\texttt{NSIDE} = 1024$ --- the same choice as that of the lensing field --- whereas the maps from current SPT datasets are made at $\texttt{NSIDE} = 8192$ \citep{Bleem:2022:tSZ}. Thus, there is more usable information in the tSZ field that we have not accessed in our datavectors given we limit our maps (and therefore our analysis) to a coarser resolution. This is done primarily in the interest of minimizing the computation cost, as generating a tSZ map of $\texttt{NSIDE} = 8192$ would take around $64\times$ longer than generating one of $\texttt{NSIDE} = 1024$. For the lensing analysis, the choice of $\texttt{NSIDE} = 1024$ is realistic, with some potential to extend to higher resolution if the systematics can be controlled robustly.

The 1D marginal distributions in Figure \ref{fig:Fisher:VaryFields} also shows a dotted blue line, which are the results from a lensing-only analysis (using 2nd and 3rd moments) but with all baryonification parameters fixed. The constraints on cosmology --- $\sigma(\Omega_m) = 0.008$ and $\sigma(\sigma_8) = 0.013$ --- are comparable to those from using all 2nd and 3rd auto/cross-moments of the lensing and tSZ field \textit{after marginalizing over} all baryonification parameters. Note that this compares lensing-only with a lensing plus tSZ analysis. In Appendix \ref{appx:fisher:params}, we consistently use a lensing plus tSZ analysis, with all baryonification parameters either fixed or varied in different stages, and find the constraints on cosmology degrades by $\approx 60\%$ if we vary all baryonification parameters.

\begin{figure}
    \centering
    \includegraphics[width=1\columnwidth]{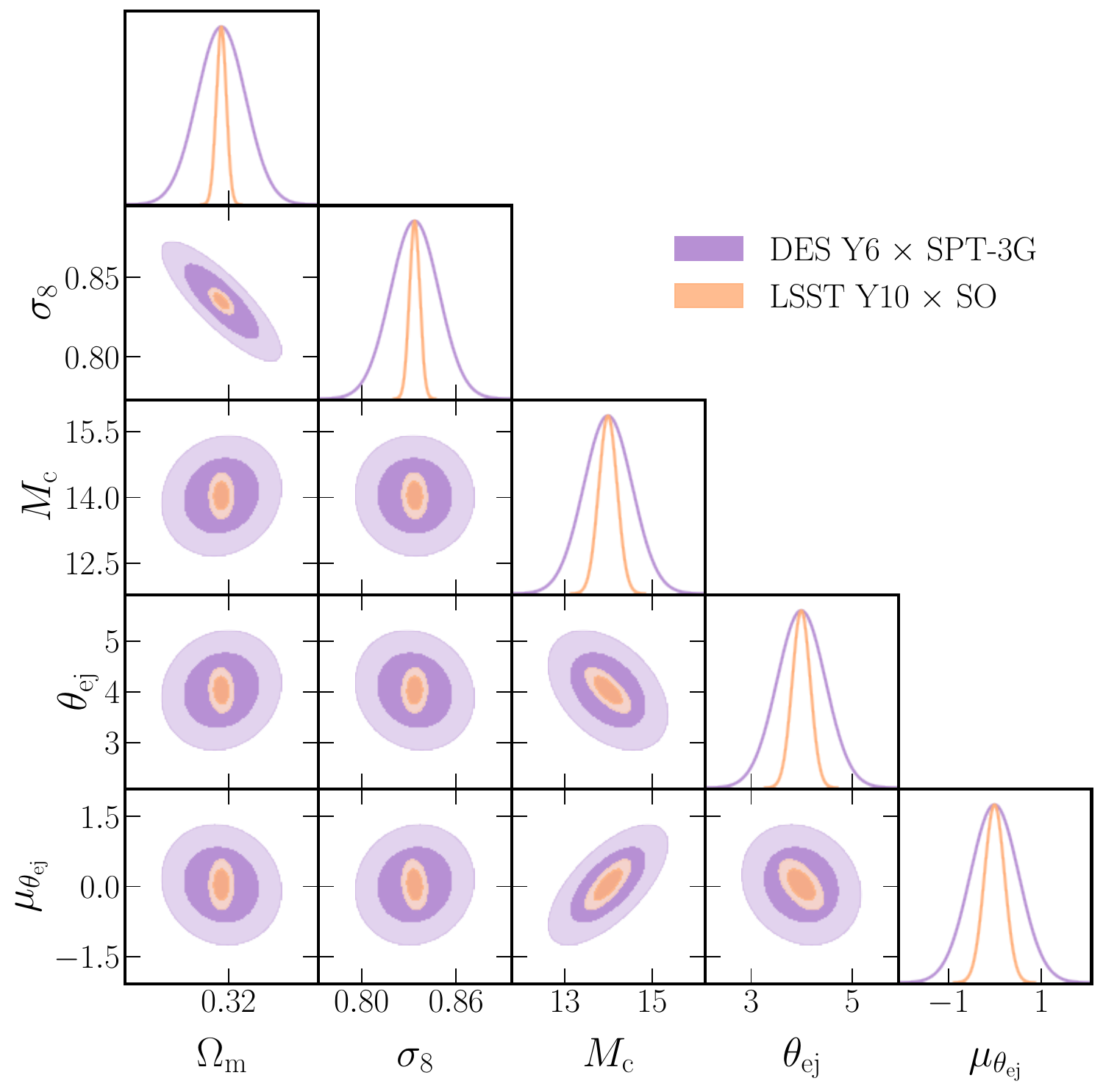}
    \caption{Fisher constraints of the baryonification parameters from different survey datasets. The datavector includes only the 2nd and 3rd moments, which contain the bulk of the signal-to-noise. The constraints from LSST Y10 and SO is between 3 to 5 times better than that from DES Y6 and SPT-3G. We have marginalized over all other parameters in the analysis. The constraints are also listed in Table \ref{tab:Fisher}.}
    \label{fig:Fisher:VarySurveys}
\end{figure}

Finally, Figure \ref{fig:Fisher:VarySurveys} compares constraints from current surveys (DES Y6 and SPT-3G) with upcoming ones (LSST Y10 and SO). We find the parameter degeneracies in the upcoming surveys are broadly similar to those in the current surveys. The extended redshift range probed by LSST Y10 (compared to DES Y6) has not changed the degeneracy directions. The parameter constraints themselves improve by factors of 3 to 5 depending on the specific parameter. While these results are simplistic in their accounting of systematic uncertainties --- we have not marginalized over intrinsic alignments for the lensing field, and we include \textit{no} systematic model to be marginalized over for the tSZ field --- they show the gain that could be realized from the final survey datasets of the next decade, relative to constraints from ongoing surveys.

All analyses above, summarised in Table \ref{tab:Fisher}, show that a number of analysis setups result in parameter uncertainties that are of the same order as the fiducial parameter values listed in Table \ref{tab:params}: (i) the $\nu_X$ parameters, which characterize the redshift evolution of $\theta_{\rm ej}$ and $M_{\rm c}$, (ii) $\gamma$ which is the gas slope at $R \sim R_{\rm ej}$, and (iii) the mass-dependence parameters $\mu_X$. This may indicate that robust predictions of the lensing and tSZ moments can be obtained while fixing these parameters. However, a robust validation of this statement requires fitting this model (with and without varying the parameters mentioned here) to mock surveys derived from different hydrodynamical simulations. Appendix \ref{sec:higher_z} also presents results that indicate the $\nu_X$ parameters may not be necessary for jointly modelling the moments measurements across redshift. \citet{Bigwood:2024:BaryonsWLkSZ}, who use the baryonification model of \citet{Giri2021Baryon} to analyse data from DES and the Atacama Cosmology Telescope (ACT), also find qualitatively similar behavior to us for constraints on their mass-dependence parameter $\mu$ (see their Table B2).

\section{Conclusions}\label{sec:Conclusions}

Cosmology is in a golden age of data-driven science, and this is set to only improve over the upcoming decade. For example, measurements of weak lensing will extend to larger fractions of the sky, and to much higher statistical precision \citep{Spergel:2015:Roman, LSST2018SRD, Euclid}. Such measurements will access a broader range of scales than was previously available. However, the actual use of these measurements --- namely those on smaller, more non-linear scales --- for obtaining constraints is a non-trivial task due to uncertainties in modelling the physics of baryons on these scales. Addressing this issue is tantamount to better constraining $\Lambda$CDM cosmology \citep[\eg][]{Secco2022Y3Shear, Amon2022Y3shear, Amon2022S8Baryons} and also to using lensing to extract information about other cosmological phenomenon, such as inflation \citep{Anbajagane2023Inflation, Sam:2024:LensingFNL}.

We introduce here a map-level baryonification technique that jointly generates baryon-corrected density fields (and thereby lensing fields) \textit{as well as} tSZ fields, using N-body simulations as an input. This is done in a computationally efficient manner by working directly with full-sky maps from simulations rather than with three-dimensional particle snapshots. We perform various validations on this technique and characterize its performance on 2D simulation fields, as well as in mock analyses of wide-sky datasets. We perform all validations by using the moments of the fields, from 2nd to 4th order, as our summary statistics. We include all auto- and cross-correlations at these orders. Our main results are as follows:

\begin{itemize}
    \item The baryonification method can be applied directly to 2D fields (rather than 3D snapshots) in a robust manner. After accounting for the pixel window function and the line-of-sight projection scale, the residuals between the 2D and 3D version is under 2\% for most physical scales (Figure \ref{fig:PartShellBaryon}).
    \item The model predictions are fairly insensitive to secondary halo properties, such as halo concentration and ellipticity. (Figure \ref{fig:c200c:stat} and \ref{fig:elldep:Stats}).
    
    \item The model is adequately flexible and can jointly fit all moments (up to 4th order) of the total matter density and gas pressure fields, as measured in \textsc{IllustrisTNG}. The residuals are within the uncertainties in all cases (Figure \ref{fig:TNGFit}). The model can also jointly fit across multiple redshifts (Figure \ref{fig:TNGFit_z1}).
    
    \item The moments have significant differences in the halo mass their signal is most sensitive to; the total mass range spans nearly two orders of magnitude across the different moments, highlighting their complementary nature (Figure \ref{fig:MassDep}).
    
    \item A Fisher forecast of current/upcoming surveys shows the inclusion of higher-order information dramatically improves the constraining power on the baryonification model (Figure \ref{fig:Fisher:VaryStats}).
    
    \item The tSZ and lensing fields are more constraining for specific subsets of parameters, and the parameter degeneracy directions in the different fields are nearly orthogonal, leading to the combination of fields providing significantly better constraints (Figure \ref{fig:Fisher:VaryFields}). The lensing data is also more informative on redshift-dependent parameters due to the tomographic redshift information of the dataset.
\end{itemize}

This work provides a first, initial confirmation that the baryonification method is viable for jointly modelling the lensing and tSZ fields directly on the sky, across multiple orders in the fields. We stress that this is \textit{not} a final and conclusive validation of the map-level baryonification technique. As was pointed out before, a more thorough validation will require full-sky hydrodynamical simulations, such as \textsc{Millenium-TNG} \citep{Pakmor:2023:MTNG} and/or the \textsc{Flamingo} project \citep{Schaye:2023:Flamingo}, and a focus on other popular summary statistics used for analyses of weak lensing and tSZ fields.

Once the baryonification method has been more precisely validated, a limiting step in applying it to data will be the forward model of the tSZ field and its foregrounds, done consistently \textit{across multiple orders}. \citet{Omori2022Agora} have provided a precisely validated prescription for converting an N-body simulation into realistic tSZ fields (with all foregrounds consistently modelled, and then also verified at the two-point level), but this requires high fidelity simulation products which are often not available for the large simulation suites used for forward modelling the lensing field \citep{Kacprzak2023Cosmogrid, Anbajagane2023Inflation, Jeffrey:2024:LFIResult, Gatti:2024:LFIResults}. Explorations of more approximate tSZ forward models will be needed in order to analyse the tSZ field with the same simulation-based forward modelling approach used for the lensing field.

The baryonification method has rapidly become a important tool in characterizing the two-point correlation functions measured by lensing surveys. The method, however, is a general prescription on how to add baryonic fields to N-body simulations through an approximate, phenomological approach. In this work, we use this formalism to model the lensing and tSZ field. One could similarly use this prescription to model other fields that are observables of gas physics, such as the kinematic SZ effect, which has already been measured using different combinations of CMB and galaxy surveys \citep[\eg][]{Maya:2023:kSZ, Boryana:2024:kSZ}, and the X-ray luminosity, measured for a large fraction of the sky by the eROSITA survey \citep{Bulbul:2024:eROSITA}. In fact, some of these observables have already been used to constrain baryonification models \citep{Schneider2019Baryonification, Schneider:2022:BaryonConstraints, Grandis:2024:XrayLensing, Bigwood:2024:BaryonsWLkSZ}. Thus, we have only probed some aspects of the full potential provided by the baryonification formalism. 

The current state of widefield survey science is rapidly advancing towards surveys that cover $\mathcal{O}(1)$ fractions of the sky, and have overlapping areas of $\mathcal{O}(10^4) \deg^2$. In this data-rich age, the advent of multi-survey, multi-wavelength science provides powerful ways to extract robust physical inferences at ever-increasing precision. The baryonification method provides one way of performing joint-analyses across different surveys in a rapid, flexible way, and is complementary to other approaches such as dedicated hydrodynamical simulations. Our prescription for the map-level baryonification pipeline is documented and made publicly available. We hope this enables easy explorations/extensions of the baryonification pipeline that build up the synergies of the growing multi-survey landscape.

\section*{Acknowledgements}

DA is supported by the National Science Foundation Graduate Research Fellowship under Grant No. DGE 1746045. SP is supported by the Simons Collaboration on Learning the Universe. CC is supported by NSF grant AST-2306166.

We thank Marco Gatti and Yuuki Omori for discussions during early stages of this work, Srini Raghunathan for guidance on the tSZ forward models used in \citet{Srini:2022:tSZCluster, Srini:2022:tSZNoise}, and Kayla Kornoelje for details on the linear combination method used by SPT. We also thank the referee for their many helpful comments that greatly enhanced our discussion and results. The conception and pursuit of this work arose from many conversations between DA and Eric Baxter. The authors are grateful to Eric for his collaboration and friendship over the many years, and for his efforts in making our research communities vibrant avenues for discussion and discovery. He will be missed.

All analysis in this work was enabled greatly by the following software: \textsc{Pandas} \citep{Mckinney2011pandas}, \textsc{NumPy} \citep{vanderWalt2011Numpy}, \textsc{SciPy} \citep{Virtanen2020Scipy}, and \textsc{Matplotlib} \citep{Hunter2007Matplotlib}. We have also used
the Astrophysics Data Service (\href{https://ui.adsabs.harvard.edu/}{ADS}) and \href{https://arxiv.org/}{\texttt{arXiv}} preprint repository extensively during this project and the writing of the paper.

\section*{Data Availability}

All simulation datasets used in this work are publicly available, and can be found at the links provided in the relevant subsections of Section \ref{sec:sims&stats}.

The baryonification pipeline is publicly available at \url{https://github.com/DhayaaAnbajagane/BaryonForge}.

\bibliographystyle{mnras}
\bibliography{References}



\appendix

\section{Forward modelling pipeline}\label{sec:Forecast:ForwardModel}

A key result of our work is the simulation-based forecast of current and future surveys (Section \ref{sec:Forecast}). The forecast is done by forward modeling the observables from these surveys; namely, the lensing convergence field (Section \ref{sec:Forecast:ForwardModel:WL}) and the thermal Sunyaev-Zeldovich field (Section \ref{sec:Forecast:ForwardModel:tSZ}). This appendix details the exact procedures used to create our forward model for each field.

Throughout, we use maps with $\texttt{NSIDE} = 1024$. The angular scale of the pixels ($3.4\arcmin$) correspond to physical scales of $0.4 < r < 7 \mpc$ for the redshift range spanned by the simulation, $0.1 < z < 3.5$.

\subsection{Weak lensing}\label{sec:Forecast:ForwardModel:WL}

The lensing convergence field, $\kappa$, is a line-of-sight integral of the density field
\begin{equation}\label{eqn:convergence_definition}
    \kappa(\nhat, z_s) = \frac{3}{2}\frac{H_0^2\Omega_{\rm m}}{c^2}\int_0^{z_s}\!\!\!\delta(\nhat, z_j) \frac{\chi_j(\chi_s - \chi_j)}{a(z_j)\chi_s}dz_j\frac{d\chi}{dz}\bigg|_{z_j},
\end{equation}
where $z_s$ is the redshift of the ``source'' plane/galaxies being lensed, $\nhat$ is the pointing direction on the sky, $\delta$ is the overdensity field, $\chi$ is the comoving distance from an observer to a given redshift, $a$ is the scale factor, $H_0$ is the Hubble constant, $\Omega_{\rm m}$ is the matter energy density fraction at $z = 0$, and $c$ is the speed of light. Here, we have used the shorthand $\chi(z_s) \equiv \chi_s$ and $\chi(z_j) \equiv \chi_j$.

We use the same forward-modelling pipeline from \cite{Anbajagane2023CDFs, Anbajagane2023Inflation}, whose description we reproduce below for completeness. The lensing maps are generated from the density fields of the \textsc{Ulagam} simulations, and then processed to include all the observational effects found in the data. These procedures have been utilized in many analyses/forecasts of weak lensing data \citep{Fluri2019DeepLearningKIDS,  Zurcher2021WLForecast, Fluri2022wCDMKIDS, Gatti2022MomentsDESY3, Zurcher2022WLPeaks, Gatti2023SC, Anbajagane2023CDFs, Anbajagane2023Inflation, Gatti:2024:LFIValidation, Gatti:2024:LFIResults, Jeffrey:2024:LFIResult}. As mentioned in Section \ref{sec:Forecast}, we focus on two surveys: (i) the Dark Energy Survey \citep[DES,][]{DES2005}, Year 6 dataset, and the (ii) the Rubin Observatory Legacy Survey of Space and Time (LSST), year 10 dataset. We detail below our forward modeling procedure:

\textbf{Constructing lensing convergence shells.} The model starts with lightcone shells of the particle counts, which are two-dimensional, \textsc{HEALPix} maps of the (projected) particle counts at different redshifts. We convert the particle count map to an overdensity map as
\begin{equation}\label{eqn:particle_to_density}
    \delta^i = N^{i}_{\rm p}/\langle N_{\rm p} \rangle - 1,
\end{equation}
where $N^{i}_{\rm p}$ is the number of particles in pixel $i$, and the average is computed over all pixels in the shell. The density shells can then be converted into the convergence $\kappa$ using Equation \ref{eqn:convergence_definition}, after converting the integral over redshift into a discrete sum over lightcone shells. Note that our baryonification step is performed separately on each individual density shell, using the halos in just that shell. The baryonified density shells are then used in the summation of Equation \eqref{eqn:convergence_definition}.  

\textbf{Source galaxy redshift distributions.} We perform a weighted average of the convergence shells to construct the convergence within the different \textit{tomographic bins} of each survey. The weights used in this averaging are the source galaxy redshift distribution, $n(z)$, of the chosen bin and survey,
\begin{equation}
    \kappa^A(\nhat) = \sum_{j = 1}^{N_{\rm steps}} n^A(z_j)\kappa(\nhat, z_j)\Delta z,
\end{equation}
where $\kappa^A$ is the true convergence of a tomographic bin, $A$. The $n(z)$ is obtained from the following: for DES Year 6 and LSST Year 10 we use the same $n(z)$ as that used in \citet[][see their Table 1]{Zhang2022CMBLSS}. The LSST modeling in that work follows the baseline analysis choices of \citet[][see their Appendix D2.1]{LSST2018SRD}. The redshift distributions for DES Y6, and LSST Y1 and Y10 are parameterized as,
\begin{equation}
    \frac{dN}{dz} \propto z^2\exp\bigg[-\bigg(\frac{z}{z_0}\bigg)^\alpha\bigg],
\end{equation}
with parameters given in Table \ref{tab:SurveySpecs}. Once the $n(z)$ of the full survey is defined, we split it into 4 (5) tomographic bins for DES Y6 (Y10) of equal number density. Each bin is then convolved with a Gaussian of width given by the photometric redshift uncertainty, also quoted in Table \ref{tab:SurveySpecs}. The DES Y6 distribution is non-zero only between $0.2 < z < 1.3$, following \citet[][see their Table 1]{Zhang2022CMBLSS}. The LSST distributions are cut at $z < 3.5$. See Figure 1 of \citet{Anbajagane2023Inflation} for the exact distributions we use in this work.

\begin{table}
    \centering
    \begin{tabular}{c|c|c|c}
        Survey & $(z_0, \alpha)$ & $n_{\rm gal}/{\rm arcmin}^2$ & $\sigma_z$ \\
        \hline
        \hline
        DES Y6 & (0.13, 0.78) & 9 & 0.1(1 + z)\\
        LSST Y10 & (0.11, 0.68) & 27 & 0.05(1 + z)\\
        \hline
    \end{tabular}
    \caption{The redshift distribution and source galaxy number density assumed for the upcoming surveys. All numbers are taken from \citet[][see their Table 1]{Zhang2022CMBLSS}.}
    \label{tab:SurveySpecs}
\end{table}

\textbf{Constructing lensing shear shells.} Weak lensing surveys measure galaxy shapes, which primarily trace the shear field, $\gamma$, and not the convergence field, $\kappa$. However, the shear and convergence field can be transformed into each other using the Kaiser-Squires (KS) transform \citep{Kaiser1993KS}, implemented in harmonic space as
\begin{equation}\label{eqn:Kappa2Shear}
    \gamma^{\ell m}_E + i\gamma^{\ell m}_B = -\sqrt{\frac{(\ell + 2)(\ell - 1)}{\ell(\ell + 1)}} \bigg(\kappa^{\ell m}_E + i\kappa^{\ell m}_B \bigg),
\end{equation}
where $X_{\{E, B\}}$ are the E-mode and B-mode (or Q and U polarizations, in \textsc{HealPix} notation) of the field.

\textbf{Shape noise.} Upon generating the two shear fields, we add shape noise in real space. The forward-modelled field includes Gaussian shape noise with a standard deviation given as
\begin{equation}\label{eqn:shapenoise}
    \sigma_\gamma = \frac{\sigma_e}{\sqrt{n_{\rm gal}A_{\rm pix}}},
\end{equation}
where $n_{\rm gal}$ is the source galaxy number density, and $A_{\rm pix}$ is the pixel area for a given map resolution. All maps in this work use $\texttt{NSIDE}=1024$, corresponding to a pixel resolution of $3.2 \arcmin$. The per-galaxy shape noise is taken to be $\sigma_e = 0.26$.

\textbf{Survey mask \& mass map construction.} The noisy shear field --- which is the sum of the true shear fields and the shape noise fields --- is then masked according to the survey footprint. For DES Y6 we use the provided survey mask in the year 3 data release \citep{Sevilla2021Y3Gold}. For LSST Y10, we divide the full-sky into three equal-area cutouts of roughly 14,000 deg$^2$ each, which is the expected area coverage \citep[][see Appendix C1]{LSST2018SRD}. The noisy shear maps are converted to convergence using the relation in Equation \ref{eqn:Kappa2Shear}. We only use the resulting E-mode field, $\kappa_E$, for our analyses. This follows the same procedures used in \citet{Chang2018MassMap, Niall2021MassMap}.

\textbf{To summarize,} lensing convergence maps are constructed from the raw particle number count maps. The $n(z)$ distributions for a given survey are used to obtain the convergence map in a given tomographic bin. This convergence map is converted to shear maps, the relevant shape noise is added, the relevant survey mask is applied, and then the noisy shear maps are converted back to a noisy convergence map. The set of procedures listed above is the standard approach for forward modeling the lensing field \citep[\eg][]{Zurcher2021WLForecast, Zurcher2022WLPeaks, Gatti2022MomentsDESY3, Anbajagane2023CDFs, Anbajagane2023Inflation}. Thus, our final convergence maps will be an accurate representation of the survey data. We have intentionally ignored the modelling of intrinsic alignments \citep{Troxel2015IAReview, Lamman2023} --- which is a systematic effect considered in all the models cited above --- as this work focuses on a pure test of the degeneracy directions between cosmology and the baryonification paraemeters. We will explore, in future works, the relationship between the baryonification model and other systematics.

\subsection{Thermal Sunayev-Zeldovich effect}\label{sec:Forecast:ForwardModel:tSZ}

The raw measurements from a CMB survey are the temperature or power measured at different frequencies. At a given frequency, the observations contain contributions from a wide variety of astrophysical and cosmological sources, of which one is the tSZ field. This tSZ field can be filtered out of these maps through a ``Linear Combination'' (LC) procedure, which is a weighted sum of frequency maps performed in harmonic space \citep[\eg][]{Matt:2020:tSZACTDR4, Bleem:2022:tSZ}. The exact method used to calculate the weights results in a variety of LC methods, such as the \textit{internal} Linear combination (ILC), Needlet ILC etc. We use the LC algorithm of \citet{Bleem:2022:tSZ}, which uses a theoretically estimated covariance matrix for the angular power spectra of the frequency maps. This estimate requires models for the other, ``foreground'' fields that contribute power to the different frequency maps. For this, we follow the approach of \citet{Srini:2022:tSZCluster, Srini:2022:tSZNoise}, who base their model on the measurements of \citet{George:2015:foregrounds} from the South Pole Telescope (SPT) data. 

\begin{figure}
    \centering
    \includegraphics[width=\columnwidth]{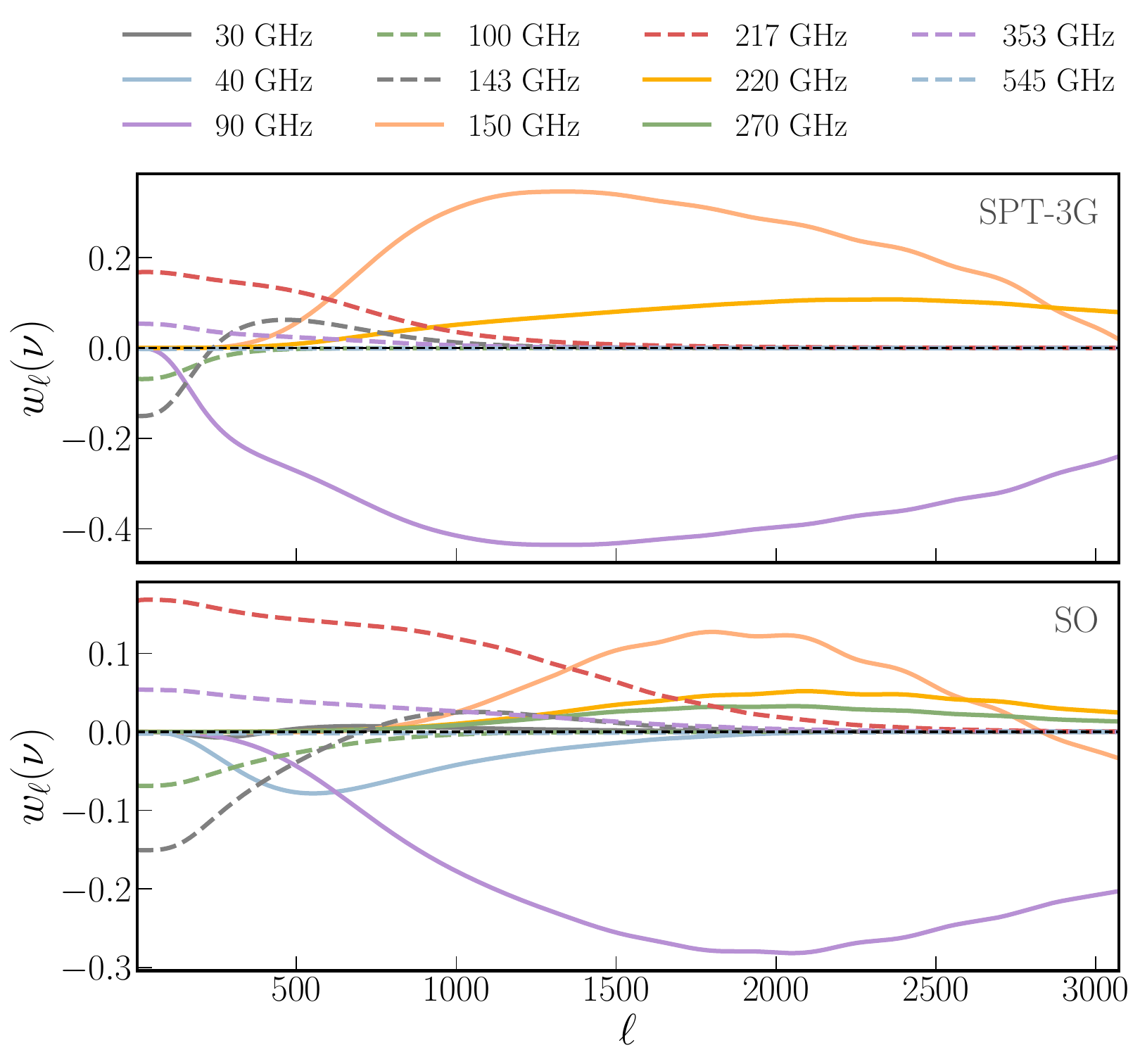}
    \caption{The harmonic-space weights used in the linear combination procedure --- see Equation \eqref{eqn:tSZ:ILC_weights} --- to combine the different frequency maps into a minimum-variance tSZ map, shown for both SPT-3G (top) and Simons Observatory (bottom). The weights for the five \textit{Planck} maps are shown as dashed lines. A black dotted line shows $w = 0$ for reference.}
    \label{fig:ILC_weights}
\end{figure}

As mentioned above in Section \ref{sec:Forecast}, we consider two CMB surveys: SPT-3G \citep{Benson2014SPT3G} and the Simons Observatory \citep[SO, ][]{Simons:2019:Experiment}. For both surveys, we assume their analyses are supplemented by \textit{Planck} data, following the procedures used in existing surveys \citep{Matt:2020:tSZACTDR4, Bleem:2022:tSZ}. As a result, we forward model the \textit{Planck} maps in this work. We follow \citet{Bleem:2022:tSZ} in using data from only the high-frequency instrument (HFI) of \textit{Planck}. We also additionally use the $545 \GHz$ band (which is not used in \citet{Bleem:2022:tSZ}), but as we will show later in Figure \ref{fig:ILC_weights}, the weights for the map is such that it provides no information to the final data product. The characteristics of the surveys (noise properties, beam, frequency bands etc.) are given in Table \ref{tab:tSZ:surveyspec}. We now detail our forward modelling procedure for the tSZ field:

\begin{table}[]
    \centering
    \begin{tabular}{ccccccccc}
         & 30 & 40 & 90 & 150 & 220 & 270 & 353 & 545 \\
    \hline
    \multicolumn{9}{c}{SPT-3G}\\
    \hline
    $\theta_{\rm FWHM}$ & & 1.7 & & 1.2 & 1.0 & & & \\
    $\sqrt{N_{\rm white}}$ & & 3.0 & & 2.2 & 8.8 & & & \\
    $\alpha_{\rm knee}$ & & 3.0 & & 4.0 & 4.0 & & & \\
    $\ell_{\rm knee}$ & & 1200 & & 2200 & 2300 & & & \\
    \hline
    \multicolumn{9}{c}{Simons Observatory (SO)}\\
    \hline
    $\theta_{\rm FWHM}$ & 7.4 & 5.1 & 2.2 & 1.4 & 1.0 & 0.9 & & \\
    $\sqrt{N_{\rm white}}$ & 71 & 36 & 8 & 10 & 22 & 54 & & \\
    $\sqrt{N_{\rm atm}}$ & 14.9 & 9.3 & 22.6 & 57.6 & 194 & 262 & & \\
    $\alpha_{\rm knee}$ & \multicolumn{8}{c}{3.5}\\
    $\ell_{\rm knee}$ &  \multicolumn{8}{c}{1000}\\
    \hline
    \multicolumn{9}{c}{\textit{Planck}}\\
    \hline
    $\theta_{\rm FWHM}$ & & & 9.69 & 7.30 & 5.02 & & 4.94 & 4.83 \\
    $\sqrt{N_{\rm white}}$ & & & 77.4 & 33.0 & 46.8 & & 153.6 & 818.2 \\
    \end{tabular}
    \caption{The noise and beam properties of the different surveys in different frequency bands (in units of $\GHz$). For \textit{Planck}, we list the 100, 143, and 217 properties under the 90, 150, 220 columns for brevity; our actual analysis makes \textit{Planck} frequency maps at 100, 143, and 217 $\GHz$. The $\alpha_{\rm knee}$ and $\ell_{\rm knee}$ of SO are constant across all bands. For SPT-3G, $N_{\rm atm} = N_{\rm white}$, and for \textit{Planck} $N_{\rm atm} = 0$. All noise estimates are in micro-kelvin-arcminute units. The beam full-width half-max, $\theta_{\rm FWHM}$, is given in arcminutes. The values for SPT-3G and SO are taken from \citet[][see their Table 1 and 2]{Srini:2022:tSZNoise} while the values for \textit{Planck} are from \citet[][see their Table 1]{Matt:2020:tSZACTDR4}. See references therein for how these values are derived.}
    \label{tab:tSZ:surveyspec}
\end{table}

\textbf{True tSZ field.} Our simulated tSZ emission is obtained by pasting pressure profiles onto a \textsc{HealPix} map. We use all halos in the lightcone to do so. The pasted profiles are already integrated over the line of sight. We therefore only need to change the units of the map through the simple rescaling,
\begin{equation}
    y(\nhat) = \frac{\sigma_T k_B}{m_ec^2} P_e(\nhat)
\end{equation}
where $P_e(\nhat)$ is the projected electron pressure in direction $\nhat$, $\sigma_T$ is the Thomson cross-section, $k_B$ is the Boltzmann constant, and $m_e c^2$ is the rest energy of an electron. As mentioned above, the raw dataproducts of a CMB survey are the temperature maps at different frequencies. We can convert the tSZ signal into these frequencies via,
\begin{equation}
    \Delta T(\nu, \nhat) = f_{\rm SZ}(\nu) T_{\rm cmb} y(\nhat)
\end{equation}
where the frequency response of the tSZ signal, $f_{\rm SZ}$, is given as,
\begin{equation}
    f_{\rm SZ}\bigg(x \equiv \frac{h\nu}{k_BT_{\rm cmb}}\bigg) = x \coth(x/2) - 4.
\end{equation}
Here, $h$ is the Planck constant and $T_{\rm cmb} = 2.7255 K$ is the temperature of the CMB.

\textbf{Foregrounds.} As we discussed previously, the temperature map at a given frequency has contributions from multiple fields, normally called ``foregrounds''. We follow \citet{Srini:2022:tSZNoise} in modelling the CMB, radio, and infrared foregrounds. The kinematic SZ emission is ignored as it is subdominant, relative to these foregrounds, by more than an order of magnitude and is thus negligible for our work \citep[][see their Figure 3]{George:2015:foregrounds}. We generate Gaussian realizations of all these foregrounds, by modelling their harmonic power spectra and using the \texttt{synfast} routine in \textsc{HealPy} to make full-sky realizations. The lensed $C_\ell^{\rm CMB}$ is modelled using \textsc{Camb} run at the best-fit cosmology from \textit{Planck} \citep{Planck2016CosmoParams}. 

\noindent In differential temperature units ($\Delta T$), the CMB template has no frequency dependence. The radio and infrared foregrounds follow \citet{George:2015:foregrounds}, and can be written as
\begin{align}
    C_{\ell, \nu_1, \nu_2}^{\rm Radio} & = A^{\rm Radio}_{G15}F^{\rm Radio}_{\nu_1, \nu_2}\bigg(\frac{\ell}{3000}\bigg)^2\bigg(\frac{2\pi}{\ell(\ell + 1)}\bigg)\label{eqn:CellRadio}\\
    C_{\ell, \nu_1, \nu_2}^{\rm CIB, poiss.} & = A^{\rm CIB, poiss.}_{G15}F^{\rm CIB, poiss.}_{\nu_1, \nu_2}\bigg(\frac{\ell}{3000}\bigg)^2\bigg(\frac{2\pi}{\ell(\ell + 1)}\bigg)\label{eqn:CellCIBP}\\
    C_{\ell, \nu_1, \nu_2}^{\rm CIB, clus.} & = A^{\rm CIB, clus.}_{G15}F^{\rm CIB, clus.}_{\nu_1, \nu_2}\bigg(\frac{\ell}{3000}\bigg)^{0.8}\bigg(\frac{2\pi}{\ell(\ell + 1)}\bigg)\label{eqn:CellCIBC}
\end{align}
The radio background is modelled as having a poisson spatial distribution (and is therefore essentially constant in $\ell$), while the CIB is modelled as a combination of a poisson-distributed component and a clustered component. The frequency response function, $F_{\rm \nu_1, \nu_2}$ is defined as,
\begin{align}
    \epsilon_{\nu_1, \nu_2} & = \bigg(\frac{B^\prime(\nu_0) B^\prime(\nu_0)}{B^\prime(\nu_1)B^\prime(\nu_2)}\bigg)\bigg|_{\rm T = T_{\rm cmb}}\label{eqn:epsnu}\\
    F^{\rm Radio}_{\nu_1, \nu_2} & = \epsilon_{\nu_1, \nu_2}\bigg(\frac{\nu_1\nu_2}{\nu_0^2}\bigg)^{-\alpha}\label{eqn:Fnu:radio}\\
    F^{\rm CIB, x}_{\nu_1, \nu_2} & = \epsilon_{\nu_1, \nu_2}\bigg(\frac{\nu_1\nu_2}{\nu_0^2}\bigg)^{\beta^x}\bigg(\frac{B(\nu_1)B(\nu_2)}{B(\nu_0)^2}\bigg)\bigg|_{\rm T = T_{\rm cib}}\label{eqn:Fnu:CIB}
\end{align}
where $B$ is the blackbody function, $B^\prime \equiv dB/dT$ its derivative with temperature, and $x \in \{\rm poiss.\,, clus.\}$. The function $\epsilon_{\nu_1, \nu_2}$ converts spectral energies to differential temperature ($\Delta T$) units. The radio component is modelled with a power-law frequency dependence, while the CIB is a modified blackbody where the modification is to the low-frequency, Rayleigh-Jeans regime of the distribution. For the CIB model in Equation \eqref{eqn:Fnu:CIB}, we use $\beta^{\rm poiss.} = 1.505$, $\beta^{\rm clus.} = 2.510$, $T_{\rm cib} = 20K$ \citep{George:2015:foregrounds}. For the radio foreground, we use $\alpha = -0.6$, following \citet{Srini:2022:tSZNoise}. 

\noindent Note that our goal in including foregrounds is only to obtain a reasonably accurate noise term for the tSZ maps. As a result, we follow \citet{Srini:2022:tSZCluster, Srini:2022:tSZNoise} in modelling these foregrounds as uncorrelated Gaussian fields, whereas in actuality these fields will be correlated. Such correlations were found to be negligible for analyses of clusters \citep[][see their Section 3.6]{Srini:2022:tSZCluster}, which dominate the information in the tSZ field. However, a more realistic forward model of the tSZ field will need to consistently connect the foregrounds to the density field and halo catalog in a given simulation realization; see \citet{Omori2022Agora}, for an example of a robust, self-consistent model.

\textbf{Instrumental and Atmospheric noise.} The primary noise contributor to the tSZ maps are the thermal noise of the detector and atmospheric noise (for ground-based surveys). This is modelled through a two-term noise component,
\begin{equation}\label{eqn:tSZ:Noise}
    N_\ell = N_{\rm white} + N_{\rm atm}\bigg(\frac{\ell}{\ell_{\rm knee}}\bigg)^{-\alpha_{\rm knee}}.
\end{equation}
The first term is the frequency-independent, ``white noise'' component arising from detector noise. The second term is a $1/f$-noise that increases towards large-scales. The values we use for the different experiments are given in Table \ref{tab:tSZ:surveyspec}. For \textit{Planck}, we set $N_{\rm atm} = 0$.\footnote{Note that \textit{Planck} still exhibits some $1/f$-noise (particularly on $\ell < 100$) arising from other, non-atmospheric components. For the purposes of this work, we ignore this term.} Equation \eqref{eqn:tSZ:Noise} is the \textit{beam-convolved} noise estimate, and to get the true noise model we deconvolve Equation \eqref{eqn:tSZ:Noise} as $N_\ell \rightarrow N_\ell/B_\ell^2$, where the beam $B_\ell$ is defined below.

\textbf{Instrument Beam.} The point-spread function (or beam) varies by experiment and detector. The values for the experiments we consider are given in Table \ref{tab:tSZ:surveyspec}. We use Gaussian beams,
\begin{equation}\label{eqn:tSZ:Beam}
    B_\ell = \exp\bigg(\frac{\sigma_{\rm FWHM}^2}{2}\ell(\ell + 1)\bigg)
\end{equation}
with $\sigma_{\rm FWHM} = \theta_{\rm FWHM}/\sqrt{8\ln(2)}$. Here $\theta_{\rm FWHM}$ is the beam full-width half-max in radians.

\textbf{Linear Combination (LC):} Finally, the (beam-deconvolved) frequency maps are summed together through an LC, similar to that of \citet{Bleem:2022:tSZ}. In this formalism, the final map is,\footnote{Under the assumption of uncorrelated Gaussian foregrounds and instrumental/atmospheric noise, the residual noise in the tSZ field is given by the spectra $N^{yy}_\ell = \boldsymbol{f}_{\rm SZ}^T C^{-1}_\ell \boldsymbol{f}_{\rm SZ}$, which is the denominator in Equation \eqref{eqn:tSZ:ILC_weights}. One can obtain a realistic, noisy tSZ field by simply generating a Gaussian field with $C_\ell = N^{yy}_\ell$ and adding it to the true tSZ field. However, our forward model still adopts the approach of decomposing the signal into different frequency bins and performing the LC, as this approach is more generalized.}
\begin{equation}
    y^{\rm LC}_{\ell, m} = \sum_i w_\ell(\nu_i) T_{\ell, m}(\nu_i)
\end{equation}
where $T_\ell(\nu_i)$ are (harmonic space) temperature maps at different frequencies and $w_\ell(\nu_i)$ are the LC weights. These weights are the optimal (minimum-variance, or MV) weights that maximimize the tSZ signal, defined as
\begin{equation}\label{eqn:tSZ:ILC_weights}
    w_\ell(\nu_i) = \frac{f^{\nu_i}_{\rm SZ} C^{-1}_\ell}{\boldsymbol{f}_{\rm SZ}^T C^{-1}_\ell \boldsymbol{f}_{\rm SZ}}.
\end{equation}
where $\textbf{f}_{\rm SZ} = \{f^{\nu_0}, f^{\nu_1}, \ldots\}$ specifies that the tSZ is our target signal, and $C^{-1}_\ell$ is the inverse of the covariance matrix between the different frequency maps at a given $\ell$. Following \citet{Bleem:2022:tSZ, Srini:2022:tSZCluster} we model the covariance matrix theoretically as,
\begin{equation}\label{eqn:tSZ:Cov}
    C^{\nu_i, \nu_j}_\ell = N_{\ell, \nu_i, \nu_j}\delta_{ij} + \sum_{a \in X} C^a_{\ell, \nu_i, \nu_j}
\end{equation}
where $N_{\ell}$ is already beam-deconvolved, and $\delta_{ij}$ ensures that instrument and atmospheric noise only affects the autocorrelation of frequency maps.\footnote{Nominally, atmospheric noise should be correlated across frequency bands, however we follow \citet{Srini:2022:tSZCluster, Srini:2022:tSZNoise} in ignoring this.} The frequency and $\ell$-dependence of the foregrounds, $X \in \{\rm CMB, Radio, CIB\}$ are defined above. The emission from these foregrounds is completely correlated across bands. Once this final map, $y^{\rm LC}_{\ell, m}$, is made, we convolve it with a Gaussian beam of $1.7 \arcmin$, which is the final resolution of the map. Note that in practice, we make maps at $\texttt{NSIDE} = 1024$, which already has a larger pixel scale ($3.4 \arcmin$) than the beam.

\textbf{In summary,} we generate true tSZ maps from our baryonification pipeline and decompose the signal into different frequency channels. We then add multiple other foregrounds to these frequency channels, including the CMB, Radio emission, and CIB. We also add the thermal noise from detectors and then the noise due to the atmospheric. The maps are linearly combined using weights that result in the minimum variance of the final tSZ map. The weights we use in this work are shown in Figure \ref{fig:ILC_weights}. We have verified our weights estimation procedure reproduces the results of \citet[][their Figures 3 and 4]{Srini:2022:tSZNoise}, and we also find our dominant frequency band at a given $\ell$ broadly follows that of \citet{Bleem:2022:tSZ};\footnote{We do not expect precise agreement given our approach is still a significantly simpler version of the analysis done in \citet{Bleem:2022:tSZ}, and given that analysis is for a previous release of the SPT data.} for example the $100 \GHz$ \textit{Planck} data dominates at low $\ell$ while the $90\GHz$ SPT data dominates at high $\ell$.

\citet{Omori2022Agora} provide a detailed, and precisely validated, prescription for consistently producing a realistic tSZ field --- including \textit{all} relevant contaminant foregrounds, lensing effects, cross-correlations etc. --- through the use of high-resolution simulation products and data-driven models. The \textsc{Ulagam} suite does not have the required resolution or fidelity of simulation products (for example, we have no merger trees or particle shells) to employ their approach. Instead, our forward model focuses solely on capturing the noise level of the tSZ maps, which is adequate for understanding the different degeneracy directions (for the different moments measurements) of the different baryonification and cosmology parameters.

\section{Impact of projection scale}\label{sec:ModelValidate:Method:projscale}

\begin{figure}
    \centering
    \includegraphics[width=\columnwidth]{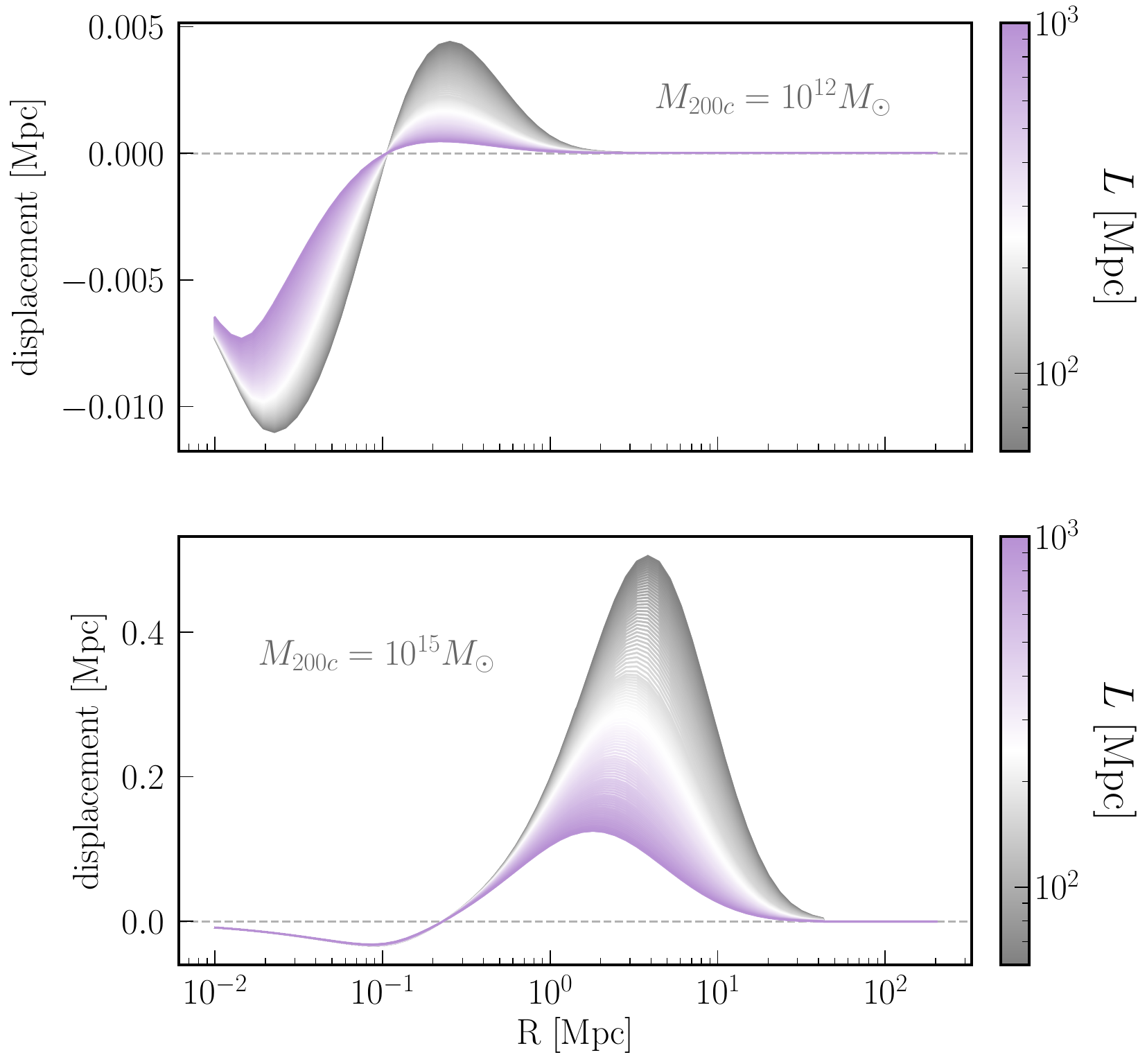}
    \caption{The impact of the chosen projection scale on the projected displacement function for halos of two different masses; see Equation \eqref{eqn:Mass_proj} and Section \eqref{sec:baryonify:density}. Increasing the limits of the line-of-sight projection integral washes out the variations in the density profile arising from baryonic imprints and lowers the amplitude of the displacements.}
    \label{fig:projection:displ}
\end{figure}

We extend on the discussions of methodology validation in Section \ref{sec:ModelValidate} by checking the sensitivity of our predictions to choices in the projection operation. As noted in Section \ref{sec:baryonify:density}, we set the limits of the projection integral according to the simulation box (or more generally, according to the thickness of the volume that we are applying baryonification on). This is in contrast to some previous works that set the the projection scale based on the projected separation $\rp$ (see Equation \eqref{eqn:Mass_proj} for definitions of the quantity). Figure \ref{fig:projection:displ} presents the impact of the projection choices on the estimated projected displacement function. In the limit of an infinite projection length, the impact of baryons will be completely washed out and there is no baryonification required. Figure \ref{fig:projection:displ} shows that increasing the projection length does indeed reduce the predicted displacements from the model. We use a minimum projection scale of $60 \mpc$, which is the width of density shells used in full-sky mocks of the lensing field \citep{Kacprzak2023Cosmogrid, Anbajagane2023Inflation}, and a maximum of $1000 \mpc$, which is twice the projection scale we use for the \textsc{Quijote}.

The displacements on the largest scales are affected most by varying the projection scale (\eg see bottom panel of Figure \ref{fig:projection:displ}) as at these radii the one-halo density profile becomes comparable/subdominant to the large-scale two-halo term. On smaller scales the impact is less prominent. Though, for adequately small halos (\eg, $M = 10^{12} \msol$, see top panel of Figure \ref{fig:projection:displ}), where the two-halo term can dominate over the one-halo term even at smaller radii, the suppression in amplitude is still seen as we increase the projection scale. This result highlights the importance in choosing a consistent, physically meaningful projection scale.

\section{Alternative choices in halo profile model}\label{sec:ModelValidate:Profile}

The halo model described in Section \ref{sec:baryonify} is not a unique parameterization of the different matter components in the halo; there are several modifications to the model that either extend it to enhance accuracy or simplify it to enhance computational cost. We test a few of these modifications below. In all cases, we show the differences at the profile level for brevity and do not present the summary statistics-level consistency checks.

\subsection{Adiabatic contraction of dark matter}\label{sec:ModelValidate:Contract}

\begin{figure}
    \centering
    \includegraphics[width = \columnwidth]{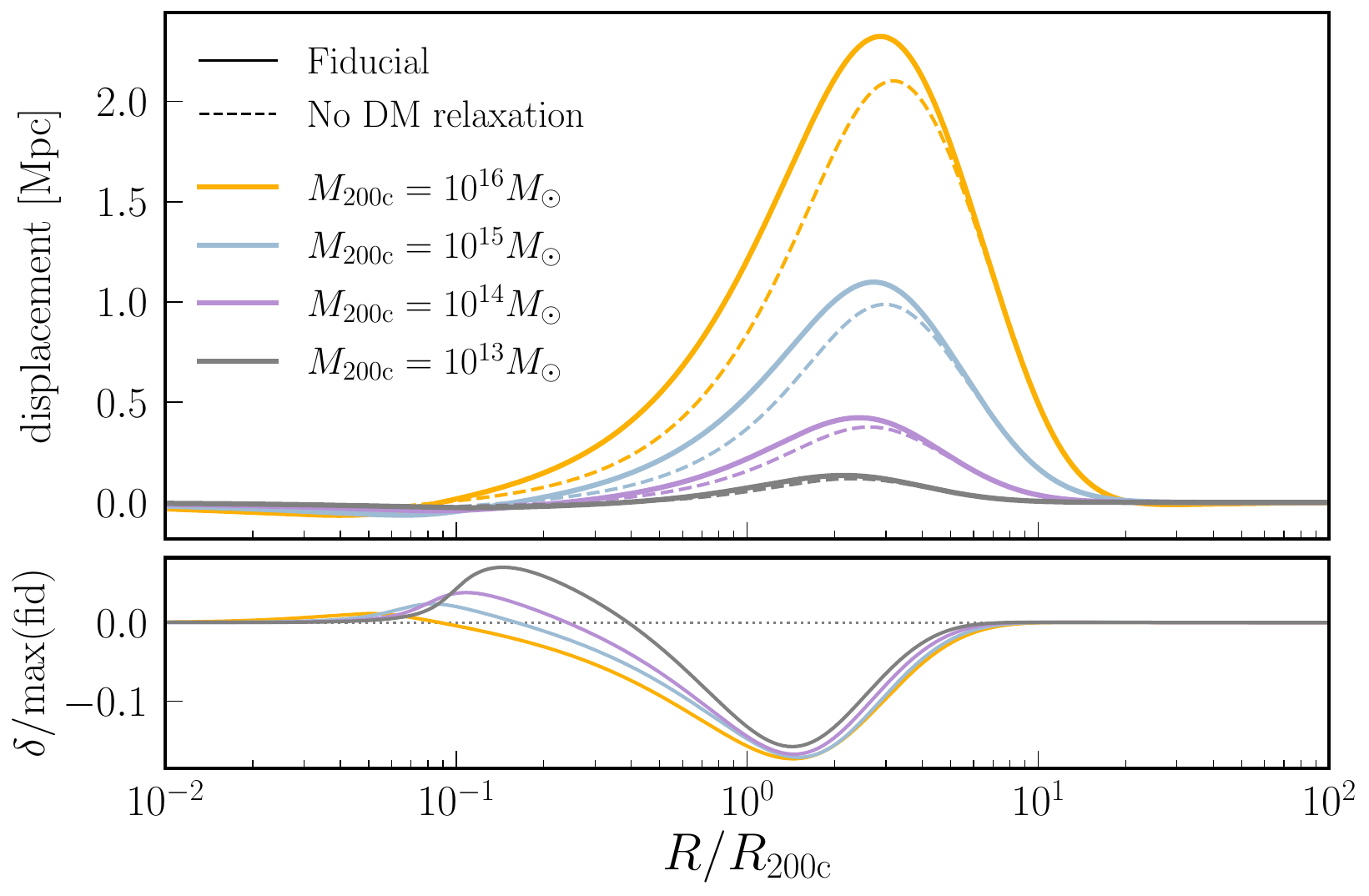}
    \caption{The dependence of the displacement function on the including/excluding the DM adiabatic relaxation. The effect changes the displacement by only 10\%, and thus, this displacement is fairly insensitive to the exact modelling choices of the DM relaxation and is instead highly sensitive to the gas distribution.}
    \label{fig:DMvsGas}
\end{figure}

The collisionless matter component of the baryonification model includes galaxies as well as DM. The distribution of the latter does not follow the same NFW profile as it did in the DMO case; this has been highlighted by many works on a multitude of simulations \citep[\eg][]{Gnedin2004AdiabaticContraction, Duffy2008Concentration, Ragagnin2019HaloConcentration, Beltz-Mohrmann2021BaryonImpactTNG, ForouharMoreno2021BaryonsConcentrationEagle, Anbajagane2022Baryons, Shao2022Baryons,Shao2023Baryons, Sorini:2024:Baryons}. The gravitational response of the DM to the presence of baryons has been studied by many works, and approximated through various (simple) relations. Our baryonification model adopts one such relation from \citet{Gnedin2004AdiabaticContraction, Abadi2010ShapesBaryons, Teyssier:2011:Contraction}. In Figure \ref{fig:DMvsGas} we show the change in the displacement function if we ignore the adiabatic relaxation of the DM. This provides an estimate of the model's sensitivity to inaccuracies in the modelling of this relaxation process. The effect is at most 15\% of the maximum displacement value. Thus, the displacement function is dominated by the form of the gas distribution rather than by the redistribution of the DM due to adiabatic relaxation. This is consistent with findings in hydrodynamical simulations \citep[\eg][see their Figure 9]{Springel2018FirstClustering}.

\subsection{Sensitivity to non-thermal pressure models}\label{sec:ModelValidate:Nth}

\begin{figure}
    \centering
    \includegraphics[width = \columnwidth]{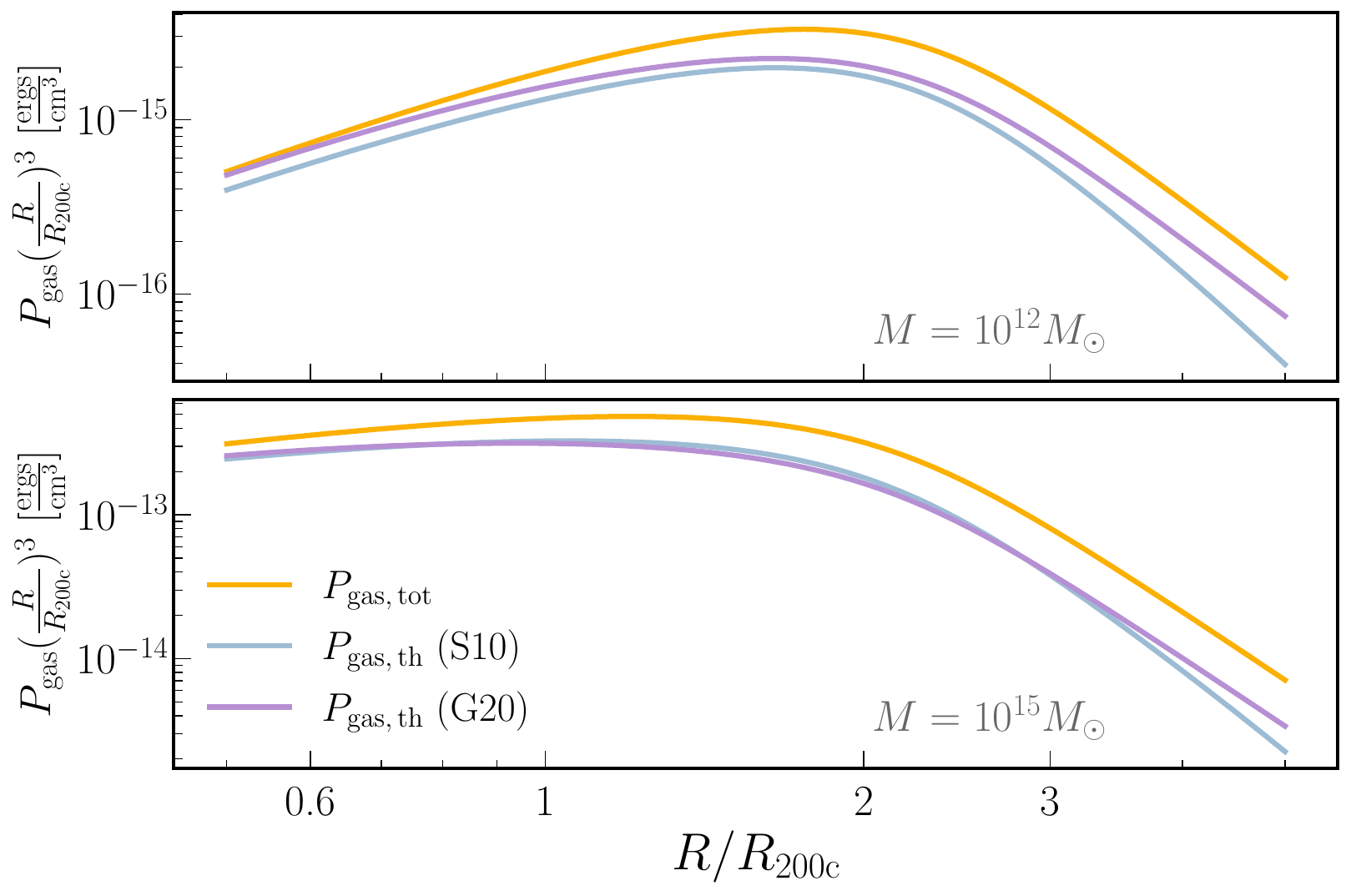}
    \caption{The profile of total gas pressure compared with only the thermal pressure, assuming some model for the non-thermal component. We show predictions from \citet{Green2020Nth}, and for the fiducial baryonification parameters (Table \ref{tab:params}) following the model of \citet{Shaw2010tSZ} for the non-thermal component as described in Section \ref{sec:baryonify:tSZ}.}
    \label{fig:NonthermalProf}
\end{figure}

The tSZ modelling described in Section \ref{sec:baryonify:tSZ} starts by assuming halos are in hydrostatic equilibrium, which means the gas distribution is supported against gravitational collapse solely due to its thermal pressure. However, simulations have shown that order $\sim 10\%$ of the pressure support comes from \textit{non-thermal} pressure, primarily constituted by turbulent motions of the gas, i.e. a velocity dispersion \citep{Nelson2014Nth, Green2020Nth}. We have corrected for this in our predictions by explicitly modelling the contribution of non-thermal pressure and varying/marginalizing the parameters of this model. Our parameterization follows \citet{Shaw2010tSZ}, which is the same choice as \citetalias{Pandey2024godmax}.

Figure \ref{fig:NonthermalProf} compares this model to other alternatives, particularly the predictions of \citet{Green2020Nth}. The latter model is an analytic prediction (see their Figure 1 for a summary of the model) that depends only on the peak height of the halo, and is validated using a set of hydrodynamic simulations from \citet{Nelson2014Nth}. For comparison, we plot thermal gas pressure profiles, using either \citet{Green2020Nth} or our prescription in Section \ref{sec:baryonify:tSZ} to estimate the non-thermal pressure. In the latter approach, predictions are made for fiducial parameter values listed in Table \ref{tab:params}. The phenomenological model, based on \citet{Shaw2010tSZ}, qualitatively matches the form predicted by \citet{Green2020Nth}, which in turn is a advanced version of previous non-thermal models, such as that of \citet{Nelson2014Nth}.

\subsection{Sensitivity to shock features}\label{sec:ModelValidate:Shock}

\begin{figure}
    \centering
    \includegraphics[width = \columnwidth]{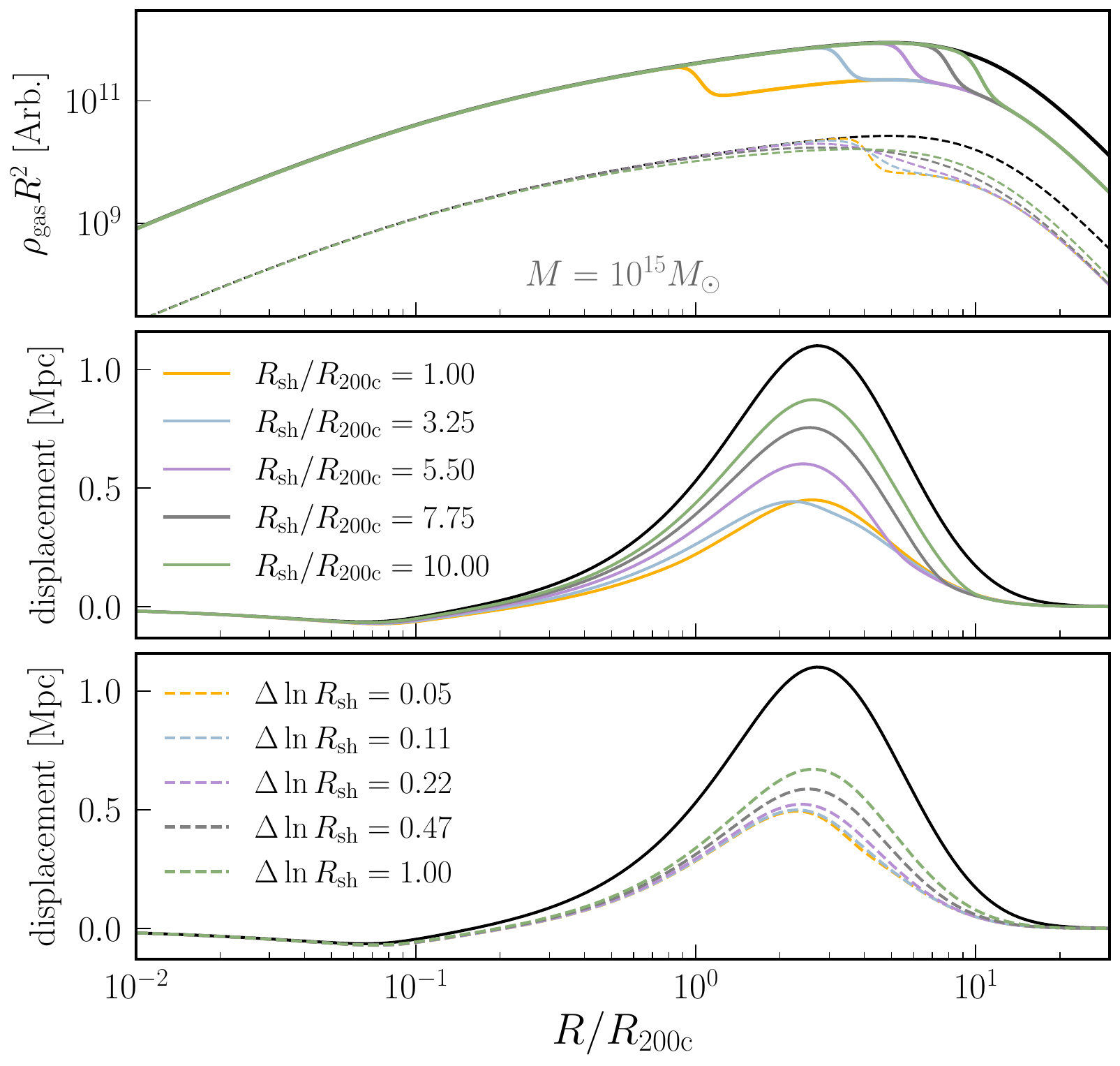}
    \caption{The dependence of the gas density profile (top) and displacement function (middle, bottom) when including features from a large-scale cosmological shock; see Equation \eqref{eqn:shockedgas}. We vary the shock location, $R_{\rm sh}$, and the shock width $\Delta \ln R_{\rm sh}$. In the top panel, the profiles varying the shock width are offset vertically for visual purposes. The bottom two panels show the shock feature changes the displacement quite significantly. However, these effects can be captured through the existing, flexible parameterization of the gas profile. See Section \ref{sec:ModelValidate:Shock} for more details. The results from profiles with no shock features is shown in black.}
    \label{fig:Shocks}
\end{figure}

The gas distribution around massive halos is expected to undergo shock heating due to the generic process of gravitational infall; cold matter from nearby large-scale structure (eg. filaments) infalls into the massive halo. Cold gas has a relatively low sound speed, but the infall velocity can exceed the sound speed by factors of $\mathcal{O}(100)$ for gas around massive halos. This causes the gas to shock heat and thereby imprint features into the gas distribution on large scales. Such features have been predicted by simulations of gas \citep[\eg][]{Quilis1998ShocksSims, Miniati2000ShocksSims, Molnar2009ShocksInSZ, Baxter2021ShocksSZ, Aung2020SplashShock}, and there are indications of these features from measurements at different wavelengths \citep[\eg][]{Hurier2019ShocksSZPlanck, Pratt2021ShocksPlanck,Zhu2021ShockXray, Anbajagane2022Shocks, Anbajagane2023Shocks, Hou2023ShockRadio}.

The change in density due to a cosmological shock can be found following the Rankine-Hugonoit jump conditions \citep{Rankine1870, Hugoniot1887}. For these shocks, which have a very high mach number --- $M \equiv v/c_s \gg 1$, where $v$ is the infall velocity and $c_s$ is the sound speed of the gas --- the density profile post-shock is transformed as $\rho \rightarrow \rho/4$. This estimate corresponds to a gas adiabatic index of $\gamma = 5/3$, which is valid under the assumption that the gas is monoatomic. We then introduce this shock feature through a modified version of the gas profile from Section \ref{sec:baryonify:dmb},
\begin{equation}\label{eqn:shockedgas}
    \rho_{\rm gas, sh}/\rho_{\rm gas} = \bigg(\frac{1 - 0.25}{1 + \exp\big[{\frac{1}{\Delta\ln R_{\rm shock}}\ln(r/R_{\rm shock})}\big]}\bigg) + 0.25.
\end{equation}
The ratio is a simple sigmoid function that asymptotes to $1/4$ for $r \gg R_{\rm shock}$ and $1$ for $r \ll R_{\rm shock}$. Note that in Equation \eqref{eqn:gas}, which defines the gas profile, the normalization of the profile is set by integrating the profile to $r \rightarrow\infty$. Given the shock feature changes the shape of the profile, it would nominally affect this normalization. However, we keep the normalization fixed and only change the large-scale behavior of the profile. This is because the shock feature only affects the physics of the large scales and is largely decoupled from the gas physics internal to the halo. Thus, it is unrealistic to couple large-scale features like shocks to changes in the density throughout the halo (the normalization is a scaling factor for the entire profile and thus changes the density at all scales).

Figure \ref{fig:Shocks} shows the profiles with shocks, and the displacement functions corresponding to such profiles. There is a significant effect both at the profile level and thereby at the displacement level. As discussed before, the feature is prominent at large-scales and so the displacement is affected only on these scales. The small-scale displacement is left largely unaltered. However, the baryonification model is flexible and therefore the large-scale changes to the gas profile can be captured through the gas profile slope parameters, $\gamma$ and/or $\delta$. The impact of such shocks on the pressure profile can also be captured by the non-thermal pressure model, as that model fundamentally captures a ``lack of thermal pressure'' in the halo, which is the signature that will be generated by the shocks \citep[][see their introduction for a description of this process]{Anbajagane2022Shocks}. We leave a dedicated study of shocks in the baryonification model to a future study, and note that Figure \ref{fig:Shocks} highlights the potential impact it can have on the total density distribution within halos.

\section{Validation at higher redshifts}\label{sec:higher_z}

\begin{figure*}
    \centering
    \includegraphics[width = 1.9\columnwidth]{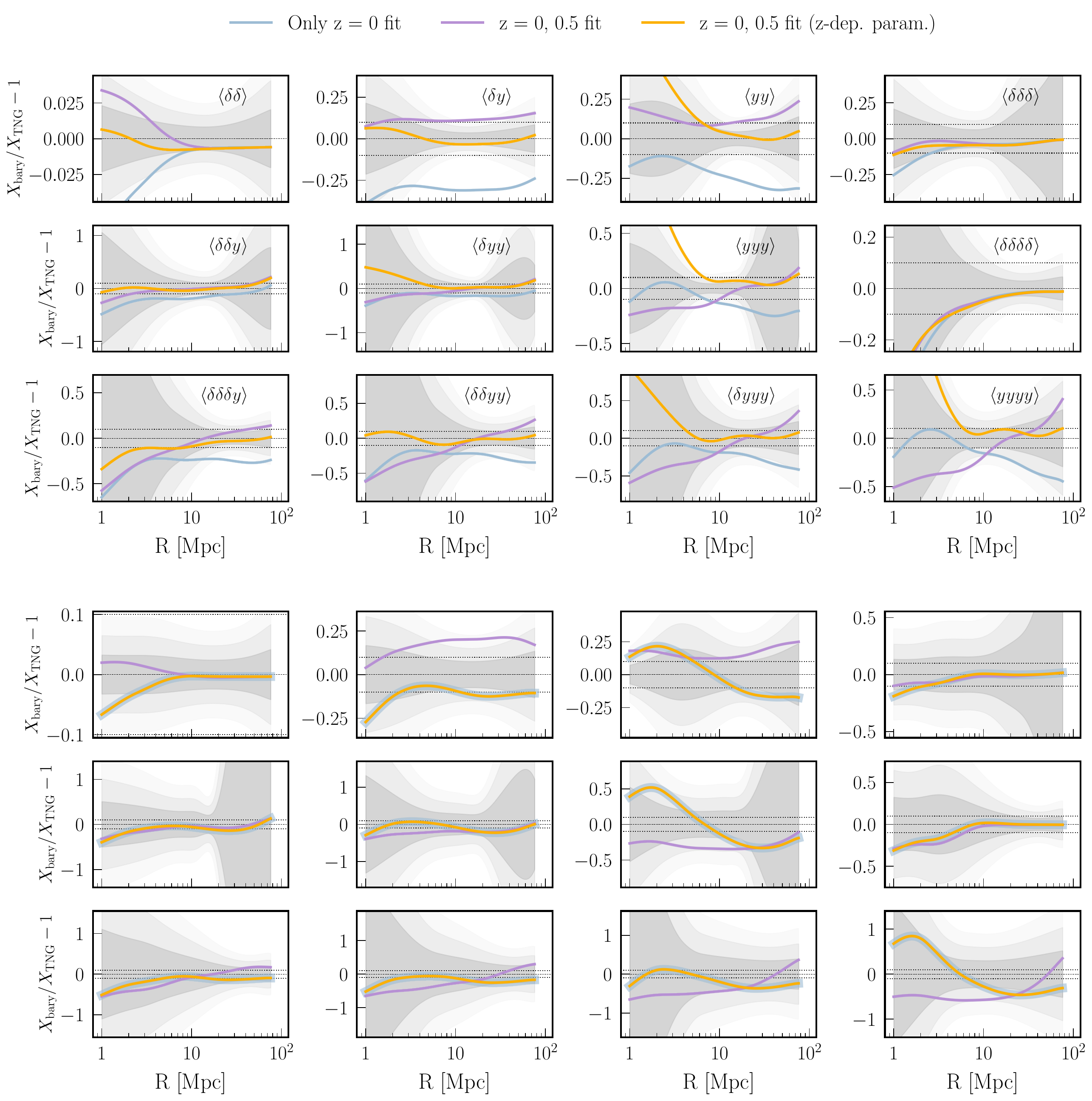}
    \caption{Similar to Figure \ref{fig:TNGFit}, but now showing a variety of a different fitting cases and their predictions for $z = 0.5$ measurements (top three rows) and for the $z = 0$ measurements previously shown in Figure \ref{fig:TNGFit} (bottom). The different curves show fits made to different combinations of redshifts. The purple line (``fixed'') is from a joint fit of $z = \{0, 0.5\}$ with all redshift dependence of the baryonification model set to $\nu_X = 0$; see Equation \eqref{eqn:gasMzcdep}. The yellow line (``varied'') is the same joint fit, but after allowing these values to be non-zero. Both the fixed and varied predictions are a decent fit to the $z = 0.5$ measurements, with the varied one providing the best fit of the two. The fit from using only the $z = 0$ measurements (blue) is shown for comparison and is the worst performing. In the bottom three rows, the blue line overlaps exactly with the yellow and so we show it as a thicker/translucent line to aid visibility.}
    \label{fig:TNGFit_z1}
\end{figure*}

While Section \ref{sec:ModelValidate:Hydro} shows that the baryonification model can fit the simulation measurements at $z = 0$, we must also verify that the model can jointly fit measurements across redshifts. Figure \ref{fig:TNGFit_z1} shows such a joint fit across all moments, up to 4th order, from $z = 0$ \textit{and} $z = 0.5$. 

The fit is done using three models: (i) the fiducial model shown in Figure \ref{fig:TNGFit}, with $\nu_X = 0$ (no redshift dependence in the gas profile parameters) and fit to the $z = 0$ data alone, then (ii) the same fiducial model now jointly fit to $z = 0$ and $z = 0.5$, and finally (iii) the same model but with those redshift-dependence parameters varied, and fit to both $z = 0$ and $z = 0.5$ data. The first model is still a decent match to the simulations, though a few predictions at $z = 0.5$ are discrepant with the simulation beyond the $3\sigma$ bound. The second model, which is a joint fit, is a better match across both redshifts. The predictions are almost always within $1-2\sigma$ of the datapoints. Finally, the third model, which now varies two redshift-dependence parameters, $\nu_{\theta_{\rm ej}}$ and $\nu_{M_{\rm c}}$, provides a similar/better match depending on the subset of measurements we focus on. There are some potential discrepancies in the higher order moments of the tSZ field, but the discrepancies are still within the $2\sigma$ uncertainties. The fractional uncertainties of the higher redshift measurements --- particularly those involving the tSZ field, or higher-order combinations of any fields --- can be larger due to the weaker signal at higher redshift. The statistical limitations of small simulation volumes, which we had discussed earlier in Section \ref{sec:ModelValidate:Hydro}, are only exacerbated at higher redshift as there are now even fewer massive halos generating the signals.

It is interesting to note that an adequate fit is achieved even when fixing the redshift-dependence parameters, $\nu_X$. This complements the results of Section \ref{sec:Forecast}, and Table \ref{tab:Fisher} in particular, which showed that all versions of our analysis were poorly constraining these parameters. We have thus seen two different analyses show that the $\nu_X$ parameters may not be a necessary feature of the model when applying baryonification to survey data. \citetalias{Pandey2024godmax} found these parameters necessary for fitting the profiles of halos across multiple redshifts. However, it may be that for summaries of the whole lensing/tSZ field measured in surveys, the mass-dependence is adequate enough without the need for any additional redshift-dependence as well. However, one limitation of our analysis here is it spans a relatively narrow range in redshift, $0 < z < 0.5$ (though our Fisher forecast in Section \ref{sec:Forecast} uses $0 < z < 3.5$) and fitting simulation measurements across a broader range could once again necessitate the redshift-dependence parameters in this particular analyses. Further analyses, using other simulations, is needed to better confirm the necessity of these parameters for analyses of survey data.

In conclusion, we find the model specified in Section \ref{sec:baryonify} adequately captures the moments of the lensing and tSZ fields, from 2nd to 4th order, and across multiple redshifts. As mentioned before, more precise validation will be enabled through the use of larger simulations. 

\section{Impact of parameter set of Fisher information}\label{appx:fisher:params}

A number of previous works have constrained the baryonification parameters using data \citep[\eg][]{Chen:2023:BaryonsWL, Arico2023BaryonY3, Grandis2023Baryons, Bigwood:2024:BaryonsWLkSZ}, and they span a variety of choices in the parameters being varied. For example \citet{Chen:2023:BaryonsWL} vary a single parameter, $M_c$, whereas \citet{Arico2023BaryonY3} varies six parameters. Additionally, \citet{Giri2021Baryon} explored the minimal set of parameters needed for baryonification to still predict the matter power spectrum from different simulations.

\begin{figure}
    \centering
    \includegraphics[width=1\linewidth]{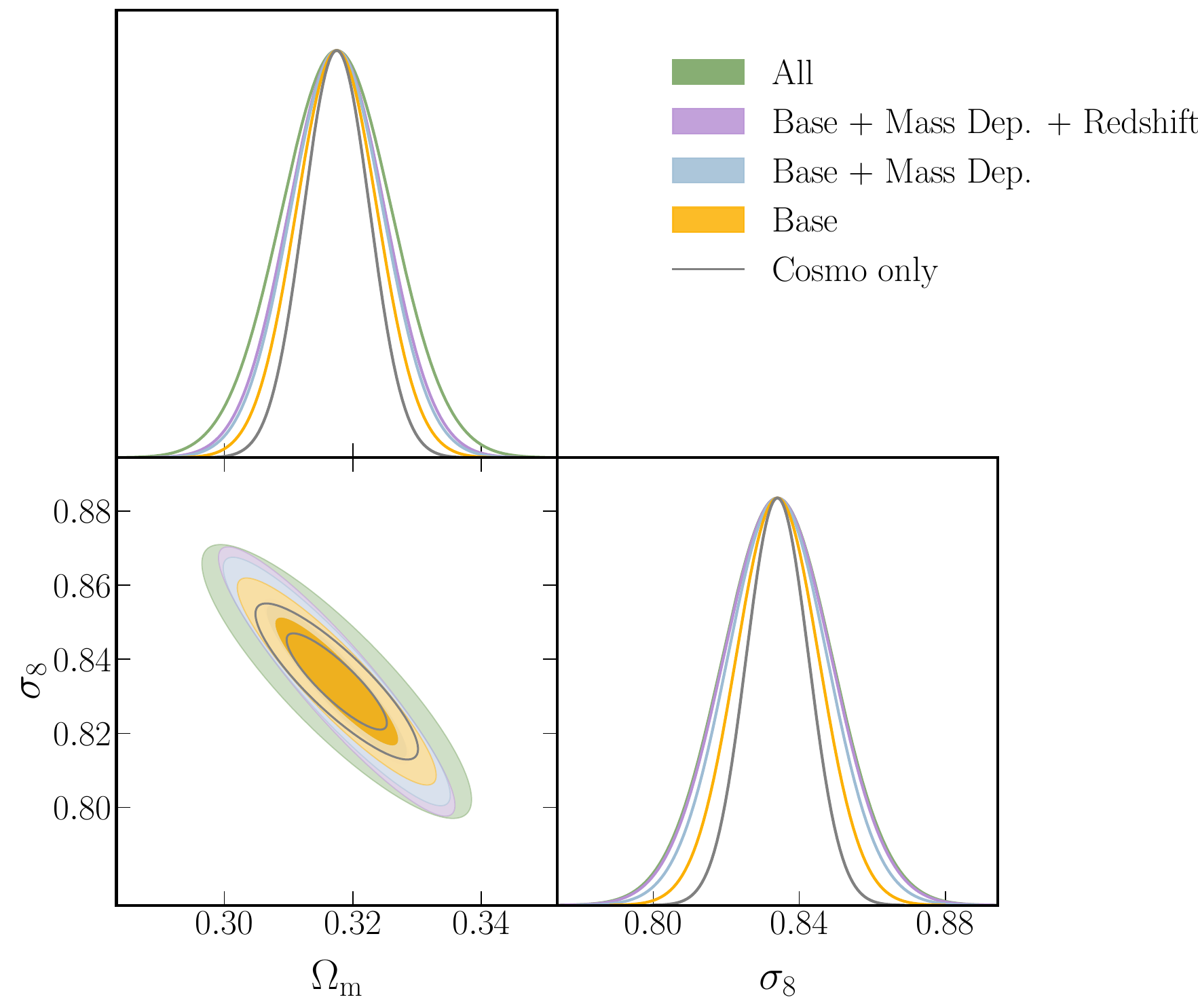}
    \caption{Constraints on two cosmology parameters as a function of the number of baryonifciation parameters varied in the analysis. We forecast for a DES Y6 x SPT 3G analysis, using the 2nd and 3rd moments of the fields. The constraints, and the number of baryonification parameters varied in each, are listed in Table \ref{tab:Fisher:ParamCompare}. The inclusion of all 11 baryonification parameters leads to a $\approx 60\%$ degradation in cosmology constraints.}
    \label{fig:Fisher:Paramcompare}
\end{figure}

Here, we show our Fisher constraints on cosmology using different sets of baryonification parameters. Ordered by increasing complexity and number of parameters, we consider: varying no baryonification parameters, varying only $M_c, \theta_{\rm ej}$ (base), varying the two mass-dependence parameters $\mu_X$, the two redshift dependence parameters $\nu_X$, and varying the four remaining parameters. See Table \ref{tab:params} for a complete list of all parameters we vary. Figure \ref{fig:Fisher:Paramcompare} shows the results of this lensing plus tSZ analysis. As expected, varying all parameters degrades cosmology constraints by roughly around $60\%$. Table \ref{tab:Fisher:ParamCompare} also lists the constraints for cosmology, as well as $M_c, \theta_{\rm ej}$, alongside the number of baryonification parameters varied per analysis.

\begin{table}[]
    \centering
    \begin{tabular}{ccccc}
        Analysis & $\Omega_{\rm m}$ & $\sigma_8$ & $M_c$ & $\theta_{\rm ej}$\\
        \hline
        Cosmo only (0) & 0.005 & 0.009 & -- & -- \\
        Base (2) & 0.006 & 0.011 & 0.102 & 0.174 \\
        Base + $M$-dep (4) & 0.007 & 0.014 & 0.132 & 0.213 \\
        Base + $M$-dep + $z$-dep (6) & 0.008 & 0.015 & 0.316 & 0.217 \\
        All (10) & 0.009 & 0.015 & 0.554 & 0.478 \\
        \hline
    \end{tabular}
    \caption{Constraints on two cosmology parameters, and the two key baryonification parameters (for lensing) as a function of the number of baryonifciation parameters varied in the analysis (also denoted with the brackets). We forecast for a DES Y6 x SPT 3G analysis, using the 2nd and 3rd moments of the fields.}
    \label{tab:Fisher:ParamCompare}
\end{table}
\section{Computational details}\label{sec:baryonify:Compute}

There are a number of salient computational choices and details in the baryonification model of Section \ref{sec:baryonify} that we list here for completeness.

\textbf{Projection:} It is computationally efficient to perform the projection integral in Fourier space, which is the choice made in the \textsc{CCL} code base. However, this increases the likelihood of aliasing, or ``ringing'' effects in the resulting profile, particularly for profiles computed in the extreme regions of parameter space. Such profiles can deviate very significantly from simple power-laws, and in these cases the \textsc{FftLog} algorithm can fail. To make our pipeline more robust, we perform all projection integrals in real-space, 
\begin{equation}\label{eqn:projprofile}
    \rho_p(r_p) = \int_0^{L/2} 2dl \rho\bigg(\sqrt{l^2 + r_p^2}\bigg).
\end{equation}
We find that using the real-space calculation, instead of the Fourier-space one, does not slow the pipeline a noticeable amount. Note that in practice we only compute the profiles once, at the start of the baryonification pipeline, to create a table. This lookup table is then used for generating profiles of all the halos in the simulations (see the ``Tabulation'' discussion below).

\textbf{Profile interpolation:} We use tabulation and interpolation for all integrals/derivatives of the profiles appearing in this work; for example, the volume integral to get the normalization in Equation \eqref{eqn:NFW_norm} and \eqref{eqn:gas_norm}. For almost all interpolation steps (the one exception is described below), we use the \texttt{PchipInterpolator} \citep{Fritsch:1984:PChip, Moler:2004:PChip} in \textsc{Scipy} as it preserves the monotonicity of the function under the presence of outliers. The \texttt{CubicSpline} interpolator, which is a popular choice and also implemented in \textsc{Scipy}, can break monotonicity if the input function has any rapid (but still monotonic) variation and this will then lead to ringing in the interpolated result. While our profiles are all well-behaved and do not contain sharp changes over radius, their \textit{pixel-convolved} versions may present such rapid deviations if the halo radius is far smaller than the pixel-scale. The \texttt{PchipInterpolator} enables the pipeline to adequately handle such extreme edge-cases as well. We use this interpolator to accurately compute integrals of functions, for example when converting density profiles to enclosed mass profiles. 

\noindent We also require interpolation for computing the derivatives ---  primarily the operation in Equation \eqref{eqn:rho_clm} to compute the collisionless matter profile --- and here we \textit{do} use the \texttt{CubicSpline}. This is because this spline guarantees continuous first \textit{and} second derivatives in the interpolated function (whereas \texttt{PchipInterpolator} only guarantees continuous first derivatives). In Equation \eqref{eqn:rho_clm}, we take the derivative of a enclosed mass function in order to obtain a density profile, and if the second derivative of the former function is discontinuous that will cause (unphysical) jumps in the resulting density profile. The \texttt{CubicSpline} prevents this by enforcing smooth second derivatives in the interpolator of the enclosed mass.

\textbf{Tabulation:} For all of our work here, we use a \textit{tabulated} profile to speed up our calculations. These profiles (or the displacement function) are computed over some specified range in $\log_{10}r$, $\log_{10}\Mtwohc$ and $\log_{10}(1 + z)$. When computing the dependence of the profiles on external parameters, such as concentration $\ctwohc$ or the projection distance $L_{\rm proj}$ (see Figures \ref{fig:c200c:displ} or \ref{fig:projection:displ}), we extend the tabulation to include this parameter. These tables are defined on regular grids, and we use simple linear interpolation of the table through the \texttt{RegularGridInterpolator} routine in \textsc{Scipy}. This approach greatly increases the computational speed of the model predictions, and leads to negligible drops in accuracy given all quantities (profiles, displacement functions etc.) are smoothly varying across radius, mass, and redshift. Grids of $\mathcal{O}(10^5)$ points (1000 points in $\log_{10}r$, 10 in $\log_{10}\Mtwohc$, and 10 in $\log_{10}(1 + z)$) take less than five minutes to build, on a single-threaded process of an Intel Broadwell CPU, and result in interpolators that are accurate to within $<0.1\%$.

\textbf{Computing runtimes:} The baryonification process is ``embarassingly parallel'' given the contributions from halos are accumulated through a loop, and each iteration in the loop can be evaluated completely disjoint from the rest. Our pipeline is parallelized across all cores in a node using the \textsc{Joblib} routines, i.e. we have implemented shared-memory parallelization. For the standard full-sky simulation used in this work --- 100 density shells of $\texttt{NSIDE} = 1024$ and halo catalogs with $M > 10^{14}\msol$; see Section \ref{sec:sims:Ulagam} --- baryonification of the density field is done in under 2 minutes on an Intel Broadwell chip with 40 cores. The profile pasting step, which uses approximately two million halos across $0 < z < 3.5$, generates the tSZ field in under 2 minutes as well. The runtime scales linearly with the number of halos in the simulation volume as well as the number of shells and number of pixels. If we used halo catalogs of $M > 10^{13}\msol$, the run time would increase by a factor of two to three; a halo catalog with $M > 10^{13}\msol$ has $10\times$ more halos than one with $M > 10^{14}\msol$ but these additional, less-massive halos will span a smaller area on the sky which in turn will reduce the computational load.


\label{lastpage}
\end{document}